\documentclass[12pt]{article}
\newlength{\abstractwidth}
\input epsf
\renewcommand{\thefootnote}{\fnsymbol{footnote}}
\renewcommand{\thanks}[1]{\footnote{#1}} 
\newcommand{\starttext}{
\setcounter{footnote}{0}
\renewcommand{\thefootnote}{\arabic{footnote}}}

\newcommand{\be}{\begin{equation}}
\newcommand{\bea}{\begin{eqnarray}}
\newcommand{\eea}{\end{eqnarray}}
\newcommand{\ee}{\end{equation}}
\newcommand{\N}{{\cal N}}
\newcommand{\<}{\langle}
\renewcommand{\>}{\rangle}
\def\ba{\begin{eqnarray}}
\def\ea{\end{eqnarray}}

\makeatletter
\newcounter{fig}
\renewcommand\thefig{\arabic{fig}}

\def\fps@fig{tbp}
\def\ftype@fig{1}
\def\ext@fig{lof}
\def\fnum@fig{\figurename~\thefig}
\newenvironment{fig}
               {\@float{fig}}
               {\end@float}
\newenvironment{fig*}
               {\@dblfloat{fig}}
               {\end@dblfloat}

\def\Im{{\rm Im}}
\def\tr{{\rm tr}}
\def\det{{\rm det}}
\def\sdet{{\rm sdet}}

\def\half{ {1\over 2}}
\def\12{{\scriptstyle {1 \over 2}}}
\def\p{\partial}
\def\tet{\vartheta}
\def\ep{{\varepsilon}}
\def\A{{\cal A}}

\def\D{{\cal D}}

\def\O{{\cal O}}
\def\M{{\cal M}}
\def\N{{\cal N}}
\def\Z{{\cal Z}}
\def\bZ{{\bf Z}}
\def\bR{{\bf R}}
\def\Dslash{{D \! \! \! \! /}}

\def\chiz{\chi _{\bar z} ^+}
\def\chiw{\chi _{\bar w} ^+}
\def\no{\nonumber}

\begin{document}
\starttext
\baselineskip=18pt
\setcounter{footnote}{0}

\begin{flushright}
UCLA/03/TEP/31 \\
Columbia/Math/03 \\
15 December  2003
\end{flushright}

\bigskip

\begin{center}
{\Large \bf TWO-LOOP SUPERSTRINGS } \\
\bigskip
{\Large \bf  ON ORBIFOLD COMPACTIFICATIONS}
\footnote{Research supported in part by a Grant-in-Aid from the Ministry
of Education, Science, Sports and Culture and grants from Keio University 
(K.A.),  and by  National Science Foundation
grants PHY-01-40151 (E.D.) and DMS-02-45371 (D.P.).}

\bigskip\bigskip

{\large Kenichiro Aoki$^a$, Eric D'Hoker$^b$ and D.H. Phong$^c$} \\

\bigskip

$^a$ {\sl Hiyoshi Department of Physics} \\
{\sl Keio University, Yokohama 223-8521,  Japan} \\
$^b$ {\sl Department of Physics and Astronomy }\\
{\sl University of California, Los Angeles, CA 90095, USA} \\
$^c$ {\sl Department of Mathematics} \\
{\sl Columbia University, New York, NY 10027, USA}

\end{center}

\bigskip\bigskip

\begin{abstract}

The two-loop chiral measure for superstring theories compactified
on $\bZ_2$ reflection orbifolds  is constructed from first
principles for even spin structures. This is achieved by a
careful implementation of the chiral splitting procedure in the twisted
sectors and the identification of a subtle worldsheet supersymmetric
and supermoduli dependent  shift  in the Prym period.
The construction is generalized to compactifications
which  involve more general  NS backgrounds preserving
worldsheet  supersymmetry. The  measures are unambiguous
and independent of the gauge slice.

 \medskip

Two applications are presented, both to superstring
compactifications where 4 dimensions are $\bZ_2$-twisted and
where the GSO projection involves a chiral summation  over spin structures.
The first is an orbifold by a single $\bZ_2$-twist; here, orbifolding  
reproduces
a supersymmetric theory and it is shown that its  cosmological constant  indeed 
vanishes.
The second model is of the type proposed by Kachru-Kumar-Silverstein and
additionally imposes a $\bZ_2$-twist by the parity of worldsheet fermion 
number;
it is shown here that the corresponding
cosmological constant does not vanish pointwise on moduli space.

\end{abstract}
\vfill\eject

\baselineskip=15pt
\setcounter{equation}{0}
\setcounter{footnote}{0}

\section{Introduction}
\setcounter{equation}{0}

Over the past few years, important progress has been made on two-loop
superstring perturbation theory.
A formula for the two-loop even spin structure superstring
measure has been constructed which is well-defined and independent of  any
choice of gauge slice \cite{I,II,III,IV,dp02}. The independence of gauge
slice choices is essential, since without it the superstring scattering
amplitudes would be ambiguous. The formula has been applied to the
evaluation of the two-loop $N$-point function for $N\leq 4$ massless
external NS-NS  string states, propagating in flat Minkowski space-time.   
In particular, the cosmological constant for flat Minkowski space-time had
been shown to vanish point by point on moduli space. 

\medskip

The original derivation of these results in
\cite{IV} was carried out using a gauge where the worldsheet gravitini are
supported at two points $q_1,q_2$ which obey the gauge condition,
$S_{\delta}(q_1,q_2)=0$, where $S_\delta$ is the Szeg\"o kernel (or
fermion propagator) for even spin structure~$\delta$.
Recently, the results of \cite{III} (which were expressed  in a general 
gauge where the measure is  formulated in terms of
meromorphic functions and forms)
were used to evaluate the superstring measure in a hyper-elliptic
gauge by Zheng, Wu and Zhu in \cite{zhu,zhu1}, 
and the vanishing of the $N\leq3$-point functions was confirmed.
The cancellations of these amplitudes are examples of non-renormalization
conjectures suggested by space-time supersymmetry \cite{martinec}.
A  bibliography of earlier work on this topic may be found in~\cite{IV}. 

\medskip

The purpose of the present paper is to derive the 
superstring measure (and the cosmological constant) for orbifold
compactification backgrounds produced by
general $\bZ_2$ reflections.
Certain orbifold models of this type provide some of the 
simplest solvable examples of
superstring theories with broken supersymmetry and are therefore of
considerable interest in particle physics. In particular, Kachru, Kumar and
Silverstein \cite{KKS} have proposed ${\bf Z}_2 \times {\bf Z}_2$ orbifolds
of Type II superstring theory
with broken supersymmetry (see also \cite{KKS1}), whose 1-loop cosmological 
constant vanishes by construction.
Clearly, the construction of any superstring
theory model with broken supersymmetry and vanishing (or very small)
cosmological constant would be of great phenomenological value 
(see also \cite{ferrara}).

\medskip

In \cite{KKS}, arguments were presented in favor of the vanishing of
the cosmological constant to two-loop order (and higher), using the earlier
string measure suggested by the picture changing operator and BRST
formalism \cite{fms,vv1}.
Shortly thereafter, arguments were given against the vanishing
of the two-loop cosmological constant in the same model \cite{ksiz},
also using the picture changing operator and BRST formalism.
In both approaches, special choices for the worldsheet
gravitini insertion points were made, the first in unitary gauge,
the second following \cite{iz}.
By now, however, the picture changing operator and BRST formalism
is well-known to be  gauge slice dependent \cite{vv1},
(see also \cite{ars1,ars2,ln,ams})
a fact that casts doubts over
the conclusions of both \cite{KKS} and \cite{ksiz}.

\medskip

A reliable calculation of the cosmological constant  requires a gauge slice
independent  measure such as the one obtained in \cite{I, II, III, IV} for
flat Minkowski  space-time.

\medskip

In the present paper, the two-loop chiral measure for superstring
theories compactified  on $\bZ_2$ reflection orbifolds  
is constructed from first
principles for even spin structures. This is achieved by a
careful implementation of the chiral splitting procedure in the twisted
sectors and the identification of a subtle worldsheet supersymmetric
and supermoduli dependent  shift  in the Prym period.
The limits of the measure are
checked under various degenerations and are found to agree with 
the behavior expected on physical grounds.

\medskip

The construction is generalized to compactifications
which  involve more general NS backgrounds, and preserve
worldsheet  supersymmetry. The  measures are shown to be 
unambiguous and independent of the gauge slice. 
Our original primary concern  was with the study of
the models of \cite{KKS}. The results obtained in this paper
for the  ${\bf Z}_2$-twisted chiral measure, however, 
will be indispensable  ingredients in the construction of 
the $\bZ_2$-twisted conformal field theory and string amplitudes
to two-loop order  in any orbifold model based on ${\bf Z}_2$  reflections.

 \medskip

Two applications are presented, both to superstring
compactifications where 4 dimensions are $\bZ_2$-twisted and
where the GSO projection involves a chiral summation  over spin structures,
i.e. a summation carried out independently on left and right movers.
The first is an orbifold by a single $\bZ_2$ twist, 
which is known to reproduce a supersymmetric theory;  
it is shown here that its  cosmological constant  vanishes.
The second model is of the type proposed by 
Kachru-Kumar-Silverstein \cite{KKS} and
additionally imposes a $\bZ_2$ twist by the parity of 
worldsheet fermion number; it is shown here that
the corresponding chiral measure is non-vanishing. 
Therefore, the associated
cosmological constant does not vanish pointwise on moduli space.
It is logically possible, though we shall argue it is unlikely, that despite
this result, a cancellation will occur for the cosmological constant
when left and right chiral blocks are
assembled and the integration over moduli is carried out.

\subsection{Outline of the construction}

In \cite{I,II,III,IV}, a detailed method was laid out for determining 
the two-loop superstring measure on general space-times. 
Explicit formulas were obtained in those papers  for the case of 
flat Minkowski space-time $M$; they involve the matter and ghost 
supercurrents, stress tensors, and partition functions.

\medskip

All compactifications discussed in this paper will only affect the 
matter fields and will leave the ghost fields unchanged.
In orbifold compactifications $C$ of flat Minkowski
space-time, the expressions for the  matter supercurrent and  stress 
tensor (in terms of the fields $x^\mu$ and $\psi _\pm ^\mu$)
are identical to those of flat Minkowski space-time and will 
be denoted by $S_{Mm}$ and $T_{Mm}$ (or simply 
$S_{m}$ and $T_{m}$ when no confusion is possible). 
The fields $x^\mu$
and $\psi_\pm ^\mu$ will be summed over untwisted and twisted 
sectors, dependent on $C$. Therefore, the partition function,
and more generally any vacuum expectation value
will depend on $C$; they will be  referred to 
by $Z_C$ and $\< \cdots \>_C$.  (On a more general
compactification background $C$, the expression for the 
supercurrent and stress tensor 
(in terms of $x^\mu$ and $\psi _\pm ^\mu$) will generally  
be  different from their expressions in flat space-time 
and will  be denoted by $S_{Cm}$ and $T_{Cm}$.)

\medskip

Thus, the main new ingredients are the construction of the 
propagators for the ${\bf Z}_2$-twisted fields 
$x^{\mu}$ and $\psi_\pm ^{\mu}$, and the identification of the 
corresponding changes in the string measure. 
These changes reflect the global geometry of the
worldsheet with a quadratic branch cut, and must be treated with care. 
We summarize here the main steps in our derivation. 

\medskip

$\bullet$ 
The first step is to determine the ${\bf Z}_2$-twisted bosonic 
propagator $\langle\p_zx(z) \p_wx(w)\rangle_\ep$ 
for given twist $\ep$. Its explicit expression will be given in terms of the
familiar Szeg\"o kernel, as well as  the Prym period $\tau_{\ep}$ 
and the Prym differential $\omega_{\ep}$ for twist $\ep$.
The propagator for the ${\bf Z}_2$-twisted fermionic field $\psi_+$ 
was identified in \cite{KKS}.

\medskip

$\bullet$ 
The next step is to carry out chiral splitting for the $\bZ_2$-twisted matter 
superfields.   Since the ${\bf Z}_2$-twisted propagators are different 
from those of flat space-time, the effective rules, derived in 
\cite{superanom,dp89} for Minkowski space-time, will not apply directly in this case.
The ${\bf Z}_2$-twisted case has to be worked
out in its own right, and its own global geometric features carefully accounted
for. In fact, it is here that a subtle and crucial supersymmetric correction 
$\Delta\tau_{\ep}$ to the super Prym period makes its appearance;

\medskip

$\bullet$
An important check is the independence of the gauge-fixed 
measure from the choice of gauge slice. In fact, a formula 
will be derived for the two-loop measure in terms of the partition
function, supercurrent and stress tensor correlators for any
compactification that does not affect the ghosts, preserves 
worldsheet supersymmetry and has matter central charge $c=15$.
The resulting measure will be shown to be unambiguous and slice 
independent under these assumptions. 
This result was announced in  \cite{I}.

\medskip

$\bullet$ 
The next task is to express the $\bZ_2$-twisted superstring
measure in terms of  $\tet$-constants. 
The procedure follows  \cite{I,II,III,IV}, but here there is a 
significant new difficulty. 
A new, internal momentum $p_\ep$-dependent, contribution arises 
from the correction $\Delta \tau _\ep$ to the super Prym period.
The explicit evaluation, in terms of $\tet$-constants, of this 
contribution is one of the more arduous steps in the paper.
The resulting chiral superstring measure  depends
on both the  spin structure $\delta$ and the twist $\ep$ and 
is denoted $d\mu _C [\delta; \ep] (p _\ep)$. 

\medskip

$\bullet$ The limit of the chiral measure 
$d\mu _C [\delta; \ep] (p _\ep)$ is computed
for a variety of degenerations and checked versus 
the limiting behavior expected on physical grounds.
The limits will also be needed to investigate the 
behavior of the cosmological constant.

\medskip

$\bullet$
Next, the behavior of the chiral measure 
$d\mu _C [\delta; \ep] (p _\ep)$ under  modular transformations 
is derived.  The full genus 2 modular group 
$Sp(4,\bZ)$ will act on the twist cycle and relate the 
measures for various twists. In carrying out the GSO projection, 
a summation over all spin structures is to be carried out, for fixed twist. 
Therefore, it is the subgroup
$H_\ep $ of all elements in $ Sp(4,\bZ)$ which leave 
$\ep$ invariant which will be of greatest importance 
in determining all possible GSO projections consistent
with modular covariance of the chiral measure.
The group $H_\ep$ is studied in detail; 
it is found to be generated by 5 independent
elements (while $Sp(4,\bZ)$ is generated by 6), 
the various orbits under $H_\ep$ for 
twists and spin structures are calculated, and 
the behavior of $d\mu _C [\delta; \ep] (p _\ep)$
is obtained.

\medskip

$\bullet$ 
As a general principle, the GSO projection is carried out by 
summing over the spin structures $\delta$ with suitable phases.
In practice, for each specific physical model, the conformal blocks 
entering the chiral string measure have to be identified and 
summed over  $\delta$ consistently with modular covariance. 
For the simplest ${\bf Z}_2$-twisted theories, where the 
orbifold group is exactly ${\bf Z}_2$ acting by reflection, 
the conformal blocks are just $d\mu_C[\delta;\ep](p_\ep)$. 
However, for more complicated orbifold
groups such as ${\bf Z}_2\times {\bf Z}_2$ in the 
models of \cite{KKS}, the conformal blocks can be
more subtle, and have to be determined with some care.
The cosmological constant for various asymmetric 
${\bf Z}_2$-orbifold models is then determined by summing 
over the spin structures $\delta$. In particular, 
the cosmological constant in the models of Kachru-Kumar-Silverstein
\cite{KKS} is shown to be non-vanishing pointwise on moduli space.

\subsection{Organization and Summary of Main Formulas}

In section \S 2, standard facts about conformal field theory 
on genus 1 and genus 2 Riemann surfaces as well as
the partition function of a $\bZ_2$-twisted boson \cite{DVV}
are reviewed.  A self-contained  construction is 
presented of the Prym differential $\omega _\ep (z)$, 
the Prym period $\tau_\ep$ and the Schottky 
relations.\footnote{Detailed discussions of  genus 1 $\tet$-functions 
and  genus 2 spin structures,  $\tet$-functions,
modular transformations and Thomae-type formulas
are collected in Appendices A and B respectively.}
The key result in section \S 2 is for the 
$\bZ_2$-twisted boson propagator on a genus 2 Riemann 
surface $\Sigma$. In a canonical homology basis with
cycles $A_I,B_I, I=1,2$, the 16 possible twists may be described
in terms of half-characteristics $\ep=(\ep_I',\epsilon_I'')$. 
The bosonic field $x(z)$ in twist sector $\ep$ satisfies
the following boundary conditions,
\be
x(z+A_I)=(-)^{2\ep_I'}x(z),
\ \ \
x(z+B_I)=(-)^{2\ep_I''}x(z)
\ee
The corresponding ${\bf Z}_2$-twisted propagator
is characterized by the same boundary 
conditions in both variables $z$ and $w$, and
the short distance singularity $(z-w)^{-2}$. 
It is obtained in terms of the Szeg\"o kernel $S_\delta$ for 
characteristic $\delta$, and the normalized 
Prym differential $\omega _\ep$,
\bea
\< \p_z x(z) \p_w x(w) \>_\ep 
= 
S_{\delta _i ^+} (z,w) S_{\delta ^- _i} (z,w)
- 4 \pi i \p_{\tau_\ep} \ln \tet _i (0,\tau_\ep)
\omega _\ep (z) \omega _\ep (w)
\eea

\noindent
The notation here is as follows. Given a twist $\ep \not=0$,
six of the ten even spin structures $\delta$ are such that
$\delta + \ep$ is also an even spin structure. 
These six spin structures may be parametrized by 
$\delta _i ^\pm$, with $i=2,3,4$ and $\delta ^- _i = \ep + \delta ^+_i$
and $\delta ^\pm _i$ modulo $\ep$ clearly map onto the 3 even
spin structures of genus 1. The $\tet_i$ are the corresponding 
genus 1 $\tet$-functions for even spin structures. The precise normalization 
of $\omega _\ep (z)$ is derived in \S 2.6.

\medskip

In section \S 3, the chiral splitting procedure is carefully implemented 
in the twisted sectors of the ${\bf Z}_2$-twisted superstring theory. 
Because of the $\bZ_2$-twisting, only a single internal loop momentum
survives (instead of the two internal loop momenta in the untwisted
sector). This is illustrated in Figure \ref{fig:1}, where 
a quadratic branch cut is applied along the cycle $C_\ep$
producing double-valued behavior along cycle $D_\ep$. The
only internal loop momentum left crosses the remaining 
independent cycle $A_\ep$.


\begin{fig}[htb]
\centering
\epsfysize=2in
\epsffile{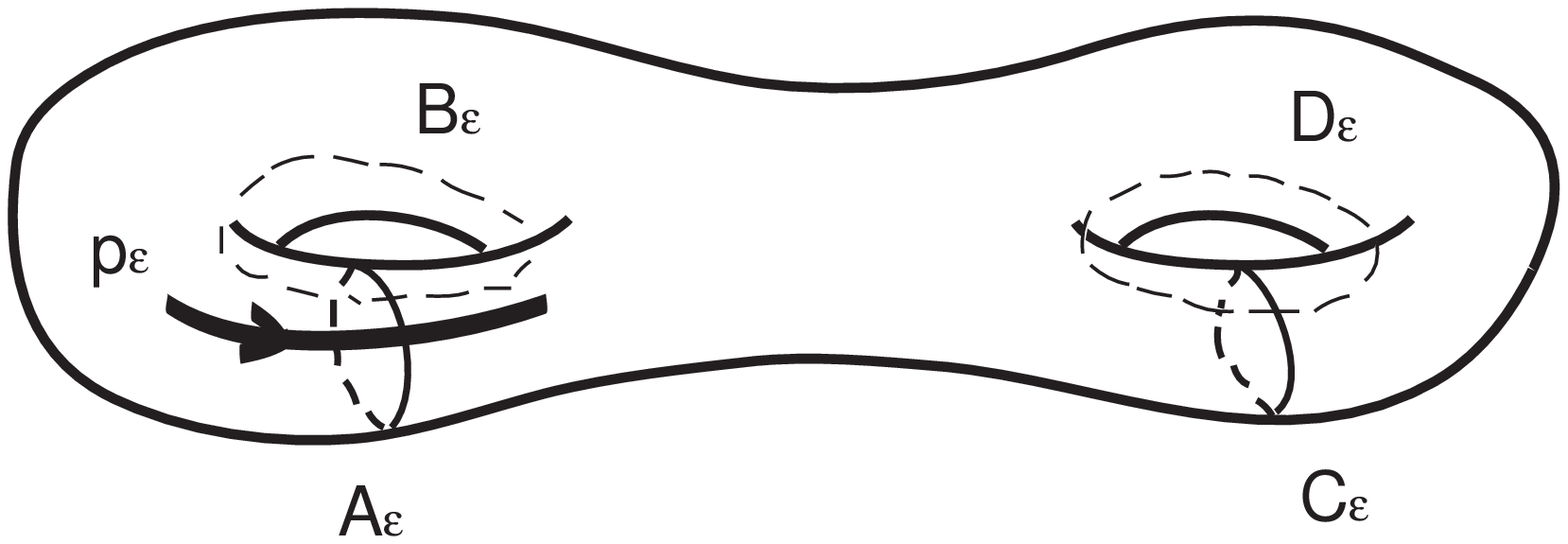}
\caption{Twisted Cycles $A_\ep$ and $B_\ep$ and the single internal 
loop momentum $p_\ep$}
\label{fig:1}
\end{fig}

 
A subtle worldsheet supersymmetric and supermoduli dependent  
shift  in the Prym period is identified. 
The results are as follows: let ${\bf A} _C [\delta]$ be
the left-right symmetric compactified amplitude  for fixed 
spin structure $\delta$ and $n$ directions twisted by $\bZ_2$.
For simplicity, only the 0-point function is considered here.
It is well-known (c.f. \cite{dp88}, p. 967) that the gauge-fixed expression
for ${\bf A}_C[\delta]$ as an integral over supermoduli space
is 
\be
\label{gaugefixed0}
{\bf A}_C [\delta]
=
\int _{s\M _2}  \prod _A |dm^A|^2 
\int D(XB\bar B C \bar C) \prod _A \left |\delta (\< H_A | B\>) \right |^2 
e^{-I_m -I_{gh}}
\ee
Here $m^A$ are parameters for a $(3|2)$-dimensional slice 
for the even and odd supermoduli, 
$H_A$ is the corresponding super Beltrami differentials, 
$X$, $B,\bar B,C,\bar C$ are respectively 
the matter and ghost superfields, and $I_m$, $I_{gh}$ are respectively 
the matter and the ghost actions. This expression is non-chiral, 
but one of its fundamental properties is  that it can be split as an
integral over internal  momenta of the norms squared of a chiral amplitude, 
$ \A_C  [\delta, \ep ](p_\ep )$,
\bea
\label{split}
{\bf A}_C [\delta ]
=
\sum _\ep \int d^{10-n} p_I  \int d^n p _\ep \int _{s\M_2} 
\prod _A |dm^A|^2  \left | \A_C  [\delta, \ep ](p_\ep ) \right |^2
|e ^{i \pi p^\mu _I \hat \Omega _{IJ} p^\mu _J}|^2
\eea
The chiral amplitude $ \A_C  [\delta, \ep ](p_\ep ^\mu)$ is superholomorphic 
on supermoduli space $s\M _2$ and may be computed in terms of the 
chiral partition functions $Z_C$ and $Z_M$ of the compactified
and Minkowskian theories respectively, 
following the method of \cite{I,II,III,IV}. 
Parametrizing  the genus 2 supermoduli space  by the super 
period matrix $\hat\Omega_{IJ}$ as bosonic coordinates and 
the two Grassmann coordinates $\zeta ^1, \zeta ^2$ as 
fermionic coordinates, one obtains,
\bea
&&
\A_C [\delta,\ep ](p_\ep )
=
\A_M [\delta] ~ {Z_C[\delta; \ep] \over Z_M[\delta]} 
\exp  \{i \pi (\hat \tau _\ep + \Delta \tau _\ep) p_\ep ^2 \}
\\ && \hskip 1in 
\times \biggl \{
1 - { 1 \over 8 \pi^2 } \int \! \! d^2 \! z \int \! \! d^2 \!  w \chiz \chiw 
\biggl  (\< S_{m} (z) S_{m}(w) \>_C - \< S_{m} (z) S_{m}(w) \>_M \biggr  )  
\no \\ && \hskip 1.4in 
+ {1 \over 2 \pi}\int d^2 z  \hat \mu (z)  
\biggl  ( \< T_{m}(z)\> _C - \< T_{m}(z)\> _M \biggr )  \biggr \}
\no
\eea
Here, the expression $\A_M[\delta]$ is the chiral superstring
measure of the Minkowski space theory,
which was evaluated in \cite{I,IV}. The expressions 
$Z_M[\delta]$ and $Z_C[\delta,\epsilon]$ are the matter partition functions for
the flat Minkowski space theory ($M$) and the ${\bf Z}_2$-twisted theory($C$). 
The expressions $S_m$ and $T_m$ are the matter supercurrent and stress tensor 
respectively. The bosonic Prym period $\tau_\ep$ is related to the bosonic period 
matrix $\Omega $, a relation that we shall denote by 
$\tau _\ep = R_\ep (\Omega)$, and is given explicitly by (for any $i,j=2,3,4$)
\bea
\label{Schottky0}
\tau _\ep = R_\ep (\Omega) \qquad \Leftrightarrow \qquad
{ \tet _i (0, \tau_\ep) ^4 \over \tet _j (0, \tau_\ep) ^4}
=
{\tet [\delta _i ^+] (0,\Omega)^2 \tet [\delta _i ^-] (0,\Omega)^2
\over
\tet [\delta _j ^+] (0,\Omega)^2 \tet [\delta _j ^-] (0,\Omega)^2}
\eea
The super-Prym period $\hat \tau _\ep$ 
and the  super-period matrix $\hat \Omega_{IJ}$ are related via this same 
equation $\hat \tau _\ep = R_\ep (\hat \Omega)$ and both are 
invariant under local worldsheet supersymmetry. The quantity $\Delta \tau_\ep$
is a subtle correction which arises from chiral splitting 
and is given by\footnote{Throughout, we shall use the notation $\p _{II}
\equiv \p / \p \Omega _{II}$, $\p _{IJ} \equiv
\half \p / \p \Omega _{IJ}$ when $I\not= J$. The heat-kernel equation for
the $\tet$-function is then simply $4 \pi i \p _{IJ}
\tet [\delta] (\zeta, \Omega) = \p _I \p_J \tet [\delta ](\zeta ,
\Omega)$, where $\p_I \equiv \p / \p \zeta ^I$.}
\bea
\Delta \tau _\ep  = - {i \over 8 \pi}
\int \! d^2 z \int \! d^2w \chiz S_\delta (z,w) \chiw \left \{
\omega _\ep (z) \omega _\ep (w) - \omega _I(z) \omega _J (w) 
\p_{IJ} R_\ep (\hat \Omega) \right \}
\eea
It is locally supersymmetric and non-vanishing.
In the expression for  $\A_C[\delta;\ep](p_\ep)$, 
all dependence on the original period matrix $\Omega_{IJ}$ has been 
eliminated in favor of $\hat\Omega_{IJ}$. Henceforth,
we view $\hat\Omega_{IJ}$ as the true  moduli, and write it simply as
$\Omega_{IJ}$.

\medskip

In section \S 4, a generalization is given 
for the chiral measure in an arbitrary background $C$
with worldsheet supersymmetry and matter central charge 15.
In the split gauge defined by $S_\delta (q_1,q_2)=0$, this expression
is evaluated for each chiral sector, denoted by $\lambda$, and takes the 
simple form,
\be
\label{gaugefixed}
{\cal A}_C [\delta;\lambda] = 
{\cal A}_M [\delta] {Z_C[\delta;\lambda]  \over Z_M[\delta]} 
\biggl \{
1 - {\zeta ^1 \zeta ^2 \over 4 \pi ^2} {\cal Z} \< S_C (q_1) S_C (q_2) \>
_{C\lambda} \biggr \}\, .
\ee
where $Z_C[\delta;\lambda]$ is the matter partition function of the compactified
theory $(C)$ for spin structure $\delta$ and chiral sector $\lambda$,
while $S_C$ is the supercurrent of the compactified theory.
This last formula was announced in \cite{I}. 
It shows that the superstring measure for an arbitrary background
reduces to the evaluation  of the matter partition function $Z_C$ 
and the supercurrent correlator $\<S_C(q_1)S_C(q_2)\>_{C\lambda}$
in sector $\lambda$. 

\medskip

In section \S 5, the two-loop ${\bf Z}_2$-twisted blocks
$\A_C[\delta;\ep](p_\ep)$ 
are computed in terms of $\tet$-functions, and a final expression
for the twisted chiral measure is obtained in terms of the 
super-period matrix $\hat \Omega _{IJ}$ (recall that
it is now denoted simply
by $\Omega _{IJ}$; similarly, the super-Prym period $\hat \tau _\ep$ 
is denoted simply by $\tau_\ep$). Defining\footnote{The 
volume element is defined as follows 
$d^3 \Omega = d\Omega _{11} d \Omega _{12} d \Omega _{22}$.}
\bea
d\mu_C[\delta;\ep](p_\ep)
& \equiv &
d^3 \Omega \int d^2\zeta \A_C[\delta;\ep](p_\ep)
\no \\
\Gamma [\delta; \ep] 
& \equiv & 
{\cal Z} \int d\zeta ^2 d\zeta ^1 \Delta \tau _\ep
\eea
we find
\bea
\label{chiralmeasure}
d\mu _C [\delta; \ep] (p _\ep)
= 
e^{i \pi \tau_\ep  p_\ep ^2}  {Z_C [ \delta, \ep] \over Z_M [\delta] }
\bigg\{   {\Xi_6[\delta] \tet[\delta]^4 \over  16\pi^6 \Psi_{10}} 
+ \biggl ( i \pi p_\ep ^2 - n \p_{\tau_\ep} \ln \tet _i (0,\tau_\ep) \biggr )
\Gamma [\delta; \ep] \bigg \}
d^3 \Omega 
\no\\
\eea
The expression $\Xi_6[\delta](\Omega)$ is the modular covariant 
form found in \cite{I,II,III,IV}. The key new expression here is
$\Gamma [\delta; \ep]$, which requires a substantial calculation,
especially the determination of its overall sign. The final result is
\bea
\Gamma [\delta ^\pm _i ,\ep] = 
\pm \< \nu _0 | \mu _i\>  {i \over (2 \pi )^7} \ {  \tet _i ^4  \over  \eta 
^{12}} 
\left ( {\tet [\delta ^\pm _i ]^2  \over   \tet [\delta ^\mp  _i ]^2 
} \right )
\left ( {\tet _j^4 \over \tet [\delta _j^+]^2 \tet [\delta _j^-]^2} \right )
\eea
Here, the $\pm$ signs are correlated with one another
on both sides of the equation.
The $\tet_i$'s are genus $1$ $\tet$-constants with respect to $\tau_{\ep}$,
while the $\tet[\delta]$'s are genus $2$ $\tet$-constants with
characteristic $\delta$. The above expression
for $\Gamma[\delta^\pm_i,\ep]$   is independent of $j$. 

\medskip

In section \S 6, modular properties in the presence of a twist
are studied. The subgroup $H_\ep$ that fixes a twist 
$\ep$ is determined: it has 48 elements, and is of the form
\bea
H_\ep = \{ I, ~ T_2 ,  ~S_2, ~S_2T_2, ~S_2^2, ~ S_2^2 M_1 T_2 \} \times 
H_\ep ^0
\eea 
The matrices $M_1$, $M_2$, $M_3$, $S_2$, $T_2$
are given in Appendix B.3; and $H_\ep^0 $ is the Abelian 
subgroup consisting of the elements $ H_\ep^0 =
\{ I, M_1, M_2, M_3, M_1M_2, M_1M_3, M_2M_3, M_1M_2M_3\}$.
The orbits under
this group of both even spin structures and twists are worked out.

\medskip

In section \S 7, some of the degeneration limits of the measure are
worked out explicitly and checked versus the results expected
on physical grounds.

\medskip

In section \S 8, applications to physically interesting models 
are studied. These models are obtained as orbifolds of
flat Minkowski space-time with symmetric or asymmetric
orbifold groups $G$. The orbifold groups considered 
here are simply generated by reflections and shifts.
The procedure of chiral splitting for 
symmetric orbifolds is reviewed. 

\medskip

A prescription is given for the construction of the string amplitudes 
for asymmetric orbifolds in terms of the chiral blocks obtained
by chirally splitting symmetric orbifold amplitudes.
The prescription  proceeds as follows.
For each model, the set of {\it chiral}  blocks 
$\{d\mu_L[\delta_L,\lambda_L](p_L)\}$, for all 
left chiral sectors $\lambda_L$,  is
determined from the symmetric orbifold theory obtained from the group
generated by all symmetric elements of the form $(f_L,f_L)$, 
where $f_L$ runs over left elements of group elements 
$f=(f_L,f_R)\in G$. Similarly, the set of antiholomorphic blocks
$\{d\mu_R[\delta_R, \lambda_R](p_R)\}$ for all right chiral sectors
$\lambda_R$, is determined from the symmetric orbifold group 
generated by all $(f_R,f_R)$. 
The GSO projection is always assumed to be implemented by 
carrying out a chiral summation over all spin structures 
independently for left and right movers,
\be
\label{GSO1}
d\mu_L[\lambda_L](p_L)
=
\sum_{\delta_L}\eta_L [\delta_L ; \lambda_L]
d\mu_L [\delta_L ,\lambda_L ](p_L)
\ee
for suitable phases $\eta _L [\delta_L; \lambda_L]$. 
The partition function for the asymmetric 
orbifold model $G$ is then given by
\be
\label{leftright}
Z_G
=
\int 
(\det\, \Im \Omega)^{-5+{n\over 2}}
\sum_{p_L,p_R}
\sum_{\lambda_L, \lambda_R}
d\mu_L[\lambda_L](p_L) \wedge
\overline{d\mu_R[ \lambda_R ](p_R)}
K (\lambda_L, \lambda_R ;p_L,p_R)
\ee
where $n$ is the number of compactified dimensions, and 
$K (\lambda_L, \lambda_R ;p_L,p_R)$ are suitable constant
coefficients, depending only on the labels
$\lambda_L ,\lambda_R,p_L,p_R$ 
for left and right chiral blocks. 
The determination of $K $
is a difficult issue, and we shall discuss it further in a forthcoming
publication \cite{adp}. In this paper, we shall examine only the
chiral models, that is, only the issue of vanishing of the chiral blocks 
$d\mu_L[\lambda_L](p_L)$ and $d\mu_R[\lambda_R](p_R)$.

\medskip

We carry out this procedure for the simplest ${\bf Z}_2$ models,
where $n=4$ and $G$ is given by $\mathbf{Z_2}$ reflections.
The chiral blocks of this theory are of course just the 
conformal blocks $d\mu_C[\delta; \ep](p_\ep) $, with index 
$\lambda= \lambda _L =\ep$, of the ${\bf Z}_2$ theory we have
just derived in (\ref{chiralmeasure}).  It is shown that there is
a unique choice of phases for each $\ep$ that is consistent 
with modular covariance, given by
\be
\label{zeetwo}
\eta [\delta; \ep]=1
\ee
Upon carrying out the GSO projection, the term involving 
$\Gamma [\delta; \ep]$ cancels due to the genus 1 Riemann identities
and the chiral block reduces to
\be
d\mu_C[\ep](p_L)
=
{ e^{i \pi \tau _\ep p^2 _L} \over 16 \pi ^6 \Psi _{10} }
{\tet [\delta _j ^+ ] ^2  \ \tet [\delta _j ^- ]^2 \over  \tet _j ^4   } d^3\Omega
\sum _\delta \ \Xi _6 [\delta ] \
\tet [\delta ]^2 \ \tet [\delta + \ep ]^2
\ee 
This expression can be shown to vanish identically, 
by using identities which involve  $\Xi_6[\delta]$ and which were
established in \cite{IV}. Thus the cosmological constant
vanishes point by point on moduli space, in agreement with the 
fact that the theory just reproduces
Type II superstrings.

\medskip

The more complicated Kachru-Kumar-Silverstein models are 
constructed in terms of an asymmetric orbifold group,
whose point group $P_G$ is isomorphic to
${\bf Z}_2\times {\bf Z}_2$. It is shown that the chiral blocks
$d\mu _C [\delta;\lambda](p_L)$
of such theories are indexed by {\it two} twists
$\lambda=\lambda _L=(\ep,\alpha)$, and given explicitly by
\be
d\mu_C[\delta;\lambda](p_L)
=
\<\alpha |\delta\>d\mu_C[\delta; \ep](p_L)
\ee
where the expression $d\mu_C[\delta; \ep](p_L)$ on the right hand 
side are the ${\bf Z}_2$ blocks of (\ref{chiralmeasure}). 
It is shown that modular covariance requires that the GSO 
projection  phases be given by $\eta_L[\delta; \lambda]=
\<\alpha |\delta\>$, that is,
the chiral blocks of the theory are given by
\bea
d\mu_C[\lambda] (p_L)
&=&
\sum_{\delta}\<\alpha |\delta\>d\mu_C[\delta; \ep](p_L)
\nonumber\\
&=&
{ e^{i \pi \tau _\ep p^2 _L} \over 16 \pi ^6 \Psi _{10} }
{\tet [\delta _j ^+ ] ^2  \ \tet [\delta _j ^- ]^2 \over  \tet _j ^4   } d^3\Omega 
\sum _\delta \ \<\alpha |\delta\>\
\Xi _6 [\delta ] \
\tet [\delta ]^2 \ \tet [\delta + \ep ]^2
\eea
For each $\ep$, the label $\alpha$ in $\lambda$ can be classified into orbits
of the modular group $H_\ep$ leaving $\ep$ invariant.
There are $4$ such orbits. 

\medskip

We find that the contributions of two 
orbits vanish identically, the contribution of a third orbit vanishes 
of order $\tau^4$ along the divisor of separating nodes,\footnote{
For brevity, the notation $\Omega _{11}=\tau_1, \Omega _{22}=\tau_2$
and $\Omega _{12}= \tau$ will be used throughout.} 
but the contribution of the remaining orbit  vanishes only
to order $\tau^2$ along the same divisor. More precisely, we find
\bea
\label{kks}
&&
\sum _{\delta }
\< \alpha  | \delta \>\,  \Xi _6 [\delta ]\,  \tet [\delta ]^2
\, \tet [\delta + \ep ]^2
\\
&&
\quad
=
- 256 \pi ^2 \tau ^2 \eta (\tau_1)^{12} \eta (\tau_2 )^{12}
\left ( \tet ^8 _3 - \tet ^8 _4 \right )(\tau_1)
\tet ^4 _2 \tet ^2 _3 \tet ^2 _4(\tau_2)
+ \O (\tau^4)
\no
\eea
for $\ep$ and $\alpha$ given by
\be
\ep = \left (\matrix{0 \cr 0\cr} \bigg | \matrix{0 \cr \12 \cr} \right ),
\ \ \ 
\alpha = \left (\matrix{0 \cr 0\cr} \bigg | \matrix{\12 \cr 0 \cr} \right ).
\ee 
Thus the cosmological constant in the KKS models does not
vanish point by point on moduli space.

\section{$\mathbf{Z_2}$ Orbifold conformal field theory}
\setcounter{equation}{0}

The main goal of this section is to derive elements
from the theory of $\bZ_2$-twisted fields which we need.
For this, we require some background on spin structures, $\tet$-functions, 
double covers, and Prym differentials, summarized in \S 2.1 and \S 2.2.
The partition function for $\bZ_2$-twisted bosonic scalar fields has been 
obtained by Dijkgraaf, Verlinde, and Verlinde \cite{DVV}. It is recalled in \S 
2.3. 
For our purpose, the main objects of interest are the propagators. 
The fermionic one is easy to find, so we concentrate on
the bosonic propagator $\<\p_zx(z)\p_wx(w)\>_{\ep}$. 
As a warm-up, it is derived for the case of genus $1$ in \S 2.4. 
The rest of the section is devoted to extending this method to the case of 
genus $h=2$.
The Schottky relations between $\tet$-functions and the Prym
differential are also derived. 


\subsection{Spin structures and $\mathbf{\tet}$-functions}

Let $\Sigma$ be an orientable compact genus $h$ Riemann surface without
boundary on which a canonical homology basis for $H_1(\Sigma, \bZ)$
of cycles $A_I$, $B_I$, $I=1, \cdots .h$, is chosen,
\bea
\label{intersectionnumber}
\# (A_I,A_J)=\# (B_I,B_J)=0
\hskip .8in
\# (A_I,B_J) = - \# (B_J,A_I) = \delta _{IJ}
\eea
Holomorphic Abelian differentials $\omega _I$ may be normalized on
$A$-cycles and their integral along $B$-cycles yields the period matrix $\Omega 
_{IJ}$,
\bea
\oint _{A_I} \omega _J = \delta _{IJ}
\hskip .8in
\oint _{B_I} \omega _J = \Omega _{IJ}
\eea
Given the intersection numbers (\ref{intersectionnumber}), the basis is unique
up to modular transformations belonging to the modular group $Sp(2h,\bZ)$.
The structure of the modular group, its action on spin structures, twists and
$\tet$-functions is summarized in Appendix B.3.

\medskip

A spinor field on $\Sigma$ requires the assignment of a spin structure on 
$\Sigma$,
corresponding to an element of $H_1 (\Sigma, \bZ_2)$. In a
given homology basis, a spin structure may be labeled by a half-integer
characteristics $\kappa \equiv (\kappa ' |\kappa '')$ where $\kappa '$ and $
\kappa ''$, each valued in $\{0,\half\}^h$, and specifies the monodromies 
around
$A$- and $B$-cycles respectively. The parity of the spin structure $\kappa$
is that of the integer $4 \kappa ' \cdot \kappa ''$, and is modular invariant.
For genus 1, the 4 spin structures  separate into  3 even and 1 odd
(respectively denoted $\mu$ and $\nu_0$). For genus 2,
the 16 spin structures separate into 10 even and 6 odd spin structures 
(respectively
denoted by $\delta$ and $\nu$).
An important pairing between characteristics  is given by the
signature,
\bea
\< \kappa |\rho \> \equiv \exp \{ 4 \pi i (\kappa ' \cdot \rho '' -
\rho ' \cdot \kappa '')\} = \pm 1
\eea

The $\tet$-function with characteristics $\kappa = (\kappa '|\kappa'')$ is
defined by
\bea
\tet [\kappa] (\zeta , \Omega) \equiv
\sum _{n \in \bZ^{h}}
\exp \{  \pi i (n+\kappa ') \Omega (n+\kappa ') + 2\pi i (n+\kappa ') (\zeta
+ \kappa '') \}
\eea
It satisfies a number of monodromy relations,
which are listed in Appendix B. For genus 1, the relations between
the $\tet$-functions with characteristics and the standard Jacobi
$\tet$-functions are as follows. For odd spin structure, we have
$\tet _1 (z|\tau) \equiv \tet [\nu_0](z| \tau)$, while for even spin 
structures,
we have $\tet _i (z|\tau) \equiv \tet [\mu_i] (z|\tau)$, with the corresponding
even spin structure characteristics given by
\bea
\label{genus1even}
\mu_2 = (\12 |0) \qquad \mu_3 = (0|0) \qquad \mu_4 = (0|\12)
\eea
Two fundamental building blocks for the theory of functions and forms on a
Riemann surface\footnote{For reviews of string perturbation theory,
see \cite{dp88,vv2,rs03}; for the original mathematical results, see
\cite{fay}.} 
 are the prime form $E(z,w)$ and the Szeg\"o kernel
$S_\delta (z,w)$ (which we shall only need for even spin structure~$\delta$),
\bea
E(z,w) =  {\tet [\nu ] (\int _w ^z \omega _I, \Omega)
\over h _\nu (z) h _\nu (w)}
\hskip 1in
S_\delta (z,w) = { \tet [\delta ] (\int _w ^z \omega _I, \Omega)
\over \tet [\delta](0, \Omega) E(z,w)}
\eea
The holomorphic 1/2 form $h_\nu (z)$ for odd spin structure $\nu$ obeys
$h _\nu (z)^2 = \p _I \tet [\nu](0) \omega _I (z)$.
The prime form is a holomorphic $-\12$ form in $z$ and $w$ on
the universal cover of $\Sigma$, which behaves as $z-w$ when $z\sim w$,
and  is independent of $\nu$. The Szeg\"o kernel for even
spin structure $\delta$ is a meromorphic $\12$ form in $z$ and $w$ on
$\Sigma$ with a simple pole at $z=w$ with unit residue,
The importance of these quantities derives from the fact that $S_\delta$ is
the chiral fermion propagator, while $\p_z \p_w \ln
E(z,w)$ is the scalar propagator on $\Sigma$.

\subsection{$\mathbf{Z_2}$ twisting ; unramified double covers}

The configurations of a scalar $x(z)$ (resp. spinor $\psi (z)$) 
field on $\Sigma$ with a
$Z_2$-twist correspond to functions (resp. sections of a spin bundle with
characteristics $\kappa$) that are double-valued on $\Sigma$. These
configurations fall into $16$ distinct topological sectors corresponding to
the elements of $H_1 (\Sigma, \bZ_2)$, and in a given homology basis may be
labeled by half characteristics $\ep \equiv (\ep ' | \ep '') $, where $\ep'$
and $ \ep''$ are each valued in $\{0,\half\}^2$ (see \cite{fay,DVV}). 
For the scalar field
$x(z)$, and the spinor field $\psi (z)$ with spin structure $\kappa$, we have
\bea
\label{monodromies}
x(z+A_I) = (-)^{2 \ep '_I} x(z) \
& \qquad &
\psi (z+A_I) = -(-)^{2 \ep '_I + 2 \kappa ' _I} \psi (z)
\no \\
x(z+B_I) =  (-)^{2 \ep ''_I} x(z)
& \qquad &
\psi (z+B_I) = -(-)^{2 \ep ''_I + 2 \kappa ''_I} \psi (z)
\eea

\medskip

A $\bZ_2$-twisted scalar (resp. spinor) field may be viewed as a
field defined on the surface $\Sigma$ with a quadratic branch cut
along a cycle $C_\ep$. The $\bZ_2$-twisted field is then double-valued
around a conjugate cycle  $D_\ep$, which we parametrize as follows,
\bea
\label{cycleCep}
D_\ep = \sum _{I=1,2} (2 \ep '_I A_I + 2 \ep '' _I B_I)
\eea
The remaining generators of $H^1 (\Sigma, \bZ)$ will be denoted
by $A_\ep$ and $B_\ep$, and have intersection numbers,
$\# (A_\ep, B_\ep)= \# (C_\ep , D_\ep)=1$, while all others vanish.
Under the modular group, all twists transform in two orbits, one
consisting of $\ep =0$ and the other of all $\ep \not= 0$.
It will often be convenient to use the action of the 
modular group on the twists
to choose a standard reference twist. A simple choice consists of taking
\bea
\label{epchoice}
\ep = \left (\matrix{0 \cr 0\cr} \bigg | \matrix{0 \cr \12 \cr} \right )
\qquad
\Leftrightarrow
\qquad
A_\ep =A_1, ~ B_\ep = B_1, ~ C_\ep = A_2, ~ D_\ep = B_2
\eea
It will often be helpful to keep the  notation
$A_\ep, B_\ep, C_\ep, D_\ep$ for the sake of generality.

\begin{fig}[htb]
\centering
\epsfysize=3in
\epsffile{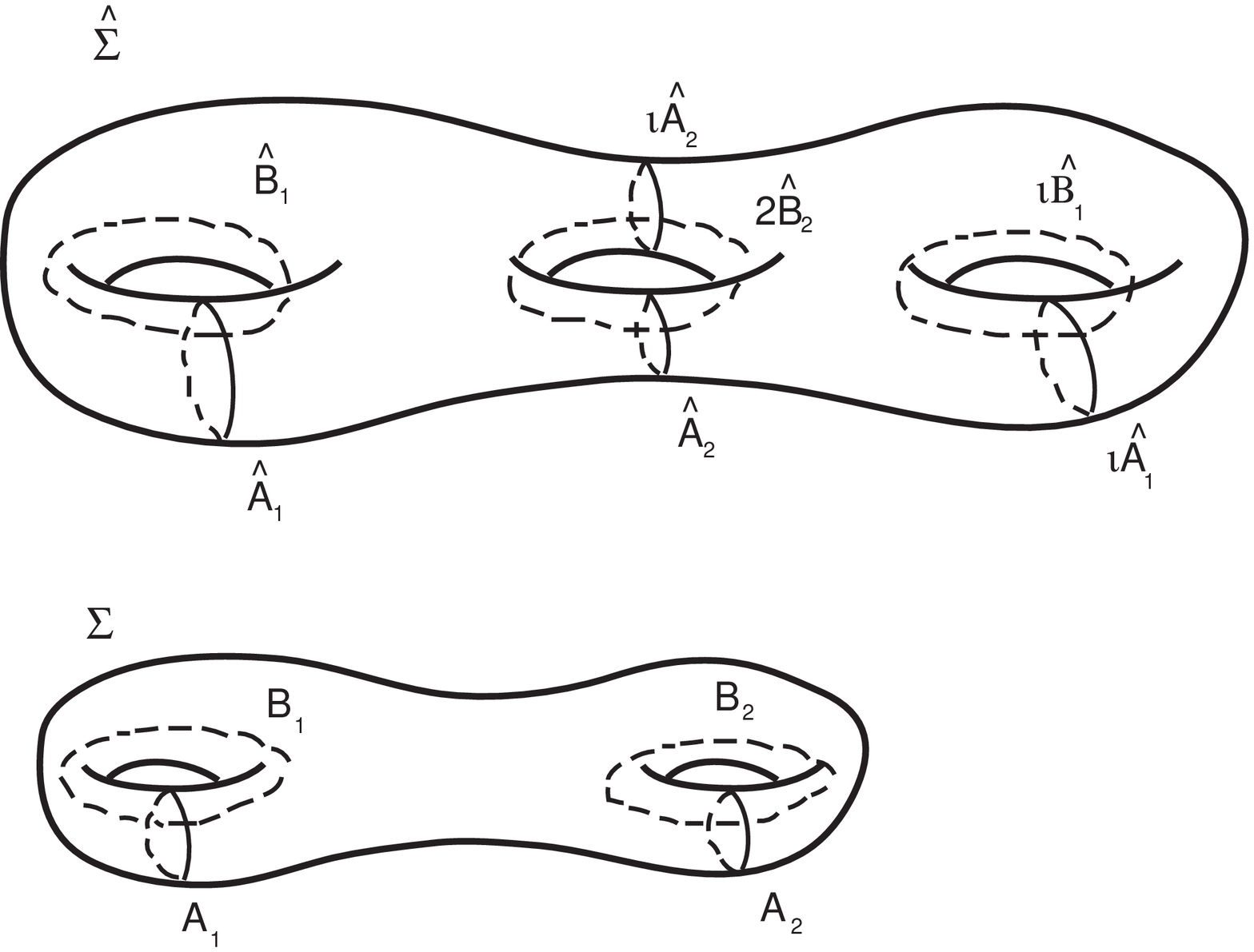}
\caption{The genus 2 Riemann surface and its unramified branched double cover}
\label{fig:2}
\end{fig}


Double valued scalar fields on a surface $\Sigma$ may also be viewed
as single valued scalar functions of the double cover $\hat \Sigma$ of
the original surface $\Sigma$. (Similarly, double valued spinor fields
may be viewed as  well-defined sections of a spin bundle of the
cover $\hat \Sigma$.)
If a function is double valued along the cycle $B_2$,
and  has a quadratic branch cut across the
conjugate cycle $A_2$, the surface $\hat \Sigma $ is
obtained by cutting open $\Sigma$ along the cycle $A_2$ and gluing
the two boundaries of this surface to the boundaries of its image under the
reflection involution $\iota$, as shown schematically in Figure \ref{fig:2}.
Given any twist $\ep \not=0$,
the double cover $\hat \Sigma$, as defined above,  is unique.

\medskip

The double cover $\hat \Sigma $ is a surface of genus 3,
whose complex structure is induced by the complex structure of $\Sigma$.
Functions and forms on $\Sigma$
that are $Z_2$-twisted (i.e. double-valued) along the cycle $B_2=D_\ep$, 
become
functions and forms on $\hat \Sigma $ that are {\sl odd} under the reflection
involution $\iota$. Of special importance is the Prym differential, 
$\omega _\ep$,
which is a holomorphic  1-form,  odd under the reflection $\iota$.
For genus 2, $\omega _\ep$ is unique up to normalization.
By construction,  its periods along the cycles $A_2=C_\ep $ 
and $B_2=D_\ep$ vanish.
Normalizing the period of $\omega _\ep$ to 1 along  $A_1=A_\ep$,
its integral along $B_1=B_\ep $ yields the Prym period $\tau_\ep$,
\bea
\label{Prym}
\oint _{A_\ep} \omega _\ep =1
\hskip 1in
\oint _{B_\ep} \omega _\ep = \tau _\ep
\eea
The Prym period is a complex number satisfying $\Im \tau _\ep > 0$.

\medskip

Specification on $\Sigma$ of a twist $\ep \not=0$, canonically separates
the even spin structures $\delta$ into two groups, according to whether
$\delta +\ep$ is even or odd. This separation is clearly modular covariant.
The group for which $\delta +\ep$ is even consists of 6 elements
which will be denoted $\delta ^+ _i$ and  $\delta _i^-, i=2,3,4$.
The group for which $\delta +\ep$ is odd consists of the remaining 4 elements.
Without loss of generality, we may assume that 
$\delta ^- _i = \delta ^+ _i + \ep$.
For the twist $\ep$ of (\ref{epchoice}), for example, we have
\bea
\label{deltapm}
\delta ^+ _i = \left ( \matrix{ \mu _i \cr 0|0 \cr}\right )
\hskip 1in
\delta ^- _i = \left ( \matrix{ \mu _i \cr 0|\12 \cr}\right )
\eea
where $\mu_i$ are the genus 1 even spin structure assignments of
(\ref{genus1even}). Clearly, a genus 2 surface with a twist $\ep\not=0$
has a canonically associated group of 6 even spin structures 
$\delta ^\pm _i$,
which in turn map 2 to 1 onto the 3 genus 1 even spin structures.

\medskip

The period matrix $\Omega_{IJ}$ of the genus 2 surface uniquely 
determines the Prym
period $\tau_{\ep}$, up to the action of generators in the Torelli
group;\footnote{The Torelli group is
the quotient of the full mapping class group, consisting of all equivalence
classes of disconnected diffeomorphisms of the surface $\Sigma$
by the modular group.} in this case, up
to the transformations $\tau _\ep \to \tau _\ep +4$ (see \cite{DVV}).
The relation between $\Omega_{IJ}$ and $\tau_{\ep}$
is given by the {\sl Schottky relations}, already introduced in
(\ref{Schottky0}). They may be expressed in an equivalent form
as follows, for any $i\not=j$ and $i,j=2,3,4$,
\bea
\label{Schottky}
{\tet [\delta _i ^+] (0,\Omega)^2 \tet [\delta _i ^-] (0,\Omega)^2
\over  \tet _i (0, \tau_\ep) ^4}
=
{\tet [\delta _j ^+] (0,\Omega)^2 \tet [\delta _j ^-] (0,\Omega)^2
\over  \tet _j (0, \tau_\ep) ^4}
\eea
Here, $\delta ^\pm _i$ are the group of 6 even spin structures
introduced in the preceding paragraph and $\tet _i$ are the
genus 1 $\tet$-functions associated with the genus 1
spin structures $\mu_i$ onto which $\delta ^\pm _i$ map.
Clearly, all three such relations are equivalent to one another.
The Schottky relations for genus 2 will be proven and their origin
will be explained in section \S 2.6.

\subsection{The Dijkgraaf-Verlinde-Verlinde partition function}

The first key result of \cite{DVV} of interest to us is the 
formula for the partition
function for a single  $\bZ_2$-twisted boson on a circle of radius $R$.
In a sector of twist $\ep$, the worldsheet instanton sum is given by
\bea
Z[\ep] (\Omega , R) = Z ^{{\rm qu}}  [\ep] (\Omega) \sum _{(p_L,  p_R) 
\in \Gamma _R } \exp
\{ \pi i (p_L^2 \tau _\ep -  p_R ^2 \bar \tau _\ep ) \}
\eea
where the momentum sum is taken over the lattice
\bea
\Gamma _R  = \left \{ (p_L, p_R) =
\left ( {n\over 2R} +  mR, {n\over 2R}  - mR \right ) \right \}
\eea
The oscillator part of the partition function is independent of the radius
and may be computed by equating the partition functions of the
$\bZ_2$-twisted model with the partition function of the circle theory
at the self-dual radius $R=1/\sqrt{2}$. The result may be expressed
in terms of the spin structures $\delta ^\pm _i$, introduced in the
preceding subsection,
\bea
\label{ZDVV}
Z^{{\rm qu}}[\ep] (\Omega ) = \left |
{\tet [\delta _i ^+] (0,\Omega) \tet [\delta _i ^-] (0,\Omega)
\over  Z(\Omega)^2 \tet _i (0, \tau_\ep) ^2} \right |
\eea
Here, $Z(\Omega)$ is the inverse of the chiral bosonic partition function
of the uncompactified circle theory.
In view of the Schottky relations (\ref{Schottky}), this expression
is independent of the choice of $i$ and therefore only depends
upon $\ep$.


\subsection{Genus one $\mathbf{Z_2}$-twisted propagators and partition functions}

On the torus $\Sigma = {\bf C}/{\bf Z}+\tau{\bf Z}$, with modulus $\tau$,
a scalar field $x(z)$ and a fermion field $\psi(z)$ with 
even spin structure $\delta$,
twisted by $\ep$, obey the following monodromy conditions,
\bea
x(z+1) = (-)^{2 \ep '} ~ x(z)
& \qquad &
\psi  (z+1) = - (-)^{2 \delta '} (-)^{2 \ep '} ~ \psi  (z)
\no  \\
x(z+\tau) = (-) ^{2 \ep ''} ~ x(z)
& \qquad &
\psi  (z+\tau) = - (-)^{2 \delta ''} (-)^{2 \ep ''} ~ \psi  (z)
\eea
The corresponding propagators satisfy the same monodromy conditions
as the fields in both $z$ and $w$. The bosonic propagator $B_\ep (z,w)$
is symmetric under interchange of $z$ and $w$, and has a double pole
at $z=w$, while the  chiral fermion  propagator $S_{\delta +\ep}(z,w)$
is antisymmetric under interchange of $z$ and $w$, and has a
simple pole at $z=w$. Matching monodromies and poles yields the
following propagators,
\bea
B_\ep (z,w) = \< \p x (z) \p x(w) \>
& = &
\p_z \p_w \ln {\tet _1 (z-w|2 \tau) \over \tet _1 (z-w+\tau |2 \tau)}
\no \\
S_{\delta +\ep} (z,w) = \< \psi _+(z) \psi _+ (w) \>
& = &
{\tet [\delta +\ep  ](z-w, \tau) ~ \tet _1 ' (0,\tau) \over 
\tet [\delta +\ep ](0, \tau)
~ \tet _1 (z-w, \tau)}
\eea
The chiral fermion propagator is just  the Szeg\"o kernel with spin structure
shifted to $\delta +\ep$.

\medskip

It is convenient to recast the twisted boson propagator in terms of the
Szeg\"o kernel, since it 
will turn out that it is these formulas which will lend themselves to
generalization to higher genus.  Using the doubling formulas for genus 1, as
well as the formulas for $z$-derivatives (see Appendix A), the following
relation is established (valid now for arbitrary twist $\ep \not=0$)
\bea
B_\ep (z,w) =
S_{\mu} (z,w) S_{\mu  + \epsilon} (z,w)
\eea
where $\mu$ is any even spin structure such that $\mu +\ep$ is also
even. There are two possible choices for $\mu$ for any given $\ep \not=0$;
e.g. if $\ep = (0|\12)$, one can have $\mu =(0|0)$ or $\mu = (0|\12)$.

\medskip

Theories with worldsheet supersymmetry will be of special interest. The
worldsheet supercurrent formed of a free scalar $x$ and a chiral fermion
$\psi _+$ with spin structure $\delta$ is given by $S=-{1\over 2}
\psi_+ \p_z x$ and is invariant under the $Z_2$ twist $\ep$.
The supercurrent two-point function is given by
\bea
\< S(z) S(w) \> = {1 \over 4} B_\ep (z,w) S_{\delta + \epsilon} (z,w)
= {1 \over 4} S_{\mu} (z,w) S_{\mu + \epsilon} (z,w)  S_{\delta +
\epsilon} (z,w)
\eea
Using the OPE for the two supercurrents yields the stress tensor for the
system,
\bea
S(z) S(w) = {\hat c /4 \over (z-w)^3 } + {\half T(w) \over z-w} +
{\rm reg}
\eea
Expanding the Szeg\"o kernels,  one find $\hat c=1$, as expected.
Using the heat equation, $ \p _z ^2 \tet [\mu] (z|\tau) = 4 \pi i \p _\tau
\tet   [\mu ] (z|\tau)$, the remaining contribution may be recast as follows,
\bea
\<T\>_{\ep, \delta} =  \pi i \p_\tau \ln \biggl ( {\tet [\mu]  \tet
[\mu +\ep]  \tet [\delta + \ep]  \over  \tet [0|0] \tet [\12| 0] \tet [0 | \12]
}\biggr )
\eea
The corresponding partition function for one boson and one fermion of spin
structure $\delta$, both twisted by $\ep$, are obtained using the relation
$\<T\>_{\ep,\delta}= 4 \pi /\sqrt{g} \delta \ln Z_{\ep,\delta} 
/\delta g^{zz}$ and are
\bea
Z_{\ep,\delta} = \left ({\tet [\delta +\ep] (0,\tau) \over \tet [\nu + \ep]
(0,\tau)} \right )^{\half}
\eea
with $\nu = (\12 |\12)$ in agreement with known results \cite{g88,dgh}.


\subsection{Genus two $\mathbf{Z_2}$-twisted propagators}

Inspection of the monodromy equations (\ref{monodromies}) reveals that
implementing a twist $\ep$ on the fermion field $\psi _+(z)$ and on its
propagator simply results in replacing the spin structure $\delta$ by
$\delta +\ep$. Of interest here are only the cases where $\delta +\ep$ is
itself an even spin structure.\footnote{The cases where $\delta +\ep$ is
and odd spin structure will not contribute to the cosmological constant
for the orbifold models we shall consider later on.}
The propagator for the twisted fermion is thus the Szeg\"o kernel for 
$\delta +\ep$,
\bea
\label{twistfermion}
\< \psi _+ (z) \psi _+ (w) \> _\ep = S_{\delta + \ep} (z,w)
\eea
Viewed on the surface $\Sigma$, this quantity is a double-valued  
section of a spin
bundle for spin structure $\delta$; it becomes a single-valued 
section of a spin
bundle on $\hat \Sigma$.

\medskip

Implementing a twist $\ep$ on a scalar field $x(z)$ is achieved by imposing
the monodromy conditions of (\ref{monodromies}).
It may be constructed in terms of the  prime form for the covering surface
$\hat \Sigma$ as was done in \cite{DVV}.
Actually, we shall only need the propagator for the twisted field $\p_z x$,
\bea
B_\ep (z,w) \equiv \< \p_z x(z) \p_w x(w) \> _\ep
\eea
It obeys the same monodromy conditions as $x$ does in both $z$ and $w$ 
and has a
double pole at $z=w$ with unit residue.

\medskip

By analogy with the case of genus 1, the propagator $B_\ep (z,w)$ may be
more simply constructed in terms of products of the Szeg\"o kernels  
for even spin
structures $\delta ^+_i$ and $\delta ^-_i$ of (\ref{deltapm}). Indeed,
the products
\bea
B_i (z,w) = S_{\delta ^+ _i} (z,w) S_{\delta ^- _i} (z,w)
\qquad
i=2,3,4
\eea
readily produce the required double pole.  Given the fact that
$\delta ^+_i +\delta ^-_i=\ep$  modulo integral periods, they
also manifestly  exhibit the required monodromy by $\ep$.

\medskip

The differences $B_i(z,w) - B_j(z,w)$ are holomorphic 1-forms in both $z$ (as
well as in $w$), and have the same non-trivial monodromy as $\p_z x$. On a
genus 2 surface, such a one-form must be proportional to the Prym differential
$\omega _\ep (z)$. Since $\omega _\ep$ is unique, there must be one linear
relation between the three $B_i$. Taking this into account,
the twisted propagator $B_\ep$ may  be expressed via any of the following
three expressions,
\bea
\label{B}
B_\ep (z,w) = S_{\delta ^+ _i} (z,w) S_{\delta ^- _i} (z,w) + b_i \omega _\ep
(z) \omega _\ep (w)
\eea
A more detailed discussion of the Prym differential and period 
will be given later.

\bigskip

\noindent
{\bf  The calculation of $b_i$}

\medskip

Below, the coefficient $b_i$ will be computed using the partition function for
the twisted boson theory, derived in \cite{DVV}
and quoted in \S 2.3.  The result, in the basis of
reference (\ref{deltapm}), is given by
\bea
\label{littleb}
b_i = - 4 \pi i \p _{\tau_\ep} \ln \tet _i (0,\tau_\ep)
\eea
The detailed calculation of $b_i$ proceeds as follows. The stress tensor for
the twisted boson $x$ is computed in two different ways; first, 
from the propagator
$B_\ep$,
\bea
\label{stressB}
T_{zz} = \half \lim _{w\to z} \biggl ( B_\ep (z,w) - {1 \over (z-w)^2} \biggr )
\eea
 and second from the variation of the partition function 
 $Z^{{\rm qu}}[\ep]$ of the twisted $x$ field,
\bea
\label{stressZ}
T_{zz} = \delta _{zz} \ln Z^{{\rm qu}}[\ep]
\qquad \qquad
\delta _{zz} \equiv { 4\pi \over \sqrt{g}} {\delta \over \delta g^{zz}}
\eea
The partition function $Z^{{\rm qu}}[\ep]$, was computed in\cite{DVV},
and was presented in the notation of (\ref{deltapm})  in (\ref{ZDVV}).
Equating (\ref{stressB}) and (\ref{stressZ}) yields the following results. In
the variation (\ref{stressZ}) of (\ref{ZDVV}), the variation of $Z$
and of the $\tet$-functions $\tet [\delta _i ^+]  \tet [\delta _i ^-]$
precisely matches the contribution to (\ref{stressB}) of the first term in
(\ref{B}). The remaining variation
gives
\bea
\label{bi}
\half \ b_i \omega _\ep (z)^2
=  - \delta _{zz} \ln \tet _i (0,\tau_\ep)
= -(\delta _{zz} \tau _\ep ) \p \ln \tet _i (0,\tau_\ep)
\eea
The variation of $\tau_\ep$ with respect to the metric is analogous to the
variation of the genus 2 period matrix, $\delta _{zz} \Omega _{IJ} = 2 \pi
i \omega _I(z) \omega _J(z)$. In the next subsection (where a detailed
construction of the Prym differential will also be given), we shall derive
the following expression,
\bea
\label{vartau}
\delta _{zz} \tau_\ep = 2 \pi i \omega _\ep (z) ^2
\eea
Combining these results  readily yields (\ref{littleb}).


\subsection{The Prym differential and the Schottky relations}

The construction of the normalized Prym differential, the Schottky relations
and the variational equation (\ref{vartau}) for $\tau_\ep$ require a more
concrete understanding of the Prym differential. It is convenient to obtain the
Prym differential from its relation to the Szeg\"o kernel,
\bea
\label{nunorm}
\omega _\ep (z) \omega _\ep (w)
\sim  \left (
S_{\delta ^+_i} (z,w) S_{\delta ^-_i} (z,w) -
S_{\delta ^+_j} (z,w) S_{\delta ^-_j} (z,w) \right )
\eea
The constants of proportionality in this relation will  be determined below.

\medskip

To do so, the hyperelliptic parametrization of genus
$2$ Riemann surfaces is used\footnote{A detailed discussion of the
map between the $\tet$-function and hyperelliptic formulations
of genus 2 Riemann surfaces was given in \cite{IV};  
see Appendix B of this paper
for a summary of the relevant facts.}. The branch points
are denoted by $p_1 , \cdots ,p_6$ and the corresponding hyperelliptic
curve is
\be
s^2=\prod_{a=1}^6(z-p_a)
\ee
Recall that the identification between even and odd spin structures and
branch points proceeds as follows,
\bea
\nu_a \quad {\rm odd} & \leftrightarrow & {\rm branch \ point} \ p_a
\\
\delta \quad {\rm even} & \leftrightarrow & {\rm partition} \ A \cup B \qquad
A = \{ p_{a_1}, p_{a_2}, p_{a_3} \}  , \qquad B = \{p_{b_1}, p_{b_2}, p_{b_3} 
\}
\no
\eea
The twist will be left general. To this end, spin structures and twists will be
parametrized by odd spin structures,
$\{\nu_a,\nu_b,\nu_c,\nu_d,\nu_e,\nu_f\}$, where $(abcdef)$ is a
permutation of $(123456)$. The ${}_6C_2=15(=16-1)$ non-zero twists $\ep$ can be
uniquely labeled by {\sl two distinct odd spin structures} $\nu_{a,b}$,
such that $  \ep=\nu_a+\nu_b \not \equiv 0$.
The same twist can also be characterized as a sum of {\sl two even
spin structures}, $\ep = - \delta^+_i + \delta^ -_i$. There are
3 distinct choices for these sets which we label here
as
\bea
\label{deltanu}
\delta ^+ _i & = & \nu_a + \nu _c + \nu _d = -(\nu _b + \nu _e + \nu _f)
\nonumber \\
\delta ^- _i & = & \nu _b + \nu _c + \nu _d = -( \nu_a + \nu _e + \nu _f )
\nonumber \\
\delta ^+ _j & = & \nu_a + \nu _d + \nu _e = -(\nu _b + \nu _c + \nu _f)
 \\
\delta ^- _j & = &  \nu _b + \nu _d + \nu _e = - (\nu_a + \nu _c + \nu _f )
\nonumber \\
\delta ^+ _k & = & \nu_a + \nu _c + \nu _e = -(\nu _b + \nu _d + \nu _f)
\nonumber \\
\delta ^- _k & = &  \nu _b + \nu _c + \nu _e = - (\nu_a + \nu _d + \nu _f )
\nonumber
\eea
We note that $a,b$ cannot belong to the same set in the partition
associated with the spin structure $\delta$ since then, $\delta+\ep$ would
be odd.

\medskip

Let $\delta$ be the even spin structure corresponding to
the partition of the $6$ branch points into two sets $A,B$
of $3$ branch points each, as in (\ref{partition}).
In the hyperelliptic representation, the Szeg\"o kernel
for the spin structure $\delta$ is given  by \cite{fay},
\bea
S_\delta (z,w) = \half { s_A(z) s_B(w) + s_A(w)  s_B(z) \over z-w}
\biggl ( {dz \over s(z)}\biggr )^\half
\biggl ( {dw \over s(w)}\biggr )^\half
\eea
where the following notation is used,
\bea
s_A (z) ^2 & = & (z-p_{a_1}) (z-p_{a_2}) (z-p_{a_3})
\nonumber \\
s_B (z) ^2 & = & (z-p_{b_1}) (z-p_{b_2}) (z-p_{b_3})
\nonumber \\
s(z) ^2 & = & (z-p_1)(z-p_2)(z-p_3)(z-p_4)(z-p_5)(z-p_6)
\eea
The combinations of Szeg\"o kernels that enter into (\ref{nunorm}) may be
expressed in this hyperelliptic formulation,
\bea
S _{\delta_j ^+} (z,w) S_{\delta _j ^- } (z,w)
- S _{\delta_k ^+ } (z,w) S_{\delta _k ^-} (z,w)
& = &   {(p_c-p_d) (p_f -p_e) \over 4 s_{cdef}(z) s_{cdef}(w)} dz dw
\nonumber \\
S _{\delta_k ^+} (z,w) S_{\delta _k ^-} (z,w)
- S _{\delta_i ^+} (z,w) S_{\delta _i ^-} (z,w)
& = &   {(p_c-p_f) (p_e -p_d) \over 4 s_{cdef}(z) s_{cdef}(w)}  dz dw
\nonumber \\
S _{\delta_i ^+ } (z,w) S_{\delta _i ^-} (z,w)
- S _{\delta_j ^+} (z,w) S_{\delta _j ^-} (z,w)
& = &   { (p_c-p_e) (p_d -p_f) \over 4 s_{cdef}(z) s_{cdef}(w)}  dz dw
\eea
The hyperelliptic form of the Prym differential is thus given by
\bea
\omega _\ep (z) \sim {dz \over s_{cdef} (z)}
\eea
This quantity involves only 4 branch points, and by $SL(2,{\bf C})$-invariance
depends on a single modulus, which is the Prym period $\tau_\ep $.

\bigskip

\noindent
{\bf The Schottky relations}

\medskip

The Schottky relations arise from relating the modulus of 
the genus 1 curve with
branch points $p_c,p_d,p_e,p_f$ to the moduli of the full genus 2 curve with
all 6 branch points. Without loss of generality, the following identification
$(e_j,e_k,e_i,\infty)=(p_c,p_d,p_e,p_f)$ can be made, which allows us to put
the elliptic curve in standard from,
\bea
y^2=4(x-e_2)(x-e_3)(x-e_4) = 4(x-p_c)(x-p_d)(x-p_e)
\eea
The genus 1 Thomae formulas then yield, (see \cite{Bat}, page 361, 
equations (7))
\bea
&&  e_j-e_k = p_c-p_d=\sigma (\mu_j,\mu_k) {\pi^2\over 4 \omega^2}\tet_i^4
\no \\
&&  e_k-e_i = p_d-p_e=\sigma (\mu_k, \mu_i) {\pi^2\over 4 \omega^2}\tet_j^4
\no \\
&&  e_i-e_j = p_e-p_c=\sigma (\mu_i,\mu_j) {\pi^2\over 4 \omega^2}\tet_k^4
\eea
The function $\sigma$ is anti-symmetric, $\sigma (\mu_i, \mu_j)
= - \sigma (\mu_j, \mu_i)$ and is normalized by
\bea
\sigma (\mu_2 ,\mu_3 )= \sigma (\mu_3, \mu_4 ) = \sigma (\mu_2,\mu_4) = +1
\eea
The consistency of these relations follows from the genus 1 Jacobi identity,
\bea
\label{Jacobi1}
\sigma (\mu_i,\mu_j)  \tet_k^4 + \sigma (\mu_j,\mu_k)  \tet_i^4
+ \sigma (\mu_k,\mu_i)  \tet_j^4 =0
\eea
Furthermore,  $2 \omega$ is the $A_\ep$ period of the elliptic curve.
Recall from \cite{IV} that, in terms of the modular object 
${\cal M}_{\nu _a \nu_b}$,
\bea
\label{calM}
\M_{\nu _a \nu _b} & \equiv &
\p_1 \tet [\nu _a] (0,\Omega) \p_2 \tet [\nu _b] (0,\Omega)
-
\p_2 \tet [\nu _a] (0,\Omega) \p_1 \tet [\nu _b] (0,\Omega)
\no \\
\M_{\nu _a \nu _b} ^2 & = & \pi ^4 \prod _{k\not= a,b} ^6 
\tet [\nu _a + \nu _b
+ \nu _k](0,\Omega)^2
\eea
we have a refined form of the genus 2 Thomae identities,
\bea
{(p_a-p_b)(p_c-p_d) \over (p_a-p_c)(p_b-p_d)}
=
{\M _{\nu _a \nu _b} \M _{\nu _c \nu _d}
\over
\M _{\nu _a \nu _c} \M _{\nu _b \nu _d}}
\eea

The identification leads to the following relations between the three cross
ratios that may be constructed out of the 4 points $c,d,e,f$,
\bea
\label{genus1-2}
  {\tet_i^8\over \tet_j^8}(0,\tau_\ep) =
  \left ({e_j-e_k\over e_k-e_i} \right )^2 =
  {\M_{cd} ^2 \M_{ef} ^2 \over  \M_{cf}^2 \M_{de} ^2}
  ={\tet^4[\delta^+_i]\tet^4[\delta_i ^-]\over
    \tet^4[\delta_j ^+]\tet^4[\delta_ j ^-]} (0,\Omega)
\eea
In particular, the relation between the separating degeneration limit of the
genus 2 periods and the Prym period is $\tet [\mu _i] (0,\Omega _{11}) ^8 =
\tet [\mu _i] (0,\tau_\ep)^8$ so that $\tau_\ep = \Omega _{11} + \O (\Omega
_{12})$ up to the possible addition of an integer multiple of 8.

\bigskip

\noindent
{\bf The Prym differential in terms of Szego kernels}

\medskip

It is now straightforward to construct the normalized Prym differential.
Still in the conventions of \cite{Bat} (pages 330 and 331),
 the $A_\ep$ cycle is chosen so that
it produces the period $2\omega$. This is achieved by taking $A_\ep$ to
be the cycle around the branch cut between the branch points $e_k$ and
$e_i$, while the $B_\ep$ cycle goes from $e_j$ to
$e_k$,
\bea
\omega = \int _{e_k} ^{e_i} {dx \over y(x)}
\qquad \qquad
\omega ' =  \int _{e_j} ^ {e_k} {dx \over y(x)}
\eea
Therefore, the normalized Prym differential is simply
\bea
\omega _\ep (x) \equiv {1 \over 2 \omega } {dx \over y(x)}
\eea
With this normalization, the relations between the normalized Prym
differential and the Szeg\"o kernel are completely fixed and are found to be
\bea
\label{PrymSzego}
S _{\delta_i ^+ } (z,w) S_{\delta _i ^-} (z,w)
- S _{\delta_j ^+} (z,w) S_{\delta _j ^-} (z,w)
=   - \sigma (\mu_i,\mu_j ) \pi ^2 \tet _k (0,\tau_\ep)^4 
\omega _\ep (z) \omega _\ep (w)
\eea
Clearly, the sum of the left terms vanishes and so does the sum of the
right terms in view of the genus 1 Jacobi identity (\ref{Jacobi1}).

\medskip

The square of the Prym differential may be deduced from the expression in
terms of the product of Szeg\"o kernels, evaluated as $w \to z$,
\bea
\omega  _\ep (z) ^2
& = &
- \sigma (\mu_i,\mu_j) {1 \over \pi ^2 \tet _k ^4} \lim _{w \to z} 
\biggl ( S _{\delta_i ^+}
(z,w) S_{\delta _i ^-} (z,w) - S _{\delta_j ^+} (z,w) S_{\delta _j ^-}
(z,w) \biggr )
\nonumber \\
& = &
- \sigma (\mu_i,\mu_j)
{1 \over 2 \pi ^2 \tet _k ^4} \sum _{I,J} \omega _I(z) \omega _J(z)
\p_I \p_J \ln \biggl ({\tet [\delta _i ^+] \tet [\delta _i ^-]
\over \tet [\delta _j ^+] \tet [\delta _j ^-]}\biggr )
\eea
Differentiating relation (\ref{genus1-2}),
with respect to $\Omega _{IJ}$, we have
\bea
(\p_{IJ} \tau_\ep ) \p _{\tau_\ep} \ln {\tet _i \over \tet _j} (0,\tau_\ep) =
\half \p_{IJ} \ln \biggl ({\tet [\delta _i^+] \tet [\delta _i^-]  \over
\tet [\delta _j^+] \tet [\delta _j^-]}\biggr )
\eea
Using the heat equation $4 \pi i \p_{IJ} \tet = \p_I \p_J \tet$,
the relation $\delta _{zz} \Omega _{IJ} = 2
\pi i \omega _I(z) \omega _J(z)$ and
\bea
\p _{\tau_\ep} \ln {\tet _i \over \tet _j} (0,\tau_\ep) =  
i {\pi \over 4} \sigma (\mu_i,\mu_j)
\tet _k (0,\tau_\ep)^4
\eea
yields the equation announced in (\ref{vartau}).

\section{Chiral Splitting for $\mathbf{Z_2}$ Orbifold Theories}
\setcounter{equation}{0}

Chiral splitting is the process of decomposition of a chirally symmetric
amplitude into a sum of terms, each of which may be written as the absolute
value square of a chiral block which is meromorphic on supermoduli space and
has meromorphic dependence on the vertex insertion data \cite{superanom, dp89}.
Each chiral block
corresponds to a single $\N=1$ superconformal family. The chiral blocks thus
obtained may be used to carry out chiral GSO projections \cite{sw},
which are required for the construction of both the Type II \cite{gs82}
and Heterotic \cite{ghmr86} superstring theories. The precise
manner in which chirally symmetric amplitudes are chirally split was worked out
long ago for flat space-time \cite{dp89} (see also \cite{dp88} and 
\cite{superanom}).
In the present section, chiral splitting will be
carried out in detail in the $\bZ_2$-twisted sectors for orbifolds of flat
space-time involving $\bZ_2$ twistings. For definiteness, the restriction to
two-loops is made from the outset.


\subsection{Chirally Symmetric Amplitudes}

The compactified space-time $C$ will be viewed as an
$\N=1$ super conformal field theory on a worldsheet $\Sigma$ of genus $2$,
coupled to a two-dimensional supergeometry with superframe $E_M{}^A$ and $U(1)$
superconnection $\Omega_M$, \cite{superg2,dp88}.
The supergeometry $(E_M {}^A, \Omega _M)$ is subject
to the Wess-Zumino constraints, indicated here by the delta function
$\delta(T)$. In Wess-Zumino gauge, we have
$E_m{}^a=e_m{}^a+\theta\gamma^a\chi_m$, omitting the auxiliary field, $e_m{}^a$
is a frame for the worldsheet metric $g_{mn}=e_m{}^ae_n{}^b\delta_{ab}$, and
$\chi_m {}^\alpha=(\chiz,\chi_z{}^-)$ is the worldsheet gravitino field. The
definition of a supergeometry requires a spin structure $\delta$ on the
worldsheet.

\medskip

For fixed spin structure $\delta$,
superstring scattering amplitudes are built out of chirally symmetric
correlations  functions of $C$ coupled to
two-dimensional supergeometry,
\be
\label{nonchiral}
{\bf A}_C [\delta]
=
\int DE_M{}^A ~ D\Omega_M ~ \delta(T) \int _C DX^\mu 
\left ( \prod _{i=1}^N V_i \right ) ~ e^{-I_m}
\ee
Here, $V_i$ are vertex operators for physical states and the subscript
$C$  stands for the functional integral evaluated in the superconformal field theory
associated with $C$. Upon factoring out the local symmetries of the theory, the
integration over supergeometries  $(E_M {}^A, \Omega _M)$ reduces to an 
integral
over supermoduli space \cite{superm, superanom}, which is defined by
\be
s{\cal M}_2=\{(E^A_M ,\Omega_M )\}/\{\rm Gauge\ Symmetries\}
\ee
This integral can be written explicitly by choosing a $(3|2)$-dimensional
slice ${\cal S}$ of supergeometries $(E_M{}^A,\Omega_M)$ (\cite{II}, eq.
(2.11)),
\bea
\label{bfA}
{\bf A}_C [\delta]
=
\int _{s\M _2} \prod _A |dm^A|^2
\int D(XB\bar B C \bar C) \prod _A \left |\delta (\< H_A | B\>) \right |^2
\left ( \prod _{i=1} ^N V_i \right ) e^{-I_m - I_{gh}}
\eea
In this formula, $m^A = (m^a,\zeta ^\alpha)$ is a set of local complex
coordinates on $s \M _2$ where $A$ runs over the $(3|2)$ complex dimensions of
$s \M _2$ and $H_A$ are the Beltrami superdifferentials for the slice chosen to
represent $s \M _2$. The superfields $B$ and $C$ represent all the superghosts;
in standard component notation, they are given by $B= \beta + \theta b$ and $C
= c +\theta \gamma$, omitting their corresponding auxiliary fields.

\medskip

For orbifold compactifications, the starting point will be the matter action
for flat space-time. The matter superfield has the following component
decomposition, 
$X^\mu = x^\mu + \theta \psi _+ ^\mu + \bar \theta \psi _-^\mu $,
up to auxiliary fields. Henceforth, we restrict to the case of
$\bZ_2$-orbifolds and we have,
\bea
I _m
& = &
{ 1 \over 4 \pi } \int d^{2|2} z ~(\sdet E_M{}^A) \D_+ X^\mu \D _- X^\mu
 \\
& = &
{1 \over 4\pi} \int d^2z \biggl (
\p_z x^\mu \p_{\bar z} x^\mu - \psi _+ ^\mu \p_{\bar z} \psi _+^\mu
- \psi _- ^\mu \p_z \psi _-^\mu
\no \\ && \hskip 1in
- \chiz S_m - \chi _z ^- \overline{S_m}
-{1 \over 4} \chiz \chi _z ^- \psi _+ ^\mu \psi _- ^\mu \biggr )
\no
\eea
while the ghost action is given by
\bea
I _{gh}
& = &
{ 1 \over 2 \pi } \int d^{2|2} z ~(\sdet E_M{}^A) \left ( B \D _- C + \bar B
\D_+ \bar C \right )
\no \\
& = &
{1 \over 2\pi} \int d^2z \left (
b \p_{\bar z} c + \beta \p_{\bar z} \gamma
+ \bar b \p_z \bar c + \bar \beta \p_{\bar z} \bar \gamma
- \chiz S_{gh} - \chi _z ^- \overline{S_{gh}} \right )
\eea
The matter and ghost supercurrents are given by
\bea
S_m & = & - \half \psi _+ ^\mu \p_z x^\mu _+
\no \\
S_{gh} & = & \half b \gamma - {3\over 2} \beta \p_z c - (\p_z\beta) c
\eea
These formulas are identical in form to those for flat space-time, with the
exception that here, both the boson $x^\mu$ and fermion $\psi _\pm^\mu $ are
$\bZ_2$-twisted by the same twist $\ep$.
This guarantees in particular that the matter supercurrent and stress tensor
are single-valued.


\subsection{Chiral splitting of general amplitudes in $\mathbf{Z_2}$-twisted
sectors}

Let $\ep\not=0$ denote the $\bZ_2$-twist which is applied to $n$ components of
the scalar field $x$ and $n$ components of the fermion fields 
$\psi  _\pm$ taken
to have spin structure $\delta$. This means that these fields satisfy the
twisted boundary conditions of  (\ref{monodromies}). The
resulting chiral scalar and chiral fermion propagators were computed in \S 2,
and are given by
\bea
\label{chiralprops}
\< \p_z x _+ (z) \p_w x _+ (w) \>_\ep
& = &
 B_\ep (z,w)
\no \\
\< \psi _+ (z) \psi _+ (w) \> _\ep
& = &
 S_{\delta + \ep} (z,w)
\eea
The functional integral over $n$ components of twisted $\psi  _\pm$ fields
is given by the determinant of the Dirac determinant, which is holomorphically
factorized (up to the Belavin-Knizhnik anomaly \cite{div}, whose
effects cancel in the full amplitude),
\bea
\left ( {\rm Det} \Dslash \right ) ^{n \over 2}
= \left | { \tet [\delta + \ep] (0,\Omega)
\over Z (\Omega)} \right |^n
\eea
The quantity $Z$ is the inverse 
of the chiral partition function of a free untwisted scalar;
its expression in terms of $\tet$-functions was given in \cite{II}, eq. (4.5)
 but will not be needed here.
The functional integral over $x^\mu$  in the twisted sector is simpler than in
the untwisted sector, because no zero-mode occurs in the twisted sectors. By 
the Belavin-Knizhnik result \cite{div}, the combination of determinants that
holomorphically factorizes involves the inner products of holomorphic one
forms. In the twisted sector there is precisely one such form, namely the Prym
differential $\omega _\ep$, normalized as in (\ref{Prym}).
The functional determinant over $n$ twisted fields $x$ is then given by
the following expression,
\bea
\label{scalardet}
\left ( {\rm Det} \Delta \right )^{-{n \over 2}}
=
{1 \over (2 \Im ~ \tau _\ep)^{n \over 2}} \left |
{\tet [\delta _j ^+] (0,\Omega) \tet [\delta ^- _j] (0,\Omega)
\over Z(\Omega)^2 ~ \tet _j (0,\tau)^2 } \right | ^n
\eea
The obstruction to its holomorphic factorization lies entirely in the prefactor
involving $\Im ~ \tau _\ep$. Chiral splitting is achieved by introducing a
single internal loop momentum $p _\ep$, in terms of which also the
prefactor may be split,
\bea
{1 \over (2 \Im ~ \tau _\ep)^{n \over 2}}
=
\int d^n p _\ep \left |
e ^{i \pi \tau _\ep  p_\ep ^2} \right |^2
\eea
The functional integral (\ref{bfA}) exhibits further obstructions to chiral
splitting. The first is through the presence of the quartic fermion term
$\chi \bar \chi \psi _+ \psi _-$ in the matter action $I_m$; the second
is through the non-holomorphic dependence of the full fermion propagator,
\bea
\label{fulltwistedboson}
\< \p_z x (z) \p_w x (w) \>_\ep
& = & B_\ep (z,w) - {2 \pi \over  \Im ~\tau _\ep } ~ \omega _\ep (z)
\omega _\ep (w)
\no \\
\< \p_z x (z) \p_{\bar w} x (w) \>_\ep
& = & -2 \pi \delta (z,w) + {2 \pi \over \Im ~\tau _\ep } ~ \omega _\ep (z)
\overline{\omega _\ep (w)}
\eea
The first line may be established by comparing the stress tensors computed
from a variation of the scalar determinant (\ref{scalardet}) and from the full
$x$-field propagator. This procedure is the same as the one used in
(\ref{stressB}) and (\ref{stressZ}), but is now applied to the full twisted
scalar partition function and propagator. The second line in
(\ref{fulltwistedboson}) is obtained by applying $\p_{\bar w}$  to the first
line and then integrating in $w$; the remaining coefficient in front of $\omega
_\ep (z) \omega _\ep (w)$ is fixed by requiring that the integration versus
$\omega _\ep (w)$ vanish.

\medskip

Next, one proceeds in parallel with the proof of chiral splitting for the case
of flat space-time \cite{dp88, dp89}. 
The result may be summarized in terms of a set of effective
rules. The final formula for the integration over the matter fields may be
recast in the following form,
\bea
\label{chiralsplit}
&&
{\bf A}_C [\delta ]
=
\sum _\ep \int d^{10-n} p_I \int d^n p _\ep \int _{s\M_2} \prod _A
|dm^A|^2  \left | \A_C  [\delta, \ep ](p_\ep ) \right |^2
|e ^{i \pi p^\mu _I \hat \Omega _{IJ} p^\mu _J}|^2
\\
&&
\A _C [\delta ,\ep](p_\ep )
=
\bigg \< \prod _A \delta (\< H_A |B\>) \prod _{i=1}^N V_i ^{{\rm chi}}
\exp \bigg \{ \int {d^2 z \over 2\pi} \chiz S (z) + p^\mu _\ep \oint
_{B_\ep} \! \! dz \p_z x _+^\mu \bigg \} \bigg \> _+
\no
\eea
Here, $S(z)$ is the total worldsheet supercurrent, given by
$S(z)= S_m (z) + S_{gh}(z)$.
The variance of the Gaussian in the first line is given by the {\sl 
super-period matrix},
which is a shift of the bosonic
period matrix by an even, Grassmann valued, and nilpotent element, 
given by \cite{dp88},
\bea
\hat \Omega _{IJ}
= \Omega _{IJ}
- { i \over 8 \pi} \int d^2z \int d^2 w ~ \omega _I (z) \chiz S_\delta (z,w)
\chiw \omega _J(w)
\eea
Furthermore, $V_i ^{{\rm chi}}$ is the chiral part of the vertex operator 
$V_i$. All
contractions in this chiral correlator $\< \cdots \>_+$ are to be carried out
with the help of the propagators for the chiral fields $x_+$ and $\psi _+$
given in (\ref{chiralprops}). In the present paper, the emphasis will be on the
chiral measure and the cosmological constant. Therefore, the precise form of 
the
operators $V_i$ and their chiral part
$V_i ^{{\rm chi}}$ will not be needed and will not be presented here.


\subsection{The chiral measure in the $\mathbf{Z_2}$-twisted sectors}

In the absence of vertex operator insertions, the $p_\ep^\mu$-dependence of the
amplitude arises from the $x^\mu _+$-contractions of the term involving $p_\ep
^\mu$ with itself and with $S_m$, yielding
\bea
&&
\bigg \<
\exp \bigg \{ \int {d^2 z \over 2\pi} \chiz S (z) + p^\mu _\ep \oint
_{B_\ep} \! \! dz \p_z x _+^\mu \bigg \} \bigg \> _{x_+,\psi _+}
\\ && \hskip .6in
=
\bigg \<  \exp \biggl \{
i \pi \tau _\ep p_\ep ^2 - \half ~p^\mu _\ep  \int d^2 z \chiz \psi _+ ^\mu
\omega _\ep (z)
\no \\ && \hskip 1in
- { 1 \over 8 \pi} \int d^2z \int d^2 w \chiz \chiw \< S (z) S (w) \>_{x_+}
\biggr \> _{\psi _+} \< 1 \>_{x_+}
\no
\eea
where $\< 1 \> _{x_+}$ denotes the $\bZ_2$-twisted chiral boson partition
function. Carrying out also the $\psi _+$ contractions of the $p_\ep$-dependent
terms, the following result is obtained,
\bea
&&
\bigg \<
\exp \bigg \{ \int {d^2 z \over 2\pi} \chiz S (z) + p^\mu _\ep \oint
_{B_\ep} \! \! dz \p_z x _+^\mu \bigg \} \bigg \> _{x_+,\psi _+}
\\ && \hskip .4in
=
 \biggl (1
- { 1 \over 8 \pi^2 } \int d^2z \int d^2 w \chiz \chiw \< S (z) S (w)
\>_{x_+} \biggr ) Z_C[\delta; \ep]  \exp \{ i \pi \tilde \tau _\ep p_\ep ^2 \}
\no
\eea
Here, the product of the $\bZ_2$-twisted chiral scalar and fermion  partition
functions $\< 1 \>_{x_+}$ and $\< 1 \>_{ \psi _+}$ is denoted by $Z_C[\delta,
\ep]$. In the above formula, all the $p_\ep$-dependence may be regrouped in
terms of a Gaussian with the following variance,
\bea
\tilde \tau \equiv \tau - {i \over 8 \pi} \int d^2z \int d^2 w ~\chiz
\omega _\ep (z) S_{\delta +\ep} (z,w) \chiw \omega _\ep (w)
\eea
This correction has the same origin as the corrections to the period matrix
that lead to the super period matrix in the uncompactified string. Its proper
interpretation here is, however, more subtle, and will be presented in detail
later.

\medskip

The contributions of the ghost partition function and supercurrent correlators
are the same as they were in flat space-time and may be taken from \cite{II}.
Assembling all contributions, the chiral measure is given by the following
expression
\bea
\label{measureA}
\A_C [\delta;\ep ](p_\ep )
& = &
{\< \prod _a b(p_a) \prod _\alpha \delta (\beta (q_\alpha)) \>_M
\over \det \Phi _{IJ+} (p_a) \cdot \det \< H_\alpha |\Phi ^* _\beta \>}
~ {Z_C [\delta; \ep] \over Z_M [\delta] }
\exp  \{i \pi \tilde \tau _\ep p_\ep ^2 \}
\\ && \hskip .6in
\times \biggl \{
1 - { 1 \over 8 \pi^2 } \int \! d^2z \int \! d^2 w \chiz \chiw \< S (z) S (w)
\>_C \biggr \}
\no
\eea
The factor of $Z_M[\delta]$ stands for the chiral matter partition function of
flat space-time; it is of course independent of the twist $\ep$. This factor
must be divided out since it was already included in the definition used in
\cite{II} for the matter-ghost correlator
$\< \prod _a b(p_a) \prod _\alpha \delta (\beta (q_\alpha)) \>_M$ in flat
space-time. Detailed definitions and explicit expressions for the various
ingredients in the above formula were given in \cite{II}. Suffice it here to
remind the reader that the ghost insertion points $p_a$ and $q_\alpha$, with
$a=1,2,3$ and
$\alpha =1,2$ are arbitrary; that $\Phi _{IJ}$ and $\Phi ^* _B$ are
superholomorphic 3/2 forms and
$H_A$ is a super Beltrami differential, all of which are subject to certain
normalization conditions, spelled out  respectively in eq. (3.18), (3.29) and
(3.28) of \cite{II}. Note that all quantities in (\ref{measureA}) are expressed
with respect to the period matrix $\Omega _{IJ}$.

\medskip

The last step in the derivation of the consistent and slice-independent
measure for the $\bZ_2$-twisted theory is the change of variables from the
bosonic period matrix $\Omega _{IJ}$ to the super-period matrix $\hat \Omega
_{IJ}$. The super-period matrix is invariant under local supersymmetry. As was
shown in \cite{II}, this guarantees the existence of consistent bosonic moduli
and permits the consistent integration over odd supermoduli.  The reformulation
of superstring amplitudes in terms of the super-period matrix is one of the key
insights into two-loop superstring perturbation theory presented in
\cite{I,II,III,IV}, and was built on earlier work in  
\cite{dp88, dp89, dp89A, dp90}.

\medskip

To carry out the change of variables from $\Omega _{IJ}$ to $\hat \Omega
_{IJ}$, one proceeds as follows. 
As explained in \cite{II}, \S 3.3-\S 3.5,
the choice of $\hat\Omega_{IJ}$
as parameters for the even supermoduli
determines the Beltrami superdifferentials $H_A$
in the gauge-fixed formulas (\ref{gaugefixed0}) and (\ref{chiralsplit}).
The remaining difficulty is that the string amplitude is still
expressed in terms of correlators of conformal field theories
with respect to a background metric with period matrix 
$\Omega_{IJ}$ instead of $\hat\Omega_{IJ}$. Now
the difference between the two period matrices
is of order two in $\chi$ and thus nilpotent. The expansion in terms of a
Beltrami differential $\hat \mu$ is thus exact to first order,
\bea
\hat \Omega _{IJ}
=\Omega _{IJ} - i \int d^2z \hat \mu \omega _I \omega _J(z)
\eea
The process will therefore only affect the terms in (\ref{measureA}) that are
independent of $\chi$. The effect of this change of variables is a change in
the worldsheet metric by means of the Beltrami differential $\hat \mu$, defined
above (up to a diffeomorphism).  In any correlator, this change may be
implemented via the insertion of the stress tensor, as was explained in
\cite{II}.  The combination of partition functions and finite-dimensional
determinants may be treated by these methods, and we obtain,
\bea
&&
{\< \prod _a b(p_a) \prod _\alpha \delta (\beta (q_\alpha)) \>_M
\over \det \Phi _{IJ+} (p_a) \cdot \det \< H_\alpha |\Phi ^* _\beta \>}
~ {Z_C[\delta; \ep] \over Z_M[\delta]}(\Omega)
\no \\ && \hskip .6in =
{\< \prod _a b(p_a) \prod _\alpha \delta (\beta (q_\alpha)) \>_M
\over \det \Phi _{IJ+} (p_a) \cdot \det \< H_\alpha |\Phi ^* _\beta \>} ~
{Z_C[\delta; \ep]
\over Z_M[\delta] }(\hat \Omega)
\times \left \{
1 +  \int {d^2 z \over 2 \pi} \hat \mu (z)  \< T (z)\> _C \right \}
\no
\eea
Substituting this result into (\ref{measureA}) yields
\bea
\label{measureB}
&&
\A_C [\delta;\ep ](p_\ep )
=
{\< \prod _a b(p_a) \prod _\alpha \delta (\beta (q_\alpha)) \>_M
\over \det \Phi _{IJ+} (p_a) \cdot \det \< H_\alpha |\Phi ^* _\beta \>} (\hat
\Omega) ~ {Z_C [\delta; \ep] \over Z_M [\delta] } (\hat \Omega)
\exp  \{i \pi \tilde \tau _\ep p_\ep ^2 \}
\\ && \hskip 1in
\times \biggl \{
1 - { 1 \over 8 \pi^2 } \int d^2z \int d^2 w \chiz \chiw \< S (z) S (w)
\>_C + \int {d^2 z \over 2 \pi} \hat \mu (z)  \< T (z)\> _C \biggr \}
\no
\eea
The ghost part of the partition function and of the supercurrent and stress
tensor correlators as well as the finite-dimensional determinants in
(\ref{measureB}) are exactly the same as those for the uncompactified theory
and given by
\bea
\label{measureC}
&&
\A_M [\delta ]
=
{\< \prod _a b(p_a) \prod _\alpha \delta (\beta (q_\alpha)) \>_M
\over \det \Phi _{IJ+} (p_a) \cdot \det \< H_\alpha |\Phi ^* _\beta \>}
\\ && \hskip .7in
\times \biggl \{
1 - { 1 \over 8 \pi^2 } \int d^2z \int d^2 w \chiz \chiw \< S (z) S (w)
\>_M + \int {d^2 z \over 2 \pi} \hat \mu (z)  \< T (z)\> _M \biggr \}
\no
\eea
Here and above, the subscripts $M$ and $C$ are used on the correlators to
indicate whether they are evaluated in the flat Minkowski theory ($M$) or in
the compactified orbifold theory ($C$). The advantage of expressing $\A_C
[\delta;\ep ](p_\ep )$ in terms of $\A _M [\delta]$ is that the latter has
already been explicitly evaluated in \cite{IV}. The result is,
\bea
\label{measureD}
\A_M [\delta ] (\Omega)
& = &
\Z + {\zeta ^1 \zeta ^2 \over 16 \pi ^6} \cdot {\Xi _6 [\delta ](\Omega) \tet
[\delta] (0,\Omega)^4  \over \Psi _{10} (\Omega) }
\eea
where the normalized partition function $\Z$ of chiral matter, ghosts and
superghosts on the bosonic surface with $\chi=0$ is given by
\bea
\label{calZee}
\Z & \equiv & {\< \prod _a b(p_a) \prod _\alpha \delta (\beta (q_\alpha)) \>_M
\over \det \omega _I \omega _J (p_a)}
 \\
& = & {\tet [\delta] (0)^5 \tet (p_1+p_2+p_3 -3 \Delta) \prod _{a<b} E(p_a,p_b)
\prod _a \sigma (p_a)^3
\over
Z^{15} \tet [\delta] (q_1+q_2-2\Delta) E(q_1,q_2) \sigma (q_1)^2 \sigma (q_2)^2
\det \omega _I \omega _J (p_a)}
\no
\eea
All quantities entering this expression were defined in \S 2, except for the
chiral scalar partition function $Z^{-1}$ and the holomorphic 1-form $\sigma (z)$,
which are defined by
\bea
Z^3 & = & {\tet (z_1+z_2-w_0-\Delta) E(z_1,z_2) \sigma (z_1) \sigma (z_2)
\over
\sigma (w_0) E(z_1,w_0) E(z_2,w_0) \det \omega _I (z_J)}
\no \\
{\sigma (z) \over \sigma (w)}
& = &
{\tet (z-z_1-z_2+\Delta) E(w,z_1) E(w,z_2)
\over
\tet (w-z_1-z_2+\Delta) E(z,z_1) E(z,z_2)}
\eea
In each formula, the points $z_1,z_2$ and $w_0$ are arbitrary.

\medskip

In terms of $\A_M [\delta]$, the following expression is obtained for the
chiral measure,
\bea
\label{measureE}
&&
\A_C [\delta,\ep ](p_\ep ^\mu)
=
\A_M [\delta] ~ {Z_C[ \delta; \ep ] \over Z_M[\delta]}
\exp  \{i \pi \tilde \tau _\ep p_\ep ^2 \}
\\ && \hskip 1in
\times \biggl \{
1 - { 1 \over 8 \pi^2 } \int \! \! d^2 \! z \int \! \! d^2 \!  w \chiz \chiw
\biggl  (\< S_{m} (z) S_{m}(w) \>_C - \< S_{m} (z) S_{m}(w) \>_M \biggr  )
\no \\ && \hskip 1.4in
+ {1 \over 2 \pi}\int d^2 z  \hat \mu (z)
\biggl  ( \< T_{m}(z)\> _C - \< T_{m}(z)\> _M \biggr )  \biggr \}
\no
\eea
Here, we have used the fact that the ghost contributions in the correlators
$\< S(z) S(w)\>_M$ and $\< S(z) S(w)\>_C $ as well as in $\< T(z)\>_M$ and
$\<T(z)\>_C$ are identical and cancel out upon taking differences, leaving only
the matter correlators in (\ref{measureE}), evaluated in the sector
twisted by $\ep$. It is understood that all parts of
(\ref{measureE}) are expressed with respect to the super-period matrix $\hat
\Omega _{IJ}$. The only ingredient in (\ref{measureE}) which needs further
clarification is the correction to the Prym period $\tilde \tau_\ep $, to be
presented in the subsequent subsection.


\subsection{The super-Prym period}

In this subsection, the role of the quantity $\tilde \tau_\ep $ in the
exponential involving the internal loop momentum $p_\ep$ in
(\ref{measureE}) is clarified. By construction, $\tilde \tau_\ep $ is invariant
under local worldsheet supersymmetry. At first sight, this property would
appear to qualify $\tilde \tau_\ep $ for the supersymmetric generalization of
the Prym period $\tau_\ep $, but this hypothesis is invalid for the following
reasons.

\medskip

Recall the Schottky relations on a bosonic Riemann surface with period matrix
$\Omega_{IJ}$, already presented in (\ref{Schottky0}) and reformulated in
(\ref{Schottky}). The solution of the Schottky relations for $\tau _\ep$ as a 
function of
$\Omega$ for a twist $\ep$ was denoted by the function $\tau_\ep = R_\ep
(\Omega)$ in (\ref{Schottky0}).
Actually, this function will be multi-valued because, for given
$\Omega$, the Schottky relations determine $\tau$ only up to a shift $\tau _\ep
\to \tau _\ep +4$. This multivaluedness is required in particular by the fact
that the Dehn twist $A_1 B_2 A_1 ^{-1} B_1 ^{-1}$, which does not act on
$\Omega$, shifts $\tau _\ep \to \tau _\ep + 4$, as shown in \cite{DVV}.

\medskip

The implications for the super-Prym period and for the quantity $\tilde \tau
_\ep$ are as follows. The genus 2 super-Riemann surface, specified by the
supermoduli $(\Omega _{IJ}, \zeta ^\alpha)$, uniquely projects to a bosonic
Riemann surface with period matrix $\hat \Omega _{IJ}$. This projection
automatically entails an associated super-Prym period $\hat \tau_\ep $, which
is defined through the bosonic Schottky relations from the super-period matrix
$\hat \Omega _{IJ}$. In summary, we have the relations,
\bea
\tilde \tau_\ep - \tau_\ep
& = &
- { i \over 8 \pi } \int \! d^2z \int \! d^2 w ~\omega  _\ep (z) \chiz
S_{\delta + \epsilon} (z,w) \chiw \omega _\ep (w)
\no \\
\hat \Omega _{IJ} - \Omega _{IJ}
& = &
- { i \over 8 \pi } \int \! d^2z \int \! d^2 w ~\omega _I (z) \chiz
S_\delta  (z,w) \chiw \omega _J (w)
\eea
as well as the defining relations,
\bea
\tau_\ep  = R_\ep  (\Omega)
\hskip 1in
\hat \tau_\ep  = R_\ep (\hat \Omega)
\eea
By their very construction, $\hat \Omega$, $\hat \tau$ and $\tilde \tau$
are supersymmetric invariant. As a result, the difference
\bea
\Delta \tau _\ep \equiv \tilde \tau_\ep - \hat \tau_\ep = \tilde \tau_\ep -
\tau_\ep  - (\hat \Omega _{IJ} - \Omega _{IJ}) \p_{IJ} R _\ep (\hat \Omega)
\eea
is also a supersymmetric invariant. This invariance may be verified directly,
using the above definitions. Since this expression is bilinear in $\chi$
already, only the supersymmetry variation of $\chi$ is required and this is
given by $\delta _\xi \chiz= - 2 \p _{\bar z}
\xi ^+$, so that
\bea
\delta _\xi \Delta \tau _\ep
& = &
-i \int \! d^2 z \xi ^+(z) \chiz \biggl [ \omega _\ep (z) \omega _\ep (z) -
\omega _I (z) \omega _J (z) \p_{IJ} R_\ep  \biggr ]
\nonumber \\
& = & -{1 \over 2 \pi} \int \! d^2 z \xi ^+(z) \chiz \biggl [ \delta _{zz}
\tau - \delta _{zz} R _\ep \biggr ]
\eea
which cancels in view of $\tau_\ep = R_\ep (\Omega)$.

\medskip

The difference $\Delta \tau _\ep$ is non-vanishing. This may
be shown by going to split gauge,
defined by $\chi_{\bar z}^+
=\sum_{\alpha=1}^2\zeta^{\alpha}\delta(z,q_{\alpha})$
with $S_{\delta}(q_1,q_2)=0$, in which $\hat \Omega = \Omega$, $\hat
\tau_\ep = \tau_\ep$, so that
\bea
\label{Deltatau}
\Delta \tau _\ep =- { i \zeta ^1 \zeta ^2 \over 4 \pi } \omega  _\ep
(q_1) S_{\delta + \epsilon} (q_1,q_2)  \omega _\ep (q_2)
\eea
but this quantity is manifestly non-vanishing when $\ep\not=0$ and $S_\delta
(q_1,q_2)=0$.

\section{The Chiral Measure for General Compactifications}
\setcounter{equation}{0}

In this section, the chiral superstring measure will be constructed for
more general compactifications than those involving $\bZ_2$ twists.
The total space-time for the compactifications considered here will again be
denoted by $C$. The chiral measure will be evaluated at fixed even spin
structure. As announced in \cite{I},
under some basic but mild assumptions, it will be shown that the
chiral blocks are independent of any choices of gauge slice, just as they were
in flat space-time. The assumptions are

\begin{enumerate}
\item The compactification only modifies the matter part of the
theory, leaving the superghost part unchanged.
\item The compactification respects local worldsheet supersymmetry,
so that the super-Virasoro algebra with matter central charge $c=15$ is
preserved.
\end{enumerate}

A simple prescription for their calculation will be given first in split gauge
and then in terms of the OPE of two supercurrents.


\subsection{The Result of chiral splitting}

Chiral splitting (the fact that the superstring amplitudes are the norms 
squared of supermeromorphic functions on supermoduli space) holds
for the contribution of individual super-conformal families. In the case of the
$\bZ_2$ orbifold measure discussed in \S 2, super-conformal families were
labelled by  the spin structure $\delta$, by the twist $\ep$ and by the
internal momenta, $p^\mu _I$ in the untwisted sector and $p_\ep ^\mu$ in the
twisted sectors. For more general compactifications, the super-conformal
family structure may be more complicated. The super-conformal families will be
labeled here by the spin structure $\delta$ and the remaining
characterization will be summarized by a label $\lambda$. 
The spin structure label is
singled out here because it must coincide with the spin structure of the ghost
part of the measure.

\medskip

All the serious complications in the derivation of the superstring measure have
to do with the gauge fixing, ghost and  finite-dimensional determinant
contributions. In view of our above assumptions, all these contributions are
sensibly the same as in flat Minkowski space-time or in the orbifolded
space-times. Thus, the general form of chiral splitting for the superstring
measure for strings moving on the compactified space-time $C$ is readily
adapted from the expression for the orbifold case in (\ref{chiralsplit}) at
two-loops,\footnote{A subscript $C$ has been appended to the supercurrent $S$
because for general compactifications, the supercurrent may not assume the flat
space-time form; the latter will henceforth be denoted by $S_M$.}
\bea
{\bf A}_C [\delta ]
& = &
\sum _\lambda \int _{s\M_2} \prod _A |dm^A|^2
\left | \A_C  [\delta;\lambda ] \right |^2
\\
\A _C [\delta;\lambda ]
& = &
\bigg \< \prod _A \delta (\< H_A |B\>) \prod _i V_i ^{{\rm chi}}
\exp \bigg \{ \int {d^2 z \over 2\pi} \chiz S_C(z)  \bigg \} \bigg \> _{\lambda,+}
\no
\eea
Here, the subscript $\lambda$ refers to the fact that the amplitude is evaluated in
the sector associated with the super-conformal family $\lambda$. Notice that, 
compared to
(\ref{chiralsplit}), no additional factor depending on internal momenta is
exhibited. The presence of internal momenta amongst the labels for
super-conformal blocks is indeed model dependent and is assumed to be part of
the definition of $\< \cdots \> _{\lambda,+}$.

\medskip

In terms of $\A_M [\delta]$, the following expression is obtained for the
chiral measure,
\bea
\label{measureF}
&&
\A_C [\delta;\lambda ]
=
\A_M [\delta] ~ {Z_C[\delta;\lambda] \over Z_M[\delta]} 
\biggl \{ 1 - {1 \over 2 \pi}\int d^2 z \hat
\mu (z)   \left  \{ \< T_{Cm}(z)\> _{C\lambda} - \< T_{Mm} (z)\> _M \right \}
\\ && \hskip .8in
 - { 1 \over 8 \pi^2 } \int d^2z \int d^2 w \chiz \chiw
\left \{ \< S_{Cm} (z) S_{Cm} (w) \>_{C\lambda} 
- \< S_{Mm} (z) S_{Mm}(w) \>_M \right \}    \biggr \}
\no
\eea
Here, we have used the fact that the ghost contributions in the correlators
$\< S_M(z) S_M(w)\>_M$ and $\< S_C(z) S_C(w)\>_{C\lambda} $ as well as 
in $\< T_M(z)\>_M$ and $\<T_C(z)\>_{C\lambda}$ are identical and 
cancel out upon  taking differences, leaving
only the matter correlators in (\ref{measureF}). It is understood that all
parts of (\ref{measureF}) are expressed with respect to the super-period matrix
$\hat \Omega _{IJ}$.


\subsection{Slice Independence of the Measure for Compactifications}

The slice independence of ${\cal A}_C[\delta; \lambda]$ may be 
deduced from the slice
independence of ${\cal A}_M[\delta]$, which was already established in \cite{II,III},
together with general properties of the supercurrent and stress tensor
correlators which enter into (\ref{measureF}). Since the ghost parts have
cancelled out of the stress tensor and supercurrent correlators, their
singularities with the ghost insertion points have also cancelled. Therefore,
both $\<T_m (z) \>_{C\lambda}$ and $\< T_m (z) \>_M$ are holomorphic and their
difference is a holomorphic 2-form that is well-defined on the Riemann surface,
and thus the formula for ${\cal A}_C[\delta;\lambda]$ is independent of the choice 
of $\hat \mu$ within a given super-conformal family $\lambda$.

\medskip

The supercurrent insertions are similarly independent of the points
$q_\alpha$. Since the ghost parts of $S_C$ and $S_M$ coincide, all the
singularities in $z$ and $w$ with the insertion points $p_a$ and $q_\alpha$ are
identical, and cancel upon taking the difference between the $C$ and $M$
contributions. Thus, the only possible singularities in the $SS$ correlator is
when $z\to w$. But this singularity is precisely cancelled by the presence of
the stress tensor contribution, as was shown in the flat case in
\cite{II,III}.

\medskip

The mutual cancellation of these singularities is also a necessary and
sufficient in order to maintain local worldsheet supersymmetry, as shown
in \cite{II}. Indeed, each singularity presents an obstruction to
supersymmetry invariance since the supersymmetry variation $\delta _\xi
\chiz = - 2 \p _{\bar z} \xi ^+$ will pick up non-vanishing contributions
at the poles. Just as in \cite{II} for flat space-time $M$, the effect
of the singularity in the supercurrent correlator at $z=w$ is precisely
cancelled by the variation $\delta _\xi \mu = \xi ^+ \chiz$ of the stress
tensor term. In summary, the insertion of 
$\< S_{Cm} (z) S_{Cm} (w) \> _{C \lambda} - 
\< S_{Mm} (z) S_{Mm} (w) \>_M$ is completely singularity free and hence 
${\cal A}_C[\delta; \lambda]$ 
is slice independent, just as ${\cal A}_M[\delta]$ was.


\subsection{The Measure for Compactifications in Split Gauge}

To evaluate explicitly the superstring measure, we
now choose pointlike insertions for $\chi$
\be
\chiz (z) = \zeta^1 \delta(z,x_1)+ \zeta^2 \delta(z,x_2)
\ee
As in \cite{III}, the slice independence of ${\cal A}_C[\delta; \lambda]$
guarantees well-defined and regular limits as
$x_{\alpha}\to q_{\alpha}$. We obtain
\bea
{\cal A}_C [\delta; \lambda ]
&= &
{\cal A}_M [\delta] {Z_C [\delta;\lambda] \over Z_M [\delta]} 
\biggl \{1- {\zeta ^1 \zeta ^2 \over 4
\pi ^2}  [\< S_{Cm} (q_1) S_{Cm} (q_2) \> _{C\lambda} - 
\< S_{Mm} (q_1) S_{Mm} (q_2) \>_M ]
\nonumber \\
&& \qquad \qquad \quad
+{1 \over 2 \pi} \sum _a \mu _a (q_1,q_2)  [\<T_{Cm} (p_a) \> _C(f) - 
\<T_{Mm} (p_a)\>_M ] \biggr \}
\eea
where the flat Minkowski space-time objects ${\cal A}_M [\delta]$ 
and $\Z$ were
given in (\ref{measureD}) and (\ref{calZee}). Here, $\mu_a$ arises 
from the Beltrami
differential $\hat\mu$ representing the shift from the period 
matrix $\Omega$ to the
superperiod matrix $\hat \Omega$, given by
\bea
\mu _a(q_1,q_2)
=
{\zeta ^1 \zeta ^2 \over 4 \pi} \varpi  _a (q_1,q_2) S_\delta
(q_1,q_2)
\eea
Here, $\varpi _a (z,w)$ is the unique form of degree 1 and holomorphic in both
$z$ and $w$ such that the 3 holomorphic 2-forms $\varpi _a (z,z)$ are
normalized by $\varpi _a (p_b,p_b)=\delta _{ab}$. Explicit forms were given in
\cite{II}, eq. (1.15).
The difference of the supercurrent and stress tensor correlators on manifolds
$C$ and $M$, given by
\bea
Q (q_1,q_2)
& \equiv &
+\< S_{Cm} (q_1) S_{Cm} (q_2) \> _{C\lambda} - 
\< S_{Mm} (q_1) S_{Mm} (q_2) \>_M
\nonumber \\ && \qquad
-\half \sum _a \varpi  _a (q_1,q_2) S_\delta (q_1,q_2)
[\<T_{Cm} (p_a) \> _{C\lambda} - \<T_{Mm} (p_a)\>_M ]
\eea
is holomorphic in both $q_1$ and $q_2$ and odd under the interchange of $q_1$
and $q_2$. Therefore, its dependence on $q_\alpha$ is determined uniquely up to
a $q_\alpha$-independent multiplicative factor.
The expression for ${\cal A}_C[\delta; \lambda]$ becomes,
\be
{\cal A}_C[\delta; \lambda] = {Z_C[\delta;\lambda] \over Z_M[\delta]} \biggl \{
{\cal Z} + {\zeta ^1 \zeta ^2 \over 16 \pi ^6} \cdot
{\tet [\delta ](0)^4 \Xi _6  [\delta ] \over \Psi _{10}}
- {\zeta ^1 \zeta ^2 \over 4 \pi ^2} {\cal Z} Q(q_1,q_2) \biggr \}
\ee
A further simplification takes place in the split gauge, defined by
the following relation between the insertion points $q_1$ and $q_2$, $S_\delta
(q_1,q_2)=0$. In this gauge, the expression for ${\cal A}_C[\delta; \lambda]$
simplifies to
\be
{\cal A}_C[\delta;\lambda] = {Z_C[\delta;\lambda] \over Z_M [\delta]} \biggl \{
{\cal Z} + {\zeta ^1 \zeta ^2 \over 16 \pi ^6} \cdot
{\tet [\delta ](0)^4 \Xi _6  [\delta ] \over \Psi _{10}}
- {\zeta ^1 \zeta ^2 \over 4 \pi ^2} {\cal Z} \< S_{Cm} (q_1) S_{Cm} (q_2) \>
_{C\lambda} \biggr \}\, .
\ee


\subsection{The Measure via a Leading Supercurrent OPE Operator}

In this subsection, an alternative formula for the chiral measure is provided
in terms of simple data that may be obtained from the supercurrent OPE.
This calculation may be carried out in terms of the operators $\O_M$ and 
$\O_C$,
defined as follows,
\bea
\label{OPE}
S_C(z) S_C(w) &=&  {1 \over 4} { T_C(z) + T_C(w) \over z-w} + (z-w) \O_C
(w)  + \O (z-w)^2
\nonumber \\
S_M(z) S_M(w) &=&  {1 \over 4} { T_M(z) + T_M(w) \over z-w} + (z-w) \O_M
(w)  + \O (z-w)^2
\eea
Notice that the leading cubic singularity cancels since the central charges for
$M$ and for $C$ are assumed to be the same, namely $c=15$.
A convenient form for the chiral measure based on the above OPE operators is
obtained by letting all points $p_a$ collapse to the point $q_2$ and
subsequently letting $q_1 \to q_2$.

\medskip

Using methods similar to those employed in \cite{III}, section \S 3.4, a
limiting formula is obtained for the summation against $\mu _a$ of the full
stress tensor  $T(p_a) \equiv \< T_{Cm} (p_a) \>_C - \< T_{Mm} (p_a) \> _M$ in
terms of holomorphic Abelian differentials $\omega ^* _\alpha $ with
normalization $\omega ^* _\alpha (q_\beta ) =\delta _{\alpha \beta}$.
In the OPE relation of (\ref{OPE}), it is customary to expand with respect
to the coordinate $q_2$ of the second operator, so we also let $p_a\to q_2$ for
all $a=1,2,3$. As a result,
\be
\lim _{p_a \to q_2} \sum _a \varpi _a (q_1,q_2) T(p_a)
=
\half {1 \over \p \omega _1 ^* (q_2)} \p T(q_2)
-
{\p \omega _2 ^* (q_2) \over \p \omega _1 ^* (q_2)}  T(q_2)
\ee
Expanding in powers of $q_2-q_1$ up to and including second order, this
limit reduces to
\be
T(q_2) + \half (q_1-q_2) \p T(q_2) + (q_1-q_2)^2 \biggl \{\half f(q_2) \p
T(q_2) -3  T_0 (q_2) T(q_2) \biggr \}
\ee
Here, we have introduced the following notations, familiar from \cite{I,II}
\bea
f(w) &=& \omega _I(w) \p _I \ln \tet (2w-w_0-\Delta) - \p _w \ln E(w,w_0)
+ \p _w \ln \sigma (w)
\nonumber \\
T_0(w) &=& \half \omega _I(w) \omega _J(w) \p _I \p _J \ln \tet (2w
-w_0-\Delta) -{1 \over 6} \p f(w) + {1 \over 6} f(w)^2 - T_1(w)
\nonumber \\
T_{1/2}(w) &=& \half \omega _I(w) \omega _J(w) \p _I \p_J \ln \tet
[\delta ](0) - T_1(w)
\eea
Collecting these results, the expansion up to order $\O (q_1-q_2)^2$ of $Q$ is
given by
\bea
Q(q_1,q_2) &=&
+\< S_C (q_1) S_C (q_2) \> _C - \< S_M (q_1) S_M (q_2) \>_M
\\ &&
- \half {\<T(q_2)\> \over q_1-q_2} - {1\over 4} \<\p T(q_2)\>
-\half (q_1-q_2) \biggl [(T_{1/2} - 3 T_0 )\<T\> + \half
f \< \p T\> \biggr ](q_2)\, .
\nonumber
\eea
It is easy to check that, within this approximation, $Q(q_1,q_2)$ is indeed a
form of weight $(3/2, 3/2)$, even though individual terms in its expression
above do not transform covariantly under conformal reparametrizations $z \to
z'= \varphi (z)$. To check this, notice that $S(z)$ and $T(z)$ are tensors of
weights 3/2 and 2 respectively, while $f$, $T_1$, $T_{1/2}$ and $T_0$
transform as connections,
\bea
\varphi ' (z)        f'(z') &=& f(z) - {3 \over 2} { \varphi ''(z)
\over \varphi '(z)}
\\
\varphi ' (z) ^2  T'_0 (z') &=& T_0(z) -{1 \over 3}
{\varphi ''(z) \over \varphi '(z)} f(z) +{1 \over 6} {\varphi
'''(z) \over \varphi '(z)}  \nonumber \\
\varphi ' (z) ^2  T'_n (z') &=& T_n (z) + {6n^2 -6n+1 \over 12} \biggl \{
{\varphi '''(z) \over \varphi '(z)} -{3 \over 2} \left ( {\varphi ''(z)
\over \varphi '(z)} \right )^2 \biggr \}
\qquad n=1/2, 1
\nonumber
\eea
Assuming that the OPE of two supercurrents is as given in
(\ref{OPE}), we get
\bea
Q(q_1,q_2) &=&  (q_1-q_2) \hat Q(q_2) + \O(q_1-q_2)^2
 \\
\hat Q (w) &=& \biggl [\< \O _C \>_C - \< \O _M
\>_M  - {1 \over 8} \<\p^2 T\> -\half (T_{1/2} - 3 T_0 ) \<T\> - {1
\over 4} f \< \p T\> \biggr ](w)
\nonumber
\eea
It remains to evaluate ${\cal Z} Q(q_1,q_2)$. Since this quantity is
independent of both $q$'s, we first let $q_1 \to q_2$ and then set
$q_2=p_3=\Delta +\nu _3$. The quantity may now be evaluated using the
methods developed for flat space-time, and we find
\be
{\cal Z} Q(q_1,q_2) =  {\tet [\delta](0, \Omega)^8 \over \M _{\nu_1
\nu_2}^2}
\cdot {\hat Q (p_3) \over \omega _{\nu _1}(p_3)^2 \omega _{\nu_2}(p_3)^2}
\ee
Substituting this result into the measure factor, we obtain the chiral measure
\be
{\cal A}_C = {Z_C \over Z_M} \biggl \{
{\cal Z} + {\zeta ^1 \zeta ^2 \over 16 \pi ^6} \cdot
{\tet [\delta ]^4 \Xi _6  [\delta ] \over \Psi _{10}}
- {\zeta ^1 \zeta ^2 \over 4 \pi ^2}  \cdot
{\tet [\delta]^8 \over \M _{\nu_1 \nu_2}^2}
\cdot {\hat Q (p_3) \over \omega _{\nu _1}(p_3)^2 \omega _{\nu_2}(p_3)^2}
\biggr \}\, ,
\ee
as well as the contribution to the cosmological constant from the
left-moving sector, by integrating over $\zeta ^\alpha$.

\section{Calculation of two-loop chiral blocks for $\mathbf{Z_2}$ twists}
\setcounter{equation}{0}

The starting point is the chiral measure of (\ref{measureE}), evaluated in
split gauge defined by $S_\delta (q_1,q_2)=0$. In this gauge, the following
simplifications occur : $\hat \Omega _{IJ}=\Omega _{IJ}$, and thus $\hat
\mu=0$ and the matter supercurrent correlator in $M$ vanishes because the
fermion propagator $S_\delta$ is evaluated between the points $q_1$ and $q_2$.
The remaining expression is given by,\footnote{For $\bZ_2$-twisting, 
the supercurrents
$S_C$ and $S_M$  take on the same functional form (as do the stress tensors
$T_C$ and $T_M$); thus, the subscripts $M$ and $C$ will be dropped from the
operators. The subscript $C$ on the correlator will be replaced with
the twist $\ep$ for each twisted sector.}
\bea
\label{orb}
{\cal A} _C [\delta; \ep]( p_\ep )
=
{Z_C [\delta; \ep] \over Z_M[\delta]  } e^{i \pi (\tau_\ep + \Delta \tau_\ep)
p_\ep ^2}
\bigg\{ \Z  + {\zeta^1\zeta^2\over 16\pi^6}
 {\Xi_6[\delta] \tet [\delta] (0)^4 \over  \Psi_{10}} -
{\zeta^1\zeta^2\over 4\pi^2}  \Z \<S_m(q_1)S_m(q_2)\>_\ep \bigg\}
\no\\
\eea
where $\Z$ was defined in (\ref{calZee}).  As this formula is written in split
gauge, we have $\hat \Omega _{IJ}=\Omega _{IJ}$, $\hat \tau _\ep =\tau_\ep$ and
the expression for $\Delta \tau _\ep$, derived in (\ref{Deltatau}), is
\bea
\Delta \tau _\ep
=
- { i \zeta ^1 \zeta ^2 \over 4 \pi } \omega  _\ep
(q_1) S_{\delta + \epsilon} (q_1,q_2)  \omega _\ep (q_2)
\eea
The focus of this paper will be on the chiral measure and the cosmological
constant, both of which receive contributions only from the top term in $\zeta
^1 \zeta ^2$. The resulting chiral measure takes the following form,
\bea
d\mu _C [\delta; \ep] (p _\ep)
& \equiv &
\int d\zeta ^2 d \zeta ^1 {\cal A} _C [\delta; \ep]( p_\ep )
\\
& = &
 e^{i \pi \tau_\ep  p_\ep ^2}  {Z_C [ \delta, \ep] \over Z_M [\delta] }
\bigg\{   {\Xi_6[\delta] \tet[\delta](0)^4 \over  16\pi^6 \Psi_{10}} -
{\Z \over 4\pi^2} \<S_m(q_1)S_m(q_2)\>_\ep + i \pi p_\ep ^2 \Gamma [\delta,
\ep] \bigg\}
\no
\eea
where the following definition has been made,
\bea
\label{Gamma}
\Gamma [\delta; \ep]  \equiv
\Z \int d\zeta ^2 d \zeta ^1  \Delta \tau _\ep
=
- { i \Z \over 4 \pi } \omega  _\ep
(q_1) S_{\delta + \epsilon} (q_1,q_2)  \omega _\ep (q_2)
\eea
It remains to calculate the various terms in the above expression in terms of
$\tet$-functions, which will be the subject of the remainder of this section.

\medskip

It will be assumed that 
$n$ dimensions are being $\bZ_2$-twisted,
leaving $10-n$ dimensions untwisted,
and that $n$ is at most $8$. In this case, only even spin structures
$\delta$ need to be taken into account. In particular, this will be the case
for the models of \cite{KKS}. The chiral partition functions $Z_M$ and
$Z_C$ are well-known, and given by
\bea
Z_M [\delta] & = &
\biggl ( {\tet [\delta ](0,\Omega ) \over Z ^3 } \biggr ) ^5
\\
Z_C [\delta; \ep] & = &
\biggl ( {\tet [\delta ](0,\Omega ) \over Z^3} \biggr ) ^{5-n/2}
\biggl ( {\tet [\delta ^+ _j](0,\Omega) \tet [\delta ^- _j](0,\Omega) \tet
[\delta + \ep] (0,\Omega) \over Z^3 \tet _j (0,\tau_\ep)^2}
\biggr ) ^{n/2}
\no
\eea
for any pair $\delta ^\pm_j$ such that $\ep = - \delta ^+_j +\delta ^-_j$.


\subsection{Calculation of the supercurrent correlator}

The supercurrent correlator may be calculated in terms of the twisted scalar
and fermion propagators evaluated in section \S 2,
\bea
\<S_m (q_1) S_m(q_2)\>_\ep  = {n \over 4} B_\ep (q_1,q_2) S_{\delta + \ep}
(q_1,q_2)
\eea
where the twisted scalar propagator is
\bea
B_\ep (z,w) = S _{\delta _j ^+} (z,w) S_{\delta _j ^-} (z,w)
+ b_j \omega _\ep (z) \omega  _\ep (w)
\eea
for any pair $\delta _j ^\pm$ such that $\ep = - \delta ^+_j + \delta ^-_j$.

\medskip

For given $\ep\not=0$, non-vanishing contributions will arise only from
$[\delta; \ep]$ where both $\delta$ and $\delta +\ep$ are even characteristics.
Upon choosing $\delta = \delta _i ^+$ (the choice $\delta = \delta _i ^-$ leads
to the same result), we have
\bea
B_\ep (q_1,q_2) = S _{\delta} (q_1,q_2) S_{\delta +\ep} (q_1,q_2)
+ b_i \omega _\ep (q_1) \omega  _\ep (q_2)
\eea
As $q_1,q_2$ obey the split gauge relation $S_\delta (q_1,q_2)=0$, the first
term on the rhs above vanishes and we have $ B_\ep (q_1,q_2) =  b_i \omega _\ep
(q_1) \omega  _\ep (q_2)$. Using the explicit expression for $b_i$, computed in
(\ref{littleb}), we have
\bea
\<S_m (q_1) S_m(q_2)\>_\ep
= - i \pi  n  \omega _\ep (q_1) \omega  _\ep (q_2)
S_{\delta _i ^-} (q_1,q_2) \p _ {\tau _\ep} \ln \tet _i (0,\tau_\ep)
\eea
The combination of this correlator with the factor of $\Z$  may be
re-expressed conveniently in terms of $\Gamma [\delta; \ep]$,
\bea
\Z \<S_m (q_1)S_m(q_2)\> _\ep
=
4 \pi^2 n \Gamma [\delta; \ep] \p_ {\tau_\ep} \ln \tet _i (0,\tau_\ep)
\eea
This leads to the following formula for the chiral measure in terms of $\Gamma
[\delta; \ep]$,
\bea
\label{finalmeasure}
d\mu _C [\delta; \ep] (p _\ep)
=
e^{i \pi \tau_\ep  p_\ep ^2}  {Z_C [ \delta, \ep] \over Z_M [\delta] }
\bigg\{   {\Xi_6[\delta] \tet[\delta](0)^4 \over  16\pi^6 \Psi_{10}}
+ \biggl ( i \pi p_\ep ^2 - n \p_{\tau_\ep} \ln \tet _i (0,\tau_\ep) \biggr )
\Gamma [\delta; \ep] \bigg \}
\eea
 It only remains to evaluate $\Gamma [\delta; \ep]$.


\subsection{Calculation of $\Gamma [\delta; \ep]$}

In the present subsection, $\Gamma [\delta; \ep]$ will be evaluated in terms of
$\tet$-constants. The calculation will be carried out in split gauge, just as
for the chiral measure in flat space-time \cite{IV}. The key challenge
presented by the calculation of $\Gamma [\delta; \ep]$ is its overall sign.
This sign is uniquely fixed by the definition of $\Gamma [\delta; \ep]$, but
during the course of the evaluation, a number of non-intrinsic signs appear and
need to be determined. For example, the $\tet$-constant itself $\tet
[\kappa](0)$ may change sign when a full period is added to $\kappa$.

\medskip

Careful choices for the twist and spin structure assignments are
needed, not just for their expression congruent mod 1, but including
full periods if they arise as well. The choices made for this calculation are
those given in (\ref{deltanu}) where
\bea
\ep = - \nu _a + \nu _b
\hskip 1in
\delta & = & \delta ^+ _i = \nu _a + \nu _c + \nu _d
\nonumber \\
\delta +\ep & = & \delta ^- _i = \nu _b + \nu _c + \nu _d
\eea
By choosing $(abcdef)$ to be an appropriate permutation of $(123456)$,
any twist and spin structure assignment may be reached by these conventions.

\medskip

The general expression for $\Gamma [\delta; \ep]$ was given in (\ref{Gamma});
it involves the partition function $\Z$ which was given in (\ref{calZee}).
Using the expression for $\Z$ in split gauge, calculated in eq (3.15) of
\cite{IV}, we obtain
\bea
\label{GammaZee}
\Gamma [\delta , \ep] & \equiv &
- {i \over 4 \pi} \Z \omega _\ep (q_1) S_{\delta _i ^-} (q_1,q_2) \omega
_\ep (q_2)
\nonumber \\
\Z & = & - {C \over C_r ^2 C_s ^2} \cdot
{\tet [\delta ]^5 E(p_r ,p_s )^4 \sigma (p_r )^2 \sigma (p_s )^2
\over
\tet [\delta] (q_1 + q_2 - 2 \Delta) E(q_1,q_2) \sigma (q_1)^2 \sigma
(q_2)^2} \cdot {1 \over \M_{\nu _r \nu _s}^2}
\eea
where $p_r = \nu _r + \Delta$ and $p_s = \nu _s + \Delta$ are two
arbitrary branch points. The exponential factors were also introduced in
\cite{IV} and are given by
\bea
C & = & - \exp \{ -4 \pi i \nu _l ' (2 \Omega \nu _r ' + 2 \nu _r '') \}
=  - \exp \{ -8 \pi i \nu _s '  \Omega \nu _r '  \}
\nonumber \\
C _{r,s } & =  & \exp \{ - \pi i \nu _{r,s}' \Omega \nu _{r,s}' - 2 \pi i
\nu _{r,s} ' \nu _{r,s}'' \}
\eea
We shall make use of the following expression, derived in (\ref{PrymSzego}), 
for
the product of the Prym differentials at two different points $z,w$,
\bea
\omega _\ep (z) \omega _\ep (w)
=
- \sigma (\mu_i,\mu_j) {1 \over \pi ^2 \tet _k ^4} \biggl ( S _{\delta_i ^+} 
(z,w) S_{\delta _i
^-} (z,w) - S _{\delta_j ^+} (z,w) S_{\delta _j ^-} (z,w) \biggr )
\eea
In view of the earlier choice $\delta = \delta _i ^+$, and the split gauge 
condition
$S_\delta (q_1,q_2)=0$,
the above formula simplifies when $z=q_1$ and $w=q_2$,
\bea
\label{Prymsplit}
\omega _\ep (q_1) \omega _\ep (q_2)
=
\sigma (\mu_i,\mu_j) {1 \over \pi ^2 \tet _k ^4} S _{\delta_j ^+} (q_1,q_2) 
S_{\delta _j ^-}
(q_1,q_2)
\eea
Expressions in terms of $\tet$-constants are most easily obtained by placing
insertion points at branch points. If $q_1,q_2$ are in split gauge, then their
limits to branch points are such that $q_1$ and $q_2$ belong to the same set in
the partition of all six branch points associated with $\delta$. The key
difficulty in the evaluation of $\Gamma [\delta; \ep]$ is that $\Z$ diverges as
the points $q_1$ and $q_2$, are taken to branch points, since $\tet [\delta
](q_1+q_2-2 \Delta)$ vanish. It turns out that one of the three Szego kernels
arising in $\Gamma [\delta; \ep]$ also vanishes, thereby making their ratio
finite.

\bigskip

\subsubsection{Linearization of split gauge around branch points}

\medskip

To circumvent the above problem, we parametrize the points $q_1$ and
$q_2$ as follows (here the argument is similar to
\cite{IV}, \S 3.6.2),
\bea
\label{splitgauge}
q_1 (t) & = & p_c + t \dot q_1  + \O (t^2)
\nonumber \\
q_2 (t) & = & p_d + t \dot q_2  + \O (t^2)
\eea
The split gauge relation between $q_1$ and $q_2$ is clearly obeyed at the
point $t=0$, since $\tet [\delta](\nu _c - \nu_d)\sim \tet [\nu_a]=0$. The
vanishing factors in the numerator and denominator are
\bea
S_{\delta +\epsilon}(p_c,p_d)  \sim \tet
[\delta +\ep ](\nu_c - \nu _d) \sim \tet [\nu _b] & = & 0
\nonumber \\
\tet [\delta ](p_c+p_d-2 \Delta) = \tet [\delta](\nu_c + \nu _d)
\sim \tet [\nu_a] & = & 0
\eea
The split gauge condition to linear order in $t$ yields a non-trivial condition
on the growths $\dot q_1$ and $\dot q_2$, given by
\bea
\dot q_1 \omega _{\nu _a} (p_c)   - \dot q_2 \omega _{\nu _a} (p_d)  =0
\eea
Here and below, the following notation is used for holomorphic Abelian
differentials with double zeros (at the branch points),
\bea
\label{qdots}
\omega _\nu (z) \equiv \omega _I (z) \p_I \tet [\nu ](0)
\eea
This notation was introduced in \cite{IV}, eq (2.40). It was also shown there
that ratios of these differentials for different $\nu$'s evaluated at the same
branch point may be expressed in terms of $\tet$-constants, via the relation
\bea
\label{omegaratio}
{\omega _{\nu_i} (p_k) \over \omega _{\nu _j} (p_k)}
=
{\M _{\nu _i \nu _k} \over \M_{\nu _j \nu _k}}
\eea
where $\M_{\nu \nu '} $ was 
introduced in \cite{IV} and
reproduced in (\ref{calM}). It was also shown in
\cite{IV} that $\M_{\nu\nu'}$ has the following expression in terms of 
$\tet$-constants
for {\sl even} spin structures,
\bea
\M_{\nu _1,\nu_2}^2 = \pi ^4   \prod _{k=3,4,5,6}
\tet [\nu _1 + \nu _2 + \nu _k] (0)^2
\eea
where $\nu _i$, $i=1,\cdots,6$ are all six distinct odd spin structures.

\bigskip

\subsubsection{Evaluation at the branch points}

\medskip

The ratio of the two vanishing factors may be computed in the limit of
vanishing $t$, with the following result,
\bea
{ \tet [\delta +\ep] (q_1-q_2) \over \tet [\delta ](q_1+q_2-2 \Delta) }
& = &
{ \{\dot q_1 \omega _I (p_c)   - \dot q_2 \omega _I (p_d) \} \p_I \tet
[\delta + \ep ] (\nu _c - \nu _d)
\over
\{\dot q_1 \omega _I (p_c)   + \dot q_2 \omega _I (p_d) \} \p_I \tet
[\delta ] (\nu _c + \nu _d)}
\\
& = &
{ \{ \omega _{\nu _a} (p_d) \omega _I (p_c) - \omega _{\nu_a} (p_c)
\omega _I (p_d) \} \p_I \tet [\delta + \ep ] (\nu _c - \nu _d)
\over
\{\omega _{\nu _a} (p_d)  \omega _I (p_c)   + \omega _{\nu_a} (p_c)
\omega _I (p_d) \} \p_I \tet [\delta ] (\nu _c + \nu _d)}
\no
\eea
The passage from the first to the second line in the above formula is made
using the relation between $\dot q_1$ and $\dot q_2$ of (\ref{qdots}).
Next, we have the following relations,
\bea
\omega _I(z) \p_I \tet [\delta + \ep ] (\nu _c - \nu _d) & = &
K_1 \omega _{\nu _b}(z)
\nonumber \\
\omega _I(z) \p_I \tet [\delta  ] (\nu _c + \nu _d) & = &
K_2 \omega _{\nu _a}(z)
\eea
where the exponential factors $K_1$ and $K_2$ will be computed later.
In terms of these quantities, we have
\bea
\label{linearized}
{ \tet [\delta +\ep] (q_1-q_2) \over \tet [\delta ](q_1+q_2-2 \Delta) }
=
{K_1 \over 2 K_2} \biggl (
{\omega _{\nu _b}(p_c) \over \omega _{\nu _a} (p_c)} -
{\omega _{\nu _b}(p_d) \over \omega _{\nu _a} (p_d)} \biggr )
=
- {K_1 \over 2 K_2} {\M_{ab} \M _{cd} \over \M_{ac} \M_{ad}}
\eea
To pass to the last line, (\ref{omegaratio}) has been used to express the
ratio of $\omega$'s in terms of $\M_{\nu\nu'}$'s, 
as well as the following algebraic
relation, $\M _{bc} \M_{ad} - \M_{ac} \M_{bd} = - \M_{ab} \M_{cd}$,
which readily follows from the definition of $\M_{\nu\nu'}$, 
given in (\ref{calM}).

\bigskip

\subsubsection{Expression in terms of $\tet$-constants}

\medskip

Assembling the expression in (\ref{GammaZee}) and (\ref{Prymsplit}), $\Gamma
[\delta; \ep]$ takes the following form
\bea
\Gamma [\delta; \ep]
 & = &
\sigma (\mu_i,\mu_j)  {i \over 4 \pi ^3 \tet _k ^4} \cdot 
{C \over C_r ^2 C_s ^2} \cdot
{\tet [\delta ]^5 E(p_r ,p_s )^4 \sigma (p_r)^2 \sigma (p_s)^2
\over
\tet [\delta] (q_1 + q_2 - 2 \Delta) E(q_1,q_2) \sigma (q_1)^2 \sigma
(q_2)^2}
\nonumber \\ && \qquad \times
 {1 \over \M_{\nu _r \nu _s}^2}  S _{\delta_j ^+} (q_1,q_2) S_{\delta _j
^-} (q_1,q_2) S_{\delta _i ^-}(q_1,q_2)
\eea
Using (\ref{linearized}), the limit $q_1 \to p_c =p_r$, $q_2 \to p_d=p_s$ may
be safely taken. Next, the expression for the Szeg\"o kernel in terms of
$\tet$-functions and the prime form is used. All factors of $\sigma$ and
$E(p_c,p_d)$ now cancel and one obtains,
\bea
\Gamma [\delta; \ep]
=
{i \over 8 \pi ^3 }  {\sigma (\mu_i,\mu_j) \over \tet _k ^4}
\
{C K_1 \over C_c ^2 C_d ^2 K_2}
\
{\M_{ab}  \over \M_{ac}  \M_{cd}  \M_{da}}
\
{\tet [\delta ](0)^5 \tet [\delta _j ^+] (\nu _c - \nu _d) 
\tet [\delta _j ^-] (\nu _c - \nu
_d) \over \tet [\delta _j ^+] (0) \tet [\delta _j ^-] (0) 
\tet [\delta _i ^-] (0) }
\quad
\eea
It remains to cast this expression in terms of standard $\tet$-constants.
To this end, we recast some of the factors using the following relations
between $\tet$-constants.
\bea
\tet [\delta _j ^+] (\nu _c - \nu _d) & = & K_3 \ \tet [\delta _k ^+](0)
\nonumber \\
\tet [\delta _j ^-] (\nu _c - \nu _d) & = & K_4 \ \tet [\delta _k ^-](0)
\eea
The exponential factors $K_3$ and $K_4$ will be evaluated later.
The final result for $\Gamma [\delta; \ep]$ is now obtained as follows,
\bea
\Gamma [\delta ,\ep] =
 {i \over 8 \pi ^7} \  {\sigma (\mu_i,\mu_j) \over  \tet _k ^4} 
 \cdot \kappa \cdot \kappa ' \cdot
{\tet [\delta ]^2
\over
\tet [\delta _j^+]^2 \tet [\delta _j^-]^2 \tet [\delta +\ep ]^2 }
\eea
where the following definitions have been made,
\bea
\kappa & \equiv &   {C K_1 K_3 K_4 \over C_c ^2 C_d ^2 K_2}
\\
\kappa ' & \equiv &
 {\pi ^4 \M_{ab}  \over \M_{ac}  \M_{cd} \M_{da} } \cdot
\tet [\delta ]^3 \tet [\delta _k ^+]  \tet [\delta _k ^-]
\tet [\delta _j ^+]  \tet [\delta _j ^-]  \tet [\delta _i ^-]
\no
\eea
The combinations $\kappa$ and $\kappa'$ will be evaluated in the
subsequent subsection, resulting in
$\kappa =\pm 1$ and $\kappa '=\pm 1$.

\subsection{Calculation of the overall sign of $\mathbf{\Gamma [\delta; \ep]}$}

Neither $\kappa$, nor $\kappa '$ is intrinsic, but the product
$\sigma (\mu_i,\mu_j) \kappa \kappa'$  will be, as will be evidenced
by the fact that the final expression for $\Gamma [\delta; \ep]$ is in
terms of squares of $\tet$-functions only.

\medskip

\subsubsection{Calculating $\mathbf{\kappa}$}

\medskip

Let us summarize the definitions of the factors entering into $\kappa$,
\bea
C & =  & - \exp \{ -8 \pi i \nu _c '  \Omega \nu _d '  \}
\no \\
C _c & = & \exp \{ - \pi i \nu _c ' \Omega \nu _c ' - 2 \pi i
\nu _c ' \nu _c'' \}
\no \\
C _d  & = & \exp \{ - \pi i \nu _d ' \Omega \nu _d ' - 2 \pi i
\nu _d ' \nu _d '' \}
\no \\
K_1 \omega _{\nu _b}(z)  & = &
\omega _I(z) \p_I \tet [\delta + \ep ] (\nu _c - \nu _d)
\no \\
K_2 \omega _{\nu _a}(z) &  = &
\omega _I(z) \p_I \tet [\delta  ] (\nu _c + \nu _d)
\no \\
K_3 \ \tet [\delta _k ^+](0) & = &
\tet [\delta _j ^+] (\nu _c - \nu _d)
\no \\
K_4 \ \tet [\delta _k ^-](0) & = &
\tet [\delta _j ^-] (\nu _c - \nu _d)
\eea
The $K$-factors may be computed starting from the following basic formula of
(\ref{tetrelations}),
\bea
\tet [\delta ]( z+\Omega \rho'+\rho'') =  \tet [\delta + \rho](z)
\exp \{
- i \pi \rho ' \Omega \rho ' - 2 \pi i \rho ' (z + \delta '' + \rho '') \}
\eea
One finds,
\bea
K_1 & = & - \exp \{
- i \pi (\nu _c - \nu _d)' \Omega (\nu _c - \nu _d)'
- 2 \pi i (\nu _c - \nu _d)' \nu _b '' + 4 \pi i \nu _d ' \nu _c '' + 4
\pi i \nu _b ' \nu _c ''\}
\nonumber \\
K_2 & = & - \exp \{ - i \pi (\nu _c + \nu _d)' \Omega (\nu _c + \nu _d)'
+ 2 \pi i (\nu _c + \nu _d )' \nu _a ''\}
\nonumber \\
K_3 & = & \exp \{
- i \pi (\nu _c - \nu _d)' \Omega (\nu _c - \nu _d)'
+ 2 \pi i (\nu _c - \nu _d)' (\nu _b + \nu _d + \nu _f)'' \}
\nonumber \\
K_4 & = & \exp \{
- i \pi (\nu _c - \nu _d)' \Omega (\nu _c - \nu _d)'
+ 2 \pi i (\nu _c - \nu _d)' (\nu _a + \nu _d + \nu _f)'' \}
\eea
Notice that under $c \leftrightarrow d$, $K_1 \to - K_1$ while $K_2 \to + K_2$,
even though these symmetries are not manifest. Multiplying all factors yields
\bea
\label{kappa}
\kappa =  { C K_1 K_3 K_4 \over C_c ^2 C_d^2 K_2}
=
\exp  4 \pi i (\nu _d ' \nu _c '' + \nu _c ' \nu _d ''
- \nu _d ' \nu _a '' + \nu _c ' \nu _f '' - \nu _d ' \nu _f '' + \nu _b '
\nu _c '')
\eea
which indeed takes the values $\pm 1$.

\subsubsection{Calculating  $\mathbf{\kappa'}$}

\medskip

A direct calculation of $\kappa'$ from its definition
\bea
\kappa ' \equiv { \pi^4 \M_{ab} \over \M_{ac} \M_{cd} \M_{da}}
\tet [\delta ]^2 \tet [\delta _i^+] \tet [\delta _i^-] \tet [\delta _j^+]
\tet [\delta _j^-] \tet [\delta _k^+] \tet [\delta _k^-]
\eea
involves the non-intrinsic sign factors, and must be computed case by
case. To simplify the process, we shall carry out this calculation for a
single non-trivial twist and obtain the expression for
$\Gamma[\delta; \ep]$ for all twists by modular invariance. This choice
uniquely determines $\nu_a$ and $\nu_b$ (up to interchange of $a$ and
$b$). The choice made here is the standard one,
\bea
\ep = \left (\matrix{0 \cr 0\cr} \bigg | \matrix{0 \cr \12 \cr} \right )
\hskip 1in
\nu_a =\nu_2, \qquad \nu_b=\nu_4
\eea
The even spin structures are fixed through the choice of
the odd spin structures and (\ref{deltanu}).

\medskip

The calculation of $\kappa'$ is started with the evaluation of the
product of all 6 even spin structures,
$\tet [\delta _2^+]
\tet [\delta _2^-]
\tet [\delta ^+_3]
\tet [\delta ^-_3]
\tet [\delta ^+_4]
\tet [\delta ^-_4]$.
This object is independent of the
remaining choices of $c,d,e,f$, since a permutation of these objects
simply permutes the various factors in the product. To evaluate it, we
choose $(c,d,e,f)=(1,3,5,6)$. The even spin structures are then
determined by (\ref{evenodd}) (see also (\ref{deltapm})), 
and the $\tet$-constants are expressed in
normalized form by
\bea
\tet [\delta_2 ^+] = \tet [\delta_7 + 2 \nu_d] & = & - \tet [\delta _7]
\no \\
\tet [\delta_2^-] = \tet [\delta_8 + 2 \nu_d] & = & -\tet [\delta _8]
\no \\
\tet [\delta^+_3] = \tet [\delta_1 + 2 \delta_0] & = & +\tet [\delta _1]
\no \\
\tet [\delta^-_3 ] = \tet [\delta_2 + 2 \delta_0] & = & +\tet [\delta _2]
\no \\
\tet [\delta ^+_4] = \tet [\delta_3 + 2 \nu_e] & = & +\tet
[\delta _3]
\no \\
\tet [\delta ^-_4] = \tet [\delta_4 + 2 \nu_e] & = & +\tet
[\delta _4]
\eea
Therefore, given the choice for $\ep = - \nu_2+\nu_4$, the following
product is the same for any choices of $c,d,e,f\not=2,4$, and we have
\bea
\tet [\delta _2^+]
\tet [\delta _2^-]
\tet [\delta ^+_3]
\tet [\delta ^-_3]
\tet [\delta ^+_4]
\tet [\delta ^-_4]
= +
\tet [\delta _1] \tet [\delta _2] \tet [\delta _3] \tet [\delta _4] \tet
[\delta _7] \tet [\delta _8]
\eea

\medskip

Next, the expressions for $\M_{ab}$, $a,b=1,\cdots,6$ are needed.
As was shown in \cite{IV},
these objects may be expressed as products of $\tet$-constants, times a
non-intrinsic sign factor. Normalizing the $\tet$-constants on the even spin
structures in canonical form, as given by (\ref{listeven}), the sign factors
$m_{ab}$ are tabulated in (\ref{littlem}). The quantities needed are
given by
\bea
\M _{24}
& = &
- \pi ^2 \tet [\delta_0] \tet [\delta _5] \tet [\delta_6] \tet [\delta _9]
\no \\
\M _{2c} \M_{cd} \M_{d2}
& = &
+ m_{2c} m_{cd} m_{d2} \pi ^6 \tet [\delta]^2 \prod _{i=0}^9 \tet [\delta
_i]
\eea
Combining all of the above,  we have
\bea
\kappa' =
- m_{2c} m_{cd} m_{d2}
\eea
Table \ref{table:1} below summarizes the results of the case by
case calculation of $\kappa'$.

\begin{table}[htb]
\begin{center}
\begin{tabular}{|c|c||c||c|c|c||c|} \hline
$c$ & $d$ & $\delta$ & $m_{2c}$ & $m_{cd}$ & $m_{d2}$ & $\kappa'$
\\ \hline \hline
1   & 3  & $\delta _7$     & $+$   & $-$    & $-$ & $-$ \\ \hline
1   & 5  & $\delta _3$     & $+$   & $+$    & $-$ & $+$ \\ \hline
1   & 6  & $\delta_2 $     & $+$   & $-$    & $-$ & $-$ \\ \hline
3   & 5  & $\delta _1$     & $+$   & $+$    & $-$ & $+$\\ \hline
3   & 6  & $\delta _4$     & $+$   & $-$    & $-$ & $-$\\ \hline
5   & 6  & $\delta _8$     & $+$   & $-$    & $-$ & $-$ \\ \hline  \hline
\end{tabular}
\end{center}
\caption{Calculation of $\kappa'$}
\label{table:1}
\end{table}

\bigskip

\subsubsection{\bf Calculating the sign of $\mathbf{\Gamma [\delta; \ep]}$}

\medskip

A general expression for $\kappa $ was obtained in (\ref{kappa}).
For our present purposes, with $a=2,~b=4$, this expression becomes,
\bea
\kappa
= \< \nu_c |\nu_d\>
\exp  4 \pi i (
- \nu _d ' \nu _2 '' + (\nu _c ' + \nu _d ' ) \nu _f '' + \nu _4 ' \nu _c '')
\eea

Notice that $\kappa$ depends on $f$ as well as on $c$ and $d$.
Thus, $\kappa$ does not just depend on $\ep$ and $\delta$
(i.e. only on $\nu_a$, $\nu_b$ and $\nu_c+\nu_d$), but also
on the choice of $\nu_f$ versus $\nu_e$. Furthermore, $\kappa$
is not, in general, symmetric under the interchange of $c$ and $d$;
instead, we have
\bea
{\kappa (c,d) \over \kappa (d,c)}
=
- \< \nu_c |\nu_d\>
\eea
for fixed $a,b,e,f$. Fortunately, these non-intrinsic dependences of
$\kappa$ are being compensated in the expression for $\Gamma [\delta; \ep]$
by the presence of another sign factor $\sigma (\mu _i, \mu_j)$.
Recall that for given $i$, there can be two choices for $j$. These choices
are actually correlated with the choices for $f$ in $\kappa$.

\begin{table}[t]
\begin{center}
\begin{tabular}{|c|c|c||c||c|c|c||c|c|c||c|} \hline
$c$ & $d$ & $f$ & $\delta$ & $\mu_i$ & $\mu_j$ & $\mu_k$ & $\kappa$ &
$\kappa'$ & $\sigma (\mu_i,\mu_j) $ & SIGN
  \\ \hline \hline
1 & 3 & 5 & $\delta _7$   & $\mu_2$   & $\mu_4$    & $\mu_3$
& $+$ & $-$& $+$& $-$
  \\ 
 &  & 6 &   &    & $\mu_3$    & $\mu_4$
& $+$ & $-$& $+$& 
  \\ \hline
1 & 5 & 3 & $\delta _3$   & $\mu_4$   & $\mu_2$    & $\mu_3$
& $+$ & $+$& $-$& $-$
  \\ 
 &  & 6 &    &   & $\mu_3$    & $\mu_2$
& $+$ & $+$& $-$& 
  \\ \hline
1 & 6 & 3   & $\delta_2 $   & $\mu_3$   
& $\mu_2$    & $\mu_4$ & $-$   & $-$    & $-$ & $-$
  \\
  &   & 5   &               &    
& $\mu_4$    & $\mu_2$ & $+$   & $-$    & $+$ & 
  \\ \hline
3 & 5 & 1 & $\delta _1$     
& $\mu_3$& $\mu_2$    & $\mu_4$ & $-$   & $+$    & $-$ &$+$
  \\
  &   & 6 &                 &    
   & $\mu_4$    & $\mu_2$ & $+$   & $+$    & $+$ & 
  \\ \hline
3 & 6 & 1 & $\delta _4$   & $\mu_4$   & $\mu_2$    & $\mu_3$
& $+$   & $-$    & $-$ & $+$
  \\ 
 &  & 5 &   &    & $\mu_3$    & $\mu_2$
& $+$   & $-$    & $-$ & 
  \\ \hline
5 & 6 & 1 & $\delta _8$    & $\mu_2$   & $\mu_4$    & $\mu_3$
& $-$   & $-$    & $+$ & $+$
\\ 
 &  & 3 &    &   & $\mu_3$    & $\mu_4$
& $-$   & $-$    & $+$ & 
\\ \hline  \hline
\end{tabular}
\end{center}
\caption{Calculation of the SIGN of $\Gamma [\delta; \ep]$}
\label{table:2}
\end{table}

Clearly, $f$-dependence enters only when $(\nu_c ' + \nu_d' )\nu_f ''\not=0$,
which is the case (here with $a=2$ and $b=4$) when $c=1, ~d=6$
and $c=3,~d=5$. To identify the correlation between the choice of
$f$ in $\kappa$ and that of $j$ in $\sigma (\mu_i,\mu_j)$, we need
consider only these cases.

\begin{itemize}

\item $c=1,~d=6$ implies
$$
\left \{ \matrix{\delta _i^+ =\delta _2 \cr \delta ^-_i=\delta _1 \cr} \right .
\qquad
\left \{ \matrix{
\delta _j^+ = \nu_1 + \nu _4 + \nu _f  \cr
\delta _j ^- = \nu_1 + \nu _2 + \nu _f  \cr} \right .
$$
For $f=3$, we have $\delta ^+_j=\delta _8$ and therefore $j=2$. For
$f=5$, we have $\delta ^+_j =\delta _4$ and therefore $j=4$.

\item $c=3,~d=5$ implies
$$
\left \{ \matrix{\delta _i^+ =\delta _1 \cr \delta ^-_i=\delta _2 \cr} \right .
\qquad
\left \{ \matrix{
\delta _j^+ = \nu_3 + \nu _4 + \nu _f  \cr
\delta _j ^- = \nu_2 + \nu _3 + \nu _f  \cr} \right .
$$
For $f=1$, we have $\delta ^+_j=\delta _8$ and therefore $j=2$. For
$f=6$, we have $\delta ^+_j =\delta _3$ and therefore $j=4$.

\end{itemize}
The corresponding results are summarized in Table \ref{table:2} above.

Assembling all the results with the basic expression for 
$\Gamma[\delta; \ep]$,
we obtain,
\bea
\Gamma [\delta^\sigma _i ,\ep] =
\sigma  {i \over 8 \pi ^7} \ { \< \nu _0 | \mu _i\>  \over  \tet _k ^4}
{\tet [\delta ]^2  \over
\tet [\delta _j^+]^2 \tet [\delta _j^-]^2 \tet [\delta +\ep ]^2 }
\eea
Here, $(i,j,k)$ is a permutation of the genus 1
spin structure indices $(2,3,4)$, and $\delta = \delta _i ^\sigma$
with $\sigma = \pm $. A more illuminating expression,
whose form is more manifestly covariant is obtained by factoring out
a combination of the $Z^{{\rm qu}}[\ep]$ partition function for 4 
twisted directions.
Using the familiar genus 1 relation $\tet_i \tet _j \tet _k = 2 \eta ^3$,
 this leads to the following final expression,
\bea
\label{finalGamma}
\Gamma [\delta ^\sigma _i ,\ep] =
\sigma  \< \nu _0 | \mu _i\>  {i \over (2 \pi )^7} \ 
{  \tet _i ^4  \over  \eta ^{12}}
\left ( {\tet [\delta ^\sigma  _i ]^2  \over   
\tet [\delta ^{(-\sigma)}  _i ]^2 } \right )
\left ( {\tet _j^4 \over \tet [\delta _j^+]^2 
\tet [\delta _j^-]^2} \right )
\eea
which is independent of $j$. Here, $\sigma = \pm$ according to which
spin structure $\delta ^\pm _i$ is being evaluated.

\section{Modular transformations and $\mathbf{Z_2}$ twisting}
\setcounter{equation}{0}

The structure of the genus 2 modular group $Sp(4,{\bf Z})$ and its action on
spin structures, twists and $\tet$-functions is summarized in  Appendix B.


\subsection{Subgroup preserving a given twist}

The subgroup of the modular group $Sp(4,{\bf Z})$ which acts on spin
structures or on half characteristics is $Sp(4,{\bf Z}_2)$. It is
isomorphic to the  permutation group of the six branch points and
therefore has $6! = 720$ elements.
To determine the subgroup $H_\ep$ of $Sp(4,{\bf Z}_2)$ that
leaves a given twist $\ep$ invariant, we proceed as follows. First, make a
definite choice for a twist,  so that the invariance equation
becomes
\bea
\label{standardep}
M= \left ( \matrix{A & B \cr C & D \cr} \right )
 \hskip .6in
\ep = \left (\matrix{0 \cr 0\cr} \bigg | \matrix{0 \cr \12 \cr} \right )
\hskip .6in
\left ( \matrix{\ep' \cr \ep''\cr} \right )=
 \left ( \matrix{D & -C \cr -B & A \cr} \right )
\left ( \matrix{\ep' \cr \ep''\cr} \right )
\eea
This equation puts the following restrictions on the entries
$A_{12}=C_{12}=C_{22}=0$ and $A_{22}=1$ of the matrices.
All elements of the group $H_\ep$ are found by solving the symplectic
relation $MJM^T=J$ under these restrictions.
The independent generators for  $H_\ep$ are
\bea
\mbox{generators of } H_\ep = \{ M_1, ~M_2, ~M_3, ~T_2=\Sigma T \Sigma,
~ S_2 = SM_1SM_1\}
\eea
The full group  may be parametrized in terms of
the  Abelian subgroup $M_\ep$ of ``translations'',

\bea
H_\ep^0 
& = &  
\{ I, ~M_1, ~M_2, ~M_3, ~M_1M_2, ~M_1M_3, ~M_2M_3, ~M_1M_2M_3\}
\no \\
H_\ep & = & \{ I, ~ T_2 ,  ~S_2, ~S_2T_2, ~S_2^2, ~ S_2^2 M_1 T_2 \} 
\times H_\ep^0
\eea
The matrices $M_i$'s are given explicitly in Appendix B3.
In total there are 48 elements. This number is as expected, since all 
elements of the group $Sp(4,{\bf Z}_2)$ may be decomposed as the product
of elements that move the twists from any given twist to any other twist
times an element that preserves the twist. There are 15 non-trivial
twists and hence  $ 720 = 15 \times 48$. The transformation laws of
the even spin structures under the generators of $H_\ep$ are listed in
Table \ref{table:4}.

\begin{table}[htb]
\begin{center}
\begin{tabular}{|c||c|c|c|c|c|c|c|c|c|c|} \hline
 $\delta$ & $M_1$  & $M_2$ & $M_3$ & $T_2$ & $S_2$ & $\epsilon ^2
(M_1)$ & $\epsilon ^2(M_2)$ & $\epsilon ^2 (M_3)$ & $\epsilon ^2 (T_2)$
& $\epsilon^2 (S_2)$
                \\ \hline \hline
 $\delta _1$
            & $\delta _3$
            & $\delta _2$
            & $\delta _1$
            & $\delta _1$
            & $\delta _3$
            & 1 &  1 & 1  & 1 & $i$
 \\ \hline
 $\delta _2$
            & $\delta _4$
            & $\delta _1$
            & $\delta _2$
            & $\delta _2$
            & $\delta _4$
             & 1 & 1 & 1 &  1 & $i$
 \\ \hline
 $\delta _3$
            & $\delta _1$
            & $\delta _4$
            & $\delta _3$
            & $\delta _4$
            & $\delta _7$
             & 1 & 1 &  1 &  1 & 1
 \\ \hline
 $\delta _4$
            & $\delta _2$
            & $\delta _3$
            & $\delta _4$
            & $\delta _3$
            & $\delta _8$
             & 1 & 1 &  1 &  1 & 1
 \\ \hline
 $\delta _5$
            & $\delta _6$
            & $\delta _5$
            & $\delta _6$
            & $\delta _9$
            & $\delta _6$
            & 1 & $i$  &  1 & 1 & $i$
 \\ \hline
 $\delta _6$
            & $\delta _5$
            & $\delta _6$
            & $\delta _5$
            & $\delta _0$
            & $\delta _9$
            & 1 & $i$ & 1 & 1 & 1
 \\ \hline
 $\delta _7$
            & $\delta _7$
            & $\delta _8$
            & $\delta _8$
            & $\delta _7$
            & $\delta _1$
            & $i$ & 1 & 1 & 1 & $i$
 \\ \hline
 $\delta _8$
            & $\delta _8$
            & $\delta _7$
            & $\delta _7$
            & $\delta _8$
            & $\delta _2$
            & $i$ & 1 & 1 & 1 & $i$
 \\ \hline
 $\delta _9$
            & $\delta _9$
            & $\delta _9$
            & $\delta _0$
            & $\delta _5$
            & $\delta _5$
            & $i$ & $i$ & $-1$ & 1 & $i$
 \\ \hline
 $\delta _0$
            & $\delta _0$
            & $\delta _0$
            & $\delta _9$
            & $\delta _6$
            & $\delta _0$
            & $i$ & $i$  & $-1$ & 1 & $-1$
 \\ \hline
\end{tabular}
\end{center}
\caption{The modular group $H_\ep$ acting on even spin structures }
\label{table:4}
\end{table}

\subsection{Transformations of the Schottky relations}

In (\ref{genus1-2}), a Schottky relation is derived, 
but a sharper form (namely
the precise sign involved in taking the square root of this relation)
is needed in the present study, as indicated in (\ref{Schottky}).
The square root may be fixed by checking the consistency of 
(\ref{Schottky}) with
modular transformations and with degenerations, both of which are
carried out in this subsection.

\medskip

The crucial ingredients in this calculation are the factors
$\epsilon (\delta, M)^2 $ when $M\in H_\ep$, since they will
determine the transformation rules for the genus 2 $\tet$-constants.
To compute them, we make use of the values for
$\epsilon (\delta, M_i)^2$, $i=1,2,3$, $\epsilon (\delta, S)^2$,
$\epsilon (\delta, T)^2$, and $\epsilon (\delta, \Sigma)^2$, 
given in (\ref{eps}),
as well as the following cocycle rule, which may be derived from the
definition of $\epsilon (\delta, M)$ in terms of the $\tet$-constants,
\bea
\epsilon (\kappa, MM') = \epsilon (\kappa, M') \times \epsilon (M'\kappa , M)
\eea
The results are summarized in  Table \ref{table:4}, and may be readily applied
to the calculation of the pairwise products
$\epsilon (\delta _j,M)^2 \epsilon (\delta _{j+1}, M)^2$. The results are given 
in
Table \ref{table:5}.

\begin{table}[htb]
\begin{center}
\begin{tabular}{|c||c|c|c|c|c|} \hline
$j$ & $M_1$ & $M_2$ & $M_3$ & $T'$ & $S_2$  \\ \hline \hline
1   & 1     & 1     & 1     & 1    & $-1$ \\ \hline
3   & 1     & 1     & 1     & 1    & $1$ \\ \hline
7   & $-1$  & 1     & 1     & 1    & $-1$ \\ \hline   \hline
\end{tabular}
\end{center}
\caption{The modular group $H_\ep$ acting on pairs $\delta ^\pm $ of even spin
structures }
\label{table:5}
\end{table}

For the choice of twist made here, (\ref{Schottky}) becomes,
\bea
  {\tet_4^4\over\tet_2^4}
   =
  {\tet^2[\delta_3]\tet^2[\delta_4]\over
    \tet^2[\delta_7]\tet^2[\delta_ 8]}
\qquad
  {\tet_2^4\over\tet_3^4}
 =
   {\tet^2[\delta_7]\tet^2[\delta_8]\over
    \tet^2[\delta_1]\tet^2[\delta_2]}
\qquad
  {\tet_3^4\over\tet_4^4}
=
   {\tet^2[\delta_1]\tet^2[\delta_2]\over
    \tet^2[\delta_3]\tet^2[\delta_4 ]}
\eea
By inspecting Table \ref{table:4}, it is manifest that
the transformations $M_2$, $M_3$ and $T_2$ do not act on
these pairs at all. It
remains to consider only the actions of $M_1$ and $S_2$,
\bea
M_1 \left (  {\tet_3^4\over\tet_4^4} \right )
& = & M_1 \left ( {\tet^2[\delta_1]\tet^2[\delta_2]\over
    \tet^2[\delta_3]\tet^2[\delta_4 ]} \right )
= + {\tet^2[\delta_3]\tet^2[\delta_4 ]
\over \tet^2[\delta_1]\tet^2[\delta_2]} = +{\tet_4^4 \over \tet_3^4}
\nonumber \\
M_1 \left (  {\tet_3^4\over\tet_2^4} \right )
& = & M_1 \left ( {\tet^2[\delta_1]\tet^2[\delta_2]\over
    \tet^2[\delta_7]\tet^2[\delta_8 ]} \right )
= - {\tet^2[\delta_3]\tet^2[\delta_4 ]
\over \tet^2[\delta_7]\tet^2[\delta_8]} = -{\tet_4^4 \over \tet_2^4}
\nonumber \\
S_2 \left (  {\tet_3^4\over\tet_4^4} \right )
& = & S_2 \left ( {\tet^2[\delta_1]\tet^2[\delta_2]\over
    \tet^2[\delta_3]\tet^2[\delta_4 ]} \right )
= - {\tet^2[\delta_3]\tet^2[\delta_4 ]
\over \tet^2[\delta_7]\tet^2[\delta_8]} = -{\tet_4^4 \over \tet_2^4}
\nonumber \\
S_2 \left (  {\tet_3^4\over\tet_2^4} \right )
& = & S_2 \left ( {\tet^2[\delta_1]\tet^2[\delta_2]\over
    \tet^2[\delta_7]\tet^2[\delta_8 ]} \right )
= + {\tet^2[\delta_3]\tet^2[\delta_4 ]
\over \tet^2[\delta_1]\tet^2[\delta_2]} = +{\tet_4^4 \over \tet_3^4}
\eea
The transformation properties of $\tet$-constants for genus 1,
under the two canonical generators, denoted here by
$T^{(1)}$ and $S^{(1)}$ are given by
\bea
\left \{ \matrix{T^{(1)} & : & \tau \to \tau +1 \cr
S^{(1)} & : & \tau \to -1/\tau \cr} \right .
\eea
We have
\bea
T^{(1)} \ \left \{ \matrix{
\tet _2 (\tau+1) & = & e^{i \pi /4} \tet _2 (\tau) \cr
\tet _3 (\tau+1) & = &  \tet _4 (\tau) \cr
\tet _4 (\tau+1) & = &  \tet _3 (\tau) \cr} \right .
\qquad \qquad
S^{(1)} \ \left \{ \matrix{
\tet _2 (-1/\tau) & = & \sqrt{-i\tau} \tet _4 (\tau) \cr
\tet _3 (-1/\tau) & = & \sqrt{-i\tau} \tet _3 (\tau) \cr
\tet _4 (-1/\tau) & = & \sqrt{-i\tau} \tet _2 (\tau) \cr} \right .
\eea
Therefore, it is clear that, signs and all, we have
\bea
M_1 \quad & \longrightarrow &  \quad T^{(1)} \cr
S_2 \quad & \longrightarrow &  \quad T^{(1)} S ^{(1)}
\eea
Therefore, the genus 2 modular transformations
 $M_1$ and $S_2$ indeed induce ordinary genus 1 modular
transformations on the Prym period. As a result,
all the modular transformations in $H_\ep$ induce modular
transformations on $\tau_\ep$, which belong to the
genus 1 modular group, according to the above correspondence.

 \subsection{Modular orbits   of the twists under $H_\ep$}

First, a parametrization of the twists is needed,
\bea
\label{listtwists}
2\ep_1 =\left (\matrix{0 \cr 0\cr} \bigg | \matrix{0 \cr 0\cr} \right )
\qquad \
2\ep_2 =\left (\matrix{0 \cr 0\cr} \bigg | \matrix{0 \cr 1\cr} \right )
\qquad \
2\ep_3 =\left (\matrix{0 \cr 0\cr} \bigg | \matrix{1 \cr 0\cr} \right )
\qquad \
2\ep_4 =\left (\matrix{0 \cr 0\cr} \bigg | \matrix{1 \cr 1\cr} \right )
\nonumber \\
2\ep_5 =\left (\matrix{0 \cr 1\cr} \bigg | \matrix{0 \cr 0\cr} \right )
\qquad \
2\ep_6 =\left (\matrix{0 \cr 1\cr} \bigg | \matrix{1 \cr 0\cr} \right )
\qquad \
2\ep_7 =\left (\matrix{1 \cr 0\cr} \bigg | \matrix{0 \cr 0\cr} \right )
\qquad \
2\ep_8 =\left (\matrix{1 \cr 0\cr} \bigg | \matrix{0 \cr 1\cr} \right )
\nonumber \\
2\ep_9 =\left (\matrix{1 \cr 1\cr} \bigg | \matrix{0 \cr 0\cr} \right )
\qquad
2\ep_{10} =\left (\matrix{1\cr 1\cr} \bigg | \matrix{1\cr 1\cr} \right)
\qquad
2\ep _{11} =\left (\matrix{0 \cr  1\cr} \bigg | \matrix{0 \cr 1\cr} \right )
\qquad
2\ep_{12} =\left (\matrix{1 \cr  0\cr} \bigg | \matrix{1 \cr 0\cr} \right )
\no \\
2\ep_{13} =\left (\matrix{0 \cr  1\cr} \bigg | \matrix{1 \cr 1\cr} \right )
\qquad
2\ep_{14} =\left (\matrix{1 \cr  0\cr} \bigg | \matrix{1 \cr 1\cr} \right )
\qquad
2\ep_{15} =\left (\matrix{1 \cr  1\cr} \bigg | \matrix{0 \cr 1\cr} \right )
\qquad
2\ep_{16} =\left (\matrix{1 \cr  1\cr} \bigg | \matrix{1 \cr 0\cr} \right )
\eea

Notice that the standard twist adopted previously is given by $\ep = \ep_2$,
and that $\ep_1$ is no twist at all.
 The modular transformations under the subgroup $H_\ep$ are then
 easily computed from the transformation formula for the twists.

  \begin{table}[htb]
\begin{center}
\begin{tabular}{|c||c|c|c|c|c|c|c|c|c|c|c|c|c|c|c|c|}   \hline

$M$  &
$\ep_1$  & $\ep_2$  & $\ep_3$  & $\ep_4$  &
$\ep_5$  & $\ep_6$  & $\ep_7$  & $\ep_8$  & $\ep_9$  &
$\ep_{10}$  & $\ep_{11}$  & $\ep_{12}$  & $\ep_{13}$  & $\ep_{14}$  &
$\ep_{15}$   & $\ep_{16}$
\\ \hline \hline
$M_1$  &
$\ep_1$  & $\ep_2$  & $\ep_3$  & $\ep_4$  &
$\ep_5$  & $\ep_6$  & $\ep_{12}$  & $\ep_{14}$  & $\ep_{16}$  &
$\ep_{15}$  & $\ep_{11}$  & $\ep_7$  & $\ep_{13}$  & $\ep_8$  &
$\ep_{10}$   & $\ep_9$
\\ \hline
$M_2$  &
$\ep_1$  & $\ep_2$  & $\ep_3$  & $\ep_4$  &
$\ep_{11}$  & $\ep_{13}$  & $\ep_7$  & $\ep_8$  & $\ep_{15}$  &
$\ep_{16}$  & $\ep_5$  & $\ep_{12}$  & $\ep_6$  & $\ep_{14}$  &
$\ep_9$   & $\ep_{10}$
\\ \hline
$M_3$  &
$\ep_1$  & $\ep_2$  & $\ep_3$  & $\ep_4$  &
$\ep_6$  & $\ep_5$  & $\ep_8$  & $\ep_7$  & $\ep_{10}$  &
$\ep_9$  & $\ep_{13}$  & $\ep_{14}$  & $\ep_{11}$  & $\ep_{12}$  &
$\ep_{16}$   & $\ep_{15}$
\\ \hline
$T_2$  &
$\ep_1$  & $\ep_2$  & $\ep_4$  & $\ep_3$  &
$\ep_9$  & $\ep_{10}$  & $\ep_7$  & $\ep_8$  & $\ep_5$  &
$\ep_6$  & $\ep_{15}$  & $\ep_{14}$  & $\ep_{16}$  & $\ep_{12}$  &
$\ep_{11}$   & $\ep_{13}$
\\ \hline
$S_2$  &
$\ep_1$  & $\ep_2$  & $\ep_{12}$  & $\ep_{14}$  &
$\ep_5$  & $\ep_{16}$  & $\ep_3$  & $\ep_4$  & $\ep_6$  &
$\ep_{15}$  & $\ep_{11}$  & $\ep_7$  & $\ep_{10}$  & $\ep_8$  &
$\ep_{13}$   & $\ep_9$
\\ \hline \hline
\end{tabular}
\end{center}
\caption{The modular group $H_\ep$ acting on twists }
\label{table:3}
\end{table}

Simple inspection of  Table \ref{table:3} reveals 4 orbits,
\bea
\O _0 [\ep] & = & \{ \ep _1 \}
\no \\
\O _\ep  [\ep] & = & \{ \ep _2\}
\no \\
\O _+  [\ep] & = & \{ \ep _3, \ep _4, \ep _7, \ep _8, \ep _{12}, \ep _{14} \}
\no \\
\O_-  [\ep] & = & \{ \ep _5, \ep _6, \ep _9 , \ep _{10}, \ep _{11}, \ep _{13},
\ep _{15}, \ep _{16} \}
\eea
The non-trivial orbits $\O_\pm[\ep]$ may be characterized in a 
simple modular covariant manner,
\bea
\upsilon \in \O _\pm [\ep] \qquad \Leftrightarrow \qquad \<\ep | \upsilon \>
= \pm 1 \ {\rm and } \ \upsilon \not= \ep_1, \ep _2
\eea

\subsection{Subgroups leaving two twists invariant}

The first twist may be denoted $\ep$ and chosen as in (\ref{standardep}).
The non-trivial cases arise when the second twist $\alpha$ belongs to
either orbit $\O_\pm[\ep]$. In each orbit, any representative may be taken
for the second twist $\alpha$. The stabilizer groups of two twists
$\ep$ and $\alpha$ will be denoted by $H_{\ep, \alpha}$; their 
generators may be deduced from inspection of Table \ref{table:3}, and
given by
\bea
H_{\ep, \ep_3} & \sim &  {\rm generators} \quad \{M_1, M_2, M_3\}
\no\\
H_{\ep, \ep_5} & \sim &  {\rm generators} \quad \{M_1, S_2, (M_3S_2T_2)\}
\eea
The group $H_{\ep, \ep_3}$ is Abelian, but the group $H_{\ep, \ep_5}$
is non-Abelian,

\section{Asymptotic behavior of the chiral measure}
\setcounter{equation}{0}

Given the choice of homology basis adopted throughout this paper,
the period matrix may be parametrized as follows,
\bea
\Omega = \left ( \matrix{ \tau _1 & \tau \cr \tau & \tau _2 \cr} \right )
\eea
where $\tau_{1,2}, \tau \in {\bf C}$, subject to the constraint 
$\Im \Omega >0$.
Degenerations fall into two classes  according to whether the degeneration
separates the surface into two disconnected components
or leaves the surface connected. Separating (resp. non-separating)
degenerations result from the shrinking of a homologically trivial
(resp. non-trivial) 1- cycle. The limit $\tau \to 0$ is separating,
while the limits $\tau_1 \to +i \infty$ or $\tau _2 \to +i \infty$
are non-separating.

\medskip

The flat space-time chiral measure depends on the spin
structure and the limiting behavior of the measure therefore
depends upon the inter-relation between the spin structure
the homology of the degenerating cycle. These degenerations were
worked out explicitly -- to leading order -- in \cite{IV}, section 8.

\medskip

The $\bZ_2$-twisted chiral measure (\ref{chiralmeasure})
depends on the spin structure $\delta$,
AND on the twist $\ep$. Its limiting behavior will therefore depend
upon the inter-relations between not only the spin structure and
the homology of the degenerating cycle, but also upon the twist.
Here, we shall derive the limit of only two representative cases,
one separating, the other non-separating; the other cases are
similar, but their number is simply too large to discuss usefully here.
Finally, only the case where $n=4$ dimensions are $\bZ_2$-twisted
will be discussed, since this will be the case of greatest physical
interest, as discussed in the next subsection.

\subsection{Separating degeneration}

This limit corresponds to letting $\tau \to 0$, while keeping
$\tau _{1,2}$ fixed. The leading behavior was given in \cite{IV};
the leading and subleading behaviors are as follows,
\bea
\label{seplimits}
\tet \left [\matrix{\mu_i \cr \mu_j} \right ] (0,\Omega)
& = &
\tet _i (0,\tau_1) \tet _j (0,\tau_2)
\biggl \{ 1 + 2\tau ^2  \p \ln \tet _i (0,\tau_1)
\p \ln \tet _j (0,\tau_2) \biggr \}
\nonumber \\
\Xi _6 \left [\matrix{\mu_i \cr \mu_j} \right ] (\Omega)
& = &
2^8 \<\mu_i|\nu_0\> \<\mu_j|\nu_0\>
\eta (\tau_1)^{12} \eta (\tau_2)^{12}
\biggl \{ -1
+ {\tau ^2 \over 2} \biggl [
3\p  \ln \tet _i^4 (0,\tau_1) \p \ln \tet _j^4 (0,\tau_2)
\nonumber \\ && \qquad \qquad
- \p \ln \eta (\tau _1)^{12} \p \ln \tet _j^4 (0,\tau_2)
- \p \ln \eta (\tau _2)^{12} \p \ln \tet _i ^4 (0,\tau_1)
\biggr ] \biggr \}
\nonumber \\
\Psi _{10} (\Omega)
& = &
2^{12} (2 \pi \tau)^2 \eta (\tau _1)^{24} \eta (\tau _2)^{24}
\biggl \{ 1 + 48 \tau^2 \p \ln \eta (\tau _1) \p \ln \eta (\tau_2) \biggr \}
\eea
Here, $\mu_i$ and $\mu_j$ represent the even spin structures on each
genus 1 component; the even spin structure which restricts to the
odd spin structure on each genus 1 component will not be needed here.

\medskip

First, the limiting behavior is needed for the
Prym period, since it enters into the form of the chiral measure.
Using the Schottky relation (\ref{Schottky0}) as well as the above
limiting behaviors, we find
\bea
\tau _\ep = \tau _1 + \tau ^2 \p \ln \left (\tet _3 (0,\tau_2) 
\tet _4 (0,\tau_2) \right  ) 
+ \O (\tau ^4)
\eea
up to shifts $\tau _\ep \to \tau _\ep +4$.
Second, we need the bosonic twisted partition function factor
\bea
{\tet [\delta ^+_j ] (0,\Omega)^2 \tet [\delta ^-_j ] (0,\Omega)^2
\over \tet _j  (0,\tau_\ep)^4}
= \tet _3 (0, \tau_2)^2 \tet _4 (0, \tau_2)^2 + \O (\tau ^4)
\eea
which is clearly independent of $j$ to this order.
Third, the limit of the terms involving $\Gamma [\delta; \ep]$ is needed.
When $n=4$, as is being assumed here, it is advantageous to consider
directly the quantity
\bea
{Z_C [\delta^\sigma _i, \ep ] \over Z_M[\delta^\sigma _i]}
 \Gamma [\delta ^\sigma _i ,\ep]
 =
 \< \nu _0 | \mu _i \> {i ~ \sigma \over (2 \pi )^7}
 { \tet _i (0,\tau_\ep)^4 \over \eta (\tau_\ep)^{12}}
 \qquad \qquad
 i=2,3,4,~ \sigma = \pm
 \eea
In view of the relation between $\tau_1$ and $\tau_\ep$, the limit
as $\tau \to 0$ of this quantity is regular and is obtained simply
from the above formula by replacing  $\tau _\ep $ by $\tau_1$.
Putting all together, we find
\bea
d \mu _C [\delta^\sigma _i,\ep] (p^\mu _\ep)
& = &
 e^{i \pi \tau _\ep p_\ep ^2} {\< \nu _0 | \mu _i \> \over 2^6 \pi ^8}
{\tet _i (0,\tau_1)^4 \over \eta (\tau _1)^{12}}
 \\ && \times
\left \{
{-\sigma \over \tau^2 ~ \tet _2 (0,\tau _2)^4}
- 2 \pi i (\sigma + \half) \p \ln \tet _i (0,\tau_1)
- \sigma {\pi ^2 \over 2} p_\ep ^2 \right \}
\no
\eea
The physical analysis of this limit is as follows.

\medskip

The $1/\tau^2$ singularity, which is familiar from the flat space-time
chiral measure, produces tachyon and massless scalar singularities
in the channel connecting both genus 1 components.
This is as expected, since the six spin structures $\delta ^\sigma _i$
restrict to even spin structures on each genus 1 component.
The summation over $\sigma$ corresponds to the GSO projection
imposed on states  in the NS sector which are traversing the $A_2$ cycle.
This part of the GSO projection eliminates the corresponding
tachyon intermediate state, and indeed the partially summed
measure is regular, as may be seen from
\bea
d \mu _C [\delta^+ _i,\ep] (p^\mu _\ep) + d \mu _C [\delta^- _i,\ep] 
(p^\mu _\ep)
=
 -i  e^{i \pi \tau _\ep p_\ep ^2} {\< \nu _0 | \mu _i \> \over 2^5 \pi ^7}
{\tet _i (0,\tau_1)^4 \over \eta (\tau _1)^{12}}
 \p \ln \tet _i (0,\tau_1)
\eea
Notice also that a partial GSO resummation over the spin structures
of the first genus 1 component vanishes to this order,
\bea
\sum _{i=2,3,4} d \mu _C [\delta^\sigma _i,\ep] (p^\mu _\ep) =0
\eea
in view of the Riemann identities for genus 1,
\bea
\label{Riemannid}
\sum _{i=2,3,4} \< \nu _0 | \mu _i \>  \tet _i (0,\tau_1)^4 =
\sum _{i=2,3,4} \< \nu _0 | \mu _i \>  \tet _i (0,\tau_1)^4 \p 
\ln \tet _i (0,\tau_1) = 0
\eea
Again, this is as expected in view of space-time supersymmetry.

\subsection{Non-separating Degenerations}

As mentioned in the opening paragraph to this subsection, many
cases need to be distinguished based on the inter-relation of the
spin structure, the twist and the shrinking cycle. The case considered
here corresponds to letting the cycle $B_2$ grow to infinite length;
more specifically, letting $\tau _2 \to + i \infty$ while keeping
$\tau $ and $\tau _1$ fixed. As usual, we introduce the variable
$q \equiv \exp \{ i \pi \tau_2\}$, in terms of which the limit is
given by $q \to 0$. We recall from \cite{IV} the following limits,
\bea
\tet \left [\matrix{\mu_i \cr \mu_j} \right ] (0,\Omega) = 
\tet _i (0,\tau_1) + \O (q), \quad
\qquad
\mu _j = (00), ~(0\12)
\eea
As a result, the Schottky relation giving $\tau_\ep$ in terms of the
period matrix $\Omega +{IJ}$ simply yield $\tau _\ep = \tau_1$
up to shifts in $4\bZ$. The ratio $Z_C/Z_M \to 1$ in this limit.
The following combination,
\bea
{Z_C [\delta ^\sigma _i , \ep] \over Z_M[\delta ^\sigma _i]}
\Gamma [\delta ^\sigma _i,\ep]
\to \<\nu _0 |\mu _i \> { i ~ \sigma \over (2 \pi )^7}
{\tet _i (0,\tau_1)^4 \over \eta (\tau_1)^{12}}
\eea
has a smooth limit. Putting all together, we obtain the following
limits for the full measure,
\bea
d\mu _C [\delta_i^\sigma; \ep] (p _\ep) =
e^{i\pi\tau_\ep p_\ep^2}{\sigma \over q}
{\langle \nu_0|\mu_i\rangle
  \tet _i ({\tau \over 2}, \tau _1)^4 + \tet _1({\tau \over 2} , \tau_1)^4
\over
2^8 \pi ^6 \eta (\tau_1)^6 \tet _1 (\tau,\tau_1)^2}
+{\cal O}(1)
\eea
This result coincides with the untwisted case, as expected.

\section{Applications to Physical Theories}
\setcounter{equation}{0}

In this section, specific physical superstring theories are considered
that are constructed from flat Minkowski space-time orbifolded by
groups acting by reflections and shift. When the action of the 
orbifold group is the same (resp. different) on left and right movers,
the orbifold is referred to as symmetric (resp. asymmetric).
The models of \cite{KKS}, for example, are asymmetric orbifolds 
of Type II  superstring theory. For definiteness, the present study 
will concentrate on Type II theories with six or fewer compactified 
dimensions,  but the methods may be extended to heterotic orbifolds
and/or more compactified dimensions.

\medskip

The construction of symmetric orbifolds
may be carried out directly from the functional integral by inserting 
projection operators and including twisted sectors; for completeness,
it will be briefly reviewed below. The construction of 
asymmetric orbifolds, on the other hand cannot, in 
general, be  carried out directly from the functional integral. 
Therefore, properties that are known to hold generally for
symmetric orbifolds (such as their behavior under modular
transformations) may or may not hold for asymmetric orbifolds.
Various methods have been proposed  to circumvent 
these obstacles. The method of \cite{nsv} uses a doubling 
of the left- and right moving degrees of freedom, while 
the results of \cite{dvvv} rather suggest the use of operator methods.
(For reviews and further references, see e.g.  \cite{orbrev}.)
\medskip

The approach taken here to the construction of asymmetric 
orbifold superstring theories will be based on chiral splitting.
The starting point will be the construction of the chiral
blocks, carried out separately for the left-movers and for
the right-movers. Once the chiral blocks are in hand, 
string amplitudes are obtained by assembling these
left and right chiral blocks in a manner consistent 
with the definition of the asymmetric orbifold model
as well as with modular invariance. This approach will be 
discussed in enough detail here so that the cosmological 
constant in the KKS models can be investigated.
Further study will appear in a forthcoming publication \cite{adp}.

\subsection{Chiral splitting of symmetric orbifolds}

Let $G$ be an orbifold group acting on ${\bf R}^n$, with 
$n\leq 6$, i.e. with at least four uncompactified  space-time dimensions. 
An element  $g\in G$ is an Euclidean transformation on ${\bf R}^n$,
consisting of a rotation $R_g\in O(n)$ and a shift $v_g\in {\bf R}^n$, 
so that $g=(R _g,v_g)$, possibly supplemented  by an action
on internal quantum numbers. The action on the fields is given by
$g x = R _g x + v_g$ and $g \psi _\pm = R _g \psi _\pm$,
and thus symmetric on left- and right movers.  Clearly, the 
non-trivial action of $G$ on the fermion fields and on the momentum 
and oscillator modes of the bosonic fields is only by the 
subgroup of rotations $(R _g,0)$ (obtained by simply
omitting the shift $v_g$ for each $g\in G$), which form 
the point group $P_G$. The subgroup of elements of the form
$(I, v)$ on the other hand only acts on the zero modes of $x$
and forms a lattice $\Lambda_G$; the coset 
$T_G= \bR ^n /\Lambda _G$ is a torus. The symmetric orbifold may 
be constructed in two equivalent ways; 
\bea
\bR ^n /G = T_G /\bar P_G
\eea
i.e. as the coset of $\bR^n$ by the full orbifold group $G$, 
or as the coset of the torus $T_G$ by the group 
$\bar P_G = G/\Lambda _G$, which is isomorphic to $P_G$.
The construction via $T_G/\bar P_G$ is generally more convenient
(because the groups $\bar P_G$ are usually finite, while $G$ is always infinite)
and is better suited for dealing with asymmetric orbifolds; therefore
it will be adopted throughout. The orbifold is Abelian (resp.
non-Abelian) if the point group $P_G\sim \bar P_G$ is
Abelian (resp. non-Abelian).
 
\medskip

The orbifold string theory is constructed as follows. 
Let $A_I, B_I, I=1,2$ be a canonical homology basis and define 
the fermion fields  $\psi_\pm^{\mu}$  with a spin structure $\delta$. 
To the basis cycles $(A_I,B_I)$ we associate a sector $(a_I, b_I)$ 
of elements of $\bar P_G$ which  satisfies the homotopy  relation 
$\prod_{I=1,2} a_Ib_Ia_I^{-1}b_I^{-1}=I$. (For Abelian orbifolds,
this condition is automatically fulfilled.) The sector is untwisted 
if $a_I$ and $b_I$ equal the identity in $\bar P_G$, and is twisted otherwise. 
In a given sector $\lambda =(a_I,b_I)$, the fields obey the following
monodromy conditions,\footnote{In the fermion monodromy, 
an extra sign may arise depending on the reference spin structure
chosen in the given homology basis. For simplicity, this factor 
has been omitted here.}
\bea
\label{twistedbc}
(x,\psi_\pm) (z+A_I)
&=&
( a_I x(z),(-1)^{2\delta_I'} a_I \psi_\pm (z))
\no\\
(x,\psi_\pm)(z+B_I)
&=& 
(b_I x(z),(-1)^{2\delta_I''} b_I \psi_\pm (z))
\eea
The string amplitude in the sector $\lambda $ is given by the familiar
functional integral but where the fields now obey the monodromy
conditions (\ref{twistedbc}),
\bea
{\bf A} _C [\delta;\lambda]
=
\int\,DE_M{}^A\,D\Omega_M\,\delta(T) \int _{(a_I,b_I)} DX^\mu e^{-I_m}
\eea
The same procedure of gauge-fixing and chiral splitting as 
was carried out in (\ref{chiralsplit}) may be applied here, 
and results in a set of  chiral blocks. 

\medskip

The chiral blocks are indexed by the sector labels 
$\lambda =(a_I,b_I)$ and  by the internal 
momenta associated respectively with uncompactified 
($p_u$) and compactified ($p_\lambda$) dimensions.
The effects due to the uncompactified internal momenta $p_u$ 
are the same as in flat space-time and yield the familiar 
measure factor $(\det \Im \Omega )^{-5+n/2}$; henceforth, 
the dependence on $p_u$ will suppressed. The internal
momenta $p_\lambda$ associated with the compactified directions
are discrete and 
will take values in the lattice $\Lambda _G ^*$ dual to 
$\Lambda _G$. The chiral blocks will be denoted by 
$d\mu _C [\delta ; \lambda ](p_\lambda) = 
d\mu_C[\delta; a_I,b_I](p_\lambda )$ and the amplitude 
${\bf A}_C$  take on the form,
\bea
{\bf A} _C [\delta; \lambda] = \int _{\M_2}( \det \Im \Omega )^{-5 +n/2}
 \sum _{p_\lambda \in \Lambda _G^*} 
\biggl  |d\mu_C[\delta; \lambda](p_\lambda ) \biggr |^2
\eea
The physical problem with this amplitude is the presence 
of the tachyon, which in particular causes the integration
over moduli space to diverge. 

\medskip

In the GSO projection, left and right fermions are treated 
independently (via chiral splitting, see \cite{superanom,dp89}),
and assigned spin structures $\delta_L$ and $\delta_R$
which are summed over independently. A consistent GSO 
projection will systematically eliminate
the tachyon in all superstring loops. Isolating the 
chiral blocks  introduces a phase arbitrariness. Thus,
the GSO projection requires a consistent set of 
phases $\eta_L [\delta_L;\lambda]$ and  $\eta _R [ \delta_R;\lambda]$,
which may depend upon the sector $\lambda $.
The GSO projected amplitude for the symmetric
orbifold is,
\bea 
Z_G
=
\int _{\M_2}  (\det\, \Im \,\Omega)^{-5+{n\over 2}}  \sum_{\lambda}
\sum_{p_\lambda \in \Lambda _G ^*}
d\mu_L [\lambda ](p_\lambda) \wedge 
\overline{d\mu_R [\lambda](p_\lambda)} 
\eea  
where the GSO resummed chiral blocks are given by
\bea
\label{GSOblocks}
d\mu_L[\lambda ](p_\lambda )
& = &
\sum_{\delta_L} \eta _L [\delta_L; \lambda] \ 
d\mu_C[\delta_L; \lambda ](p_\lambda)
\no \\
\overline{d\mu_R [\lambda ](p_\lambda )}
& = &
\sum_{ \delta_R}  \eta_R [\delta_R; \lambda ]
\overline{d\mu_C[\delta_R ; \lambda ](p_\lambda )}
\eea
As is familiar from flat space-time, the phase assignments
on the left and right GSO summation may be taken to be 
the same (as in Type IIA) or different (as in Type IIB) 
from one another, even though for symmetric orbifolds
the chiral blocks for fixed spin structure for right movers 
are simply the complex conjugate of those for  left movers.

\subsection{Asymmetric Orbifolds}

An asymmetric orbifold results from taking the coset
of Minkowski space-time by a group $G$ that acts asymmetrically
on left and right moving degrees of freedom. More 
precisely, an asymmetric orbifold group $G$ consists
of {\sl pairs of Euclidean transformations of $\bR^n$}
$g = (R _{gL}, v_{gL}; R _{gR}, v_{gR})$. On the genuine
conformal fields $\p x$ and $\psi _\pm$, propagating
say on a cylinder, the action of the asymmetric orbifold
group elements is clearly given by the point group $P_G$
\bea
g (\p_z x, \psi _+ )  & =  & (R_{gL} \p_z x, R_{gL}  \psi _+ )
\no \\
g (\p_{\bar z} x, \psi _- )  & =  & (R_{gR} \p_{\bar z} x, R_{gR}  \psi _- )
\eea
where $x$ is viewed as a field taking values in a torus $T_G$.
The action by the shifts on the zero mode part of $x$, however, is 
more subtle; we shall not make use of it here and 
postpone a more detailed discussion to  \cite{adp}.

\medskip

For a higher genus surface, the group elements $a_I,b_I$
assigned to the canonical homology cycles $A_I,B_I$ 
now also each come in a 
pair of left and right rotations, $a_I = (a_{LI}; a_{RI})$ and 
$b_I = (b_{LI}; b_{RI})$. One would naturally be led to 
consider fields obeying the monodromy conditions
\bea
(\p_z x,\psi_+) (z+A_I)
& = &
( a_{LI} \p_z x(z),(-1)^{2\delta_I'} a_{LI} \psi_+ (z))
\no \\
(\p _z x,\psi_+)(z+B_I)
& = & 
(b_{LI} \p_z x(z),(-1)^{2\delta_I''} b_{LI} \psi_+ (z)) 
\no \\
(\p_{\bar z} x,\psi_-) (z+A_I)
& = &
( a_{RI}  \p_{\bar z} x(z),(-1)^{2 \bar \delta_I'} a_{RI} \psi_- (z)) 
\no \\
(\p _{\bar z}  x,\psi_-)(z+B_I)
& = & 
(b_{RI} \p_{\bar z} x(z),(-1)^{2\bar \delta_I''} b_{RI} \psi_- (z)) 
\eea
These monodromy conditions are, however, only formal,
as no real field $x$ in the functional integral will exist
that satisfies conditions with $a_{LI}\not=a_{RI}$ or
$b_{LI}\not= b_{RI}$. 

\medskip

The strategy adopted here is to obtain the amplitudes
to higher loop order for the asymmetric theory from 
the  chiral blocks  of  symmetric orbifold theories.
Specifically, each element $f \in P_G$ is written as a 
pair $f = (f_L;f_R)$ of rotations of $\bR^n$.
The set of all elements $f_L$ forms a group, which will
be denoted $P_L$, while the set of all elements $f_R$
forms a group $P_R$. The group $P_G$ is thus viewed as a 
subgroup of $P_L \otimes P_R$ subject to the pairing 
relation $f=(f_L;f_R)$. The chiral blocks for the left 
movers are the holomorphic blocks of the symmetric
theory with orbifold group $P_L$, while 
the chiral blocks for the right movers are the anti-holomorphic
blocks of the symmetric theory with orbifold 
group $P_R$. The internal momenta of the 
left and right movers  do not need to match,
as is familiar from toroidal and orbifold compactifications
of the heterotic string.  They will be denoted by $p_L$ and $p_R$;
the pair $(p_L,p_R)$ belongs to a self-dual even
lattice associated with $G$. When the elements of
$\bar P_G$ have non-trivial shifts, the self-dual lattice will, 
in general, depend on the sector label $\lambda = (a_I,b_I)$.

\medskip

Thus, the block associated with the
spin structures $\delta_L$ and $\delta_R$ and
sector $\lambda = (a_I,b_I)$, where $a_I = (a_{LI};a_{RI})$ 
and $b_I=(b_{LI};b_{RI})$, and chiral sector labels
$\lambda _L = (a_{LI}, b_{LI})$,
$\lambda _R = (a_{RI}, b_{RI}) $,
is given by
\bea
d\mu _C [\delta_L; \lambda _L] (p_L) \wedge 
\overline{d \mu _C [\delta_R ; \lambda _R](p_R)}
\no
\eea
Just as in the case of flat Minkowski space-time (or,
as explained in the previous subsection, for symmetric
orbifold compactifications) a GSO projection must 
be performed to eliminate the tachyon. For most 
interesting asymmetric orbifolds, this projection is also
carried out in a chiral manner, i.e. summing 
independently over the spin structures of left and right
fermions. The relevant GSO resummed blocks are 
then precisely those of (\ref{GSOblocks}) but
now with different sector labels for left and right,
\bea
d\mu_L[\lambda _L ](p_L)
&=&
\sum_{\delta_L}\eta _L [\delta_L ; \lambda _L]\ 
d\mu_C[\delta_L;\lambda _L ](p_L)
\no\\
\overline{d\mu_R[\lambda _R ](p_R)}
&=&
\sum_{\delta_R}  \eta_R [\delta_R; \lambda _R]\ 
\overline{d\mu_C[\delta_R;\lambda _R ](p_R)}
\eea
In the next subsection, constraints will be established
on the phases $\eta _L$ and $ \eta_R$ arising from
modular symmetry.

\medskip

The proposal made here for the partition function $Z_G$ 
of the asymmetric orbifold theory with asymmetric 
orbifold group $G$ is given by the expression 
\bea
\label{ZG}
Z_G
=
\int _{\M_2} (\det\, \Im \,\Omega)^{-5+{n\over 2}}\sum_{p_L,p_R}
\sum_{\lambda _L, \lambda _R} K(\lambda _L, \lambda _R;p_L,p_R)
\,d\mu_L [\lambda _L](p_L) \wedge
\overline{d\mu_R[\lambda _R](p_R)} 
\eea 
Here,  the coefficients $K(\lambda _L, \lambda _R,p_L, p_R)$ 
may depend on both 
sets of left and right group elements and internal momenta.
Non-trivial internal momentum dependence in $K$ will arise
in particular when the group elements of $G$ involve 
non-trivial shifts.
Clearly, much of the structure of the original asymmetric group $G$ is 
encoded in these coefficients. Their evaluation can be
a subtle issue, even to one-loop, and we shall return to this in \cite{adp}. 
In this paper, we shall restrict attention to the pointwise vanishing 
of the cosmological constant, and examine only the vanishing of 
each GSO resummed chiral block
$d\mu_L[\lambda _L](p_L)$ and $\overline{d\mu_R[\lambda _R](p_R)}$.

\subsection{GSO projection phases and $\mathbf{Z_2}$-twisting}

In the models to be studied below,  the point group $P_G$ will be
Abelian and generated by chiral $\bZ_2$ reflections
(possibly together with an action on internal quantum 
numbers, such as by the operators  $(-)^F$).
In all such cases, the chiral blocks will  be given 
(up to phases)  by the ${\bf Z}_2$-twisted blocks $d\mu_C[\delta; \ep](p_\ep)$ 
which were  derived earlier. 
The sector labels will be abbreviated by $\lambda \equiv (a_I,b_I)$.
The twist $\ep$, which enters the chiral  blocks  $d\mu_C[\delta; \ep](p_\ep)$ 
will be determined by the sector labels
$\lambda$, whence  the notation  $\ep = \ep(\lambda)$. 

\medskip

It is assumed that the GSO summation over 
$\delta$ is carried out independently on left- and right-movers. 
For the models of greatest physical interest, the number of
${\bf Z}_2$-twisted dimensions is~4. This is the number
of twisted dimensions in the KKS model, as well as in the orbifolding
of the Type II theories by a pure $Z_2$ twist which
yield again the same Type II theory. Henceforth, the number of
twisted dimensions will be assumed to be 4.

\medskip

In view of the above assumptions, the relevant summation is a chiral
summation over the measure calculated in (\ref{finalmeasure}) with $n=4$.
The result is given by (we use the abbreviation $\delta = \delta _L, 
\lambda = \lambda _L$ and $p=p_L$),
\bea
\label{GSOmeasure}
d\mu _L[\lambda ] (p  )
=
\sum _{\delta} \eta _L[\delta ;\lambda] ~d\mu _C [\delta, \ep(\lambda)] (p  )
=
d\mu _L^{(1)}[\lambda]  (p ) + d\mu _L^{(2)}[\lambda]  (p)
\eea
where the partial measures are defined by (still using the notation 
$\ep = \ep(\lambda)$)
\bea
d\mu _L^{(1)}[\lambda]  (p  )
& = &
\sum _\delta \eta _L [\delta; \lambda] ~
e^{i \pi \tau_{\ep}  p ^2}  {Z_C [ \delta, \ep] \over Z_M [\delta] }
 \biggl ( i \pi p ^2 - 4 \p_{\tau_\ep} \ln \tet _i (0,\tau_\ep) \biggr )
\Gamma [\delta; \ep]
\no \\
d\mu_L^{(2)}[\lambda]  (p ) 
& = &
\sum _\delta \eta _L [\delta ;\lambda] ~
e^{i \pi \tau_\ep  p ^2}  {Z_C [ \delta, \ep] \over Z_M [\delta] }
 {\Xi_6[\delta] \tet[\delta]^4 \over  16\pi^6 \Psi_{10}}
\eea
Here, $\eta _L[\delta; \lambda]$ are the left chiral 
GSO projection phases, which remain to  be determined.

\medskip

The assumption $n=4$ actually leads to considerable simplifications.
A key ingredient in the chiral blocks is the ratio of chiral partition 
functions for the twisted space dimensions. For $n=4$, this  quantity 
simplifies,
\bea
{Z_C [\delta; \ep]  \over Z_M [\delta] }
=
{\tet [\delta _j ^+ ] ^2  \ \tet [\delta _j ^- ]^2  \ \tet
[\delta +\epsilon ]  ^2
\over
\tet _j ^4  \ \tet [\delta ]  ^2 }
\eea
since now only
squares of $\tet$-constants are involved. 

\medskip

Next, the expression for $\Gamma [\delta; \ep]$ was
calculated in (\ref{finalGamma}).
Finally, the Prym period $\tau _\ep$ is determined by the
Schottky relations in terms of the super-period matrix, which is
now simply denoted by  $ \Omega_{IJ}$. 
In our formalism,  $\Omega_{IJ}$ is interpreted as the period matrix 
of a  bosonic spin structure independent Riemann surface. 
Therefore, the Prym period $\tau _\ep$ must be viewed as independent 
of the spin structures $\delta$, and the Gaussian
involving $\tau_\ep$ may be factored out of the sum over $\delta$.
The following simplified expressions are obtained, 
\bea
d\mu _L^{(1)}[\lambda]  (p  )
& = &
{ i \over (2\pi )^7 \eta (\tau _\ep)^{12} } {\p \over \p \tau _\ep}
\left [
e^{i \pi \tau_\ep  p ^2}
\sum _{i=2,3,4} ~ \sum _{\alpha = \pm } \alpha \  
\eta _L [\delta ^\alpha _i;\lambda]
\< \nu _0 | \mu _i \> \tet _i  ^4 \right ]
\no \\
d\mu_L^{(2)}[\lambda]  (p )
& = &
{ e^{i \pi \tau _\ep p^2 } \over 16 \pi ^6 \Psi _{10} }
{\tet [\delta _j ^+ ] ^2  \ \tet [\delta _j ^- ]^2 \over  \tet _j ^4   }
\sum _\delta \eta _L[\delta ;\lambda ] \ \Xi _6 [\delta ] \
\tet [\delta ]^2 \ \tet [\delta + \ep ]^2
\eea
It remains to put constraints on the GSO phases and 
determine all possible solutions to these constraints.

\subsection{Modular Covariance constraints on GSO phases}

Constraints on the GSO phases arise from the requirement of 
modular invariance of the full string measure. Under a 
general modular transformation $M\in Sp(4,\bZ)$,
the period matrix $\Omega$, the internal momenta $p_L$,
the  spin structure $\delta$, 
the sector label $\lambda =(a_I,b_I)$ and the twist  $\ep$
are transformed to $\tilde \Omega$, $\tilde p_L$, $\tilde \delta$, 
$\tilde \lambda$ and $\tilde \ep$, according to familiar 
rules, which are summarized in Appendix B. 
Modular invariance requires the following transformation
law for the GSO resummed measure,
\bea
\label{covariance}
d \mu _L[\tilde \lambda] (\tilde p _L, \tilde \Omega )
=
\varphi (\lambda ,M) (c \tau _\ep +d)^{-2} 
d\mu _L[\lambda] (p _L, \Omega)
\eea
The factor $(c \tau _\ep +d)^{-2} $ represents the effect of
the modular transformation induced by $M$ on the Prym
period $\tau _\ep$, and $\varphi$ is a set of phases.

\medskip

The constraints on the GSO phases $\eta $ are
most easily established by restricting $M$  to the
subgroup $H_\ep$ of the modular group which leaves 
$\ep$ invariant. The two terms 
$d\mu_L^{(1)}$ and $d\mu_L^{(2)}$ are 
functionally independent, and therefore must each result 
from a modular covariant GSO summation
over $\delta$. The modular transformations of 
$d\mu_L^{(2)}$ are obtained by combining the relation
\bea
\Xi_6 [\tilde \delta] (\tilde \Omega) \tet [\tilde \delta] (0,\tilde \Omega)^4
=
\det (C \Omega + D)^8 \Xi_6 [\delta] (\Omega) \tet [\delta]  (0, \Omega)^4
\eea
familiar from \cite{IV} with the transformation law of the $\tet$-constants,
\bea
{\tet [\tilde  \delta +  \ep] (0, \tilde \Omega) \over
\tet [\tilde \delta ](0, \tilde \Omega)}
=
{\epsilon (\delta +\ep,M) \over \epsilon (\delta,M)} \
{\tet [\delta + \ep] (0, \Omega) \over  \tet [\delta ](0,\Omega)}
\eea
Therefore, modular covariance requires that
\bea
\eta _L[\tilde \delta ; \tilde \lambda ] \ {\epsilon (\delta +\ep,M)^2  \over \epsilon
(\delta,M) ^2 }  = \varphi  (\lambda, M) \eta _L[\delta ; \lambda]
\qquad {\rm for \ all} \qquad M \in H_\ep
\eea
The ratios of $\epsilon ^2$ may be computed by inspecting Table \ref{table:4}.
The symmetric pairs $(\delta, \delta+\ep)$ take the values
$(\delta _1,\delta_2)$, $(\delta _3, \delta _4)$ and $(\delta _7, \delta _8)$.
It is manifest from Table  \ref{table:4} that the $\epsilon ^2$ values for
both members of each pair coincide  for all $M \in H_\ep$. 
Thus, we have the following condition for each given $\lambda$,
\bea
\label{modcov}
\eta _L[\tilde \delta ; \tilde \lambda] = \varphi  (\lambda , M) \eta _L[\delta ;\lambda]
\qquad {\rm for \ all} \qquad M \in H_\ep
\eea
It is straightforward to show that the second part of the chiral measure
$d\mu_L^{(1)}$ is then automatically modular covariant. To solve these
constraints requires more detailed information on the structure
of the orbifold group $G$. Two specific examples will be discussed 
in the remainder of this section.

\medskip

From Table \ref{table:4}, it is clear that the modular subgroup,
as it acts on the even spin structures, has {\sl two distinct orbits}.
Actually, these orbits are distinguished by the signatures
$\< \ep | \delta \>$ in the following manner,
\bea
\label{deltaepsig}
\< \ep | \delta \> = +1 \quad & {\rm orbit} &  \quad
\{ \delta _1, \delta _2, \delta _3, \delta _4, \delta _7, \delta _8 \}
\no \\
\< \ep | \delta \> = -1 \quad & {\rm orbit} &  \quad
\{ \delta _5, \delta _6, \delta _9, \delta _0 \}
\eea
The orbit $\{ \delta _1, \delta _2, \delta _3, \delta _4, 
\delta _7, \delta _8 \}$ is
precisely the one that contains all even spin structures 
$\delta$ for which
$\delta + \ep$ is also even; these are the only spin 
structures that enter here.

\subsection{$\mathbf{Z_2}$ Orbifolds}

We now consider the model where $4$ directions are 
compactified by an orbifold group $G$ whose point $P_G$
group consists of single $\mathbf{Z_2}$ chiral reflection $r_L$
of the fields $x$ and $\psi$. Since $r_L^2=1$, all sectors
may be labeled precisely by a single twist $\ep$, i.e. $\lambda = \ep$.
As a result, the modular covariance condition (\ref{modcov}) 
further simplifies and becomes,
\bea
\label{GSOmod}
\eta _L[\tilde \delta ; \ep] = \varphi  (\ep , M) \eta _L[\delta ; \ep]
\qquad {\rm for \ all} \qquad M \in H_\ep
\eea
By a modular transformation, $\ep$ may be brought to the standard 
form (\ref{standardep}). The spin structures transform under 
$H_\ep$ in the two orbits listed in (\ref{deltaepsig}), and 
(\ref{GSOmod}) is to be solved separately in each orbit.

\medskip

Observe that if a modular transformation $M\in H_\ep$
leaves any one of the spin structures $\delta$ invariant, 
one has $\eta _L[\delta  ;\ep] = \varphi  (\ep , M) \eta _L[\delta ;\ep]$, 
and thus $\varphi (\ep,M)=1$. (The alternative, $\eta _L[\delta ;\ep]=0$
would lead to zero GSO phases throughout the orbit, and thus 
a vanishing partition function, which is excluded.) It also 
follows that any spin structures related by $M$ will have the same 
GSO phase $\eta _L[\delta ; \ep]= \eta _L[\tilde \delta ;\ep]$.

\medskip

Applying these observations to the case of  orbit 
$\{ \delta _1, \delta _2, \delta _3, \delta _4, \delta _7,  \delta _8 \}$, simple
inspection of Table \ref{table:4} reveals that $M_1, M_3$ and $T_2$
have fixed points, so that
$\varphi (\ep, M_1)=\varphi (\ep, M_3)=\varphi (\ep, T_2)=1$, and  
$\eta _L[\delta _1;\ep]= \eta _L[\delta _2;\ep]$, 
$\eta _L[\delta _3;\ep]= \eta _L[\delta _4;\ep]$, and
$\eta _L[\delta _7;\ep]= \eta _L[\delta _8;\ep]$.
From the first two relations, it follows that 
$\varphi (\ep, S_2)=1$, while from the last that $\varphi (\ep, M_2)=1$.
Using the action of $S_2$ and $M_2$ it follows that
$\eta _L[\delta ; \ep]$ is independent of $\delta$ within this entire  orbit, 
and may be  set equal to $\eta _L[\delta ;\ep] =1$. (The 
analysis in the case of the orbit 
$\{ \delta _5, \delta _6, \delta _9,\delta _0\}$ 
leads to the conclusion that $\eta _L[\delta ;\ep]$ 
is  constant throughout also this orbit.) Since the chiral
measure for $\bZ_2$ reflections is non-zero
for only the first orbit above, the GSO
resummed measure becomes simply,
\bea
d \mu _L[\ep] (p _L )
=
\sum _{\delta} \ d \mu _L [\delta , \ep ]  (p _L )
\eea
Notice that with a single twist, the assignment of GSO 
phases consistent with modular invariance (in the even
spin structure sector) is unique, a situation that is 
familiar from flat space-time \cite{IV}.

\medskip

For the above  GSO phases, the chirally summed measure
$d\mu_C^{(2)}$  vanishes pointwise on moduli space.
To prove this, the starting point is a relation, derived first in \cite{IV},
\bea
\sum _\delta \Xi _6 [\delta ] \tet [\delta ]^4 S_\delta (z,w)^2 =
\sum _\delta \Xi _6 [\delta ]  \ \tet [\delta ]^2(0)  \ \tet [\delta ]^2
\left ( \int _w ^z \omega _I \right ) =0
\eea
for any pair of points $z,w$ on the surface. 
It suffices to choose the pair to be branch points, canonically associated 
with the twist via $\ep= -\nu _a + \nu _b$ and $z = \Delta + \nu _b$
and $ w = \Delta + \nu_a$, where $\Delta$ is the Riemann class. 
Using  (\ref{tetrelations}) and the specific form of the standard
twist $\ep$,  the desired relation is readily obtained,
\bea
\sum _\delta
\Xi _6 [\delta ]  \ \tet [\delta ]^2  \ \tet [\delta +\ep ]^2   =0
\eea
which proves that $d\mu_L^{(2)}[\ep]  (p_L )=0$.
The vanishing of the measure $d\mu_L^{(1)}[\ep]  (p_L)$
is even simpler, as it results directly from the Riemann identities
for the Prym variety,
\bea
\sum _{i=2,3,4} \< \nu _0|\mu_i\> \tet _i (0,\tau_\ep)^4 =0
\eea
The vanishing of $d\mu _L ^{(1,2)}$ might have been  expected, 
on the grounds that the model obtained by orbifolding by a
single $\bZ_2$ reflection group produces an orbifold superstring 
theory that coincides with the original 
Type II theory.\footnote{We are grateful
to Eva Silverstein and Shamit Kachru for detailed 
explanations of this point.}

\subsection{$\bZ_2 \times \bZ_2$ Orbifolds :  KKS models }

The models introduced by Kachru-Kumar-Silverstein (KKS) 
in \cite{KKS} are constructed on a square torus $T_G$ at 
self-dual radius and an asymmetric orbifold
group $G$ whose point group  $\bar P_G$ is generated by
two elements, 
\bea
f & = & \left ( (r_L,s_R)^{1-4} , (1,s_R^2)^5, (s_L,s_R)^6; (-)^{F_R} \right )
\no \\
g & = & \left ( (s_L,r_R)^{1-4} , (s_L,s_R)^5, (s_L^2,1)^6; (-)^{F_L} \right )
\eea
The superscripts indicate the dimension labels on which the
corresponding generator acts. The operators $(-)^{F_L}$ and $(-)^{F_R}$
denote the parity of left and right chirality worldsheet fermion number
respectively, with value $+$ on NS states and $-$ on R states.
The operations $r_L,r_R,s_L$ and $s_R$ are chiral reflections
and shifts, defined below.

\subsubsection{Chiral Reflections and Shifts}

The chiral reflections $r_L$ and $r_R$ act on $x^\mu $ and $\psi ^\mu $ by
reflecting respectively the left and right oscillators of both fields
and the momenta of $x^\mu$. Since $r_L$ and $r_R$ are reflections,
their order is 2. The product $r=r_L r_R$ is the usual
reflection on the non-chiral  fields, given by $r x^\mu = - x^\mu $
and $r\psi ^\mu = - \psi ^\mu $. 
The operators $s_L$ and $s_R$ leave the full fields
$\psi ^\mu$, as well as the oscillators of $x^\mu $ invariant.
Their action on $x^\mu$ is solely through chiral shifts of the zero mode.
The operators may be represented in terms of the left and right
momentum operators $p_L^\mu$ and $p_R ^\mu$,
\bea
s_L = \exp \{2 \pi i \sum _\mu p_L ^\mu R/2 \}
\hskip 1in
s_R = \exp \{ 2 \pi i \sum _\mu p_R ^\mu R/2 \}
\eea
The product $s=s_L s_R$ is the usual shift by half a circumference.
At the self-dual radius $R=1/\sqrt{2}$, both operators 
$s_L$ and $s_R$ are of order 4. Compactification to a radius $R$ restricts
the momentum parameters $(p_L,p_R)$ to a discrete Lorentzian
lattice $\Gamma_R$. 
To simplify the notation, we shall not indicate this explicitly.

\subsubsection{GSO phases}

Of relevance here is the action of each generator on the left chiral blocks,
including their spin structure dependence. The action of the chiral shift
operator $s_L$ is diagonal on  the chiral blocks and will enter only when
assembling left and right chiral blocks together. Thus, when carrying out the
GSO projection by summation over spin structures in each chirality  sector
separately, the action of the chiral shift operators is immaterial. 
Effectively, the action of the orbifold group $G$ on left chiral blocks
reduces to the action of the point group $P_G$, whose generators are
\bea
f=(f_L;f_R) & \quad & f_L = r_L ^{1-4} \hskip .4in f_R=(-)^{F_R}
\no \\
g = (g_L;g_R) & \quad & g_L= (-)^{F_L} \hskip .3in  g_R = r_R ^{1-4}
\eea
Clearly, each generator is of order 2, whence the fact that on the left chiral
blocks, the action of the orbifold group is by the Abelian point 
group $P_G= \bZ _2 \times \bZ_2 $.

\medskip

All possible sectors of the $\bZ _2 \times \bZ_2 $ theory
on the genus 2 surface may  be parametrized  by two
half-characteristics $\ep$ and $\alpha$. To establish this, 
denote the assignments of elements of $\bZ _2 \times \bZ_2 $
around each  homology cycle as before, $(A_I,B_I)  \to
(a_I,b_I)$ for $I=1,2$ and $a_I,b_I \in \bZ _2 \times \bZ_2 $.
One has the following correspondence between the 
twist values $\ep, \alpha$ and the group element assignments
on the homology cycles,
\bea
a_1 = (f_L ^{2 \ep _{1} '} g_L ^{2\alpha _{1}'}; 
f_R ^{2 \ep _{1} '} g_R ^{2\alpha _{1}'})
\hskip 1in
b_1 = (f_L ^{2 \ep _{1}''} g_L ^{2\alpha _{1}''}; 
f_R ^{2 \ep _{1}''} g_R ^{2\alpha _{1}''})
\no \\
a_2 = (f_L ^{2 \ep _{2} '} g_L ^{2 \alpha _{2}'}; 
f_R ^{2 \ep _{2} '} g_R ^{2 \alpha _{2}'})
\hskip 1in
b_2 = (f_L ^{2 \ep _{2}''} g_L ^{2 \alpha _{2}''}; 
f_R ^{2 \ep _{2}''} g_R ^{2 \alpha _{2}''})
\eea
Given a spin structure $\delta$, the effect of $g_L$ is to assign a sign
factor depending on whether the state traversing the corresponding
cycle is NS or R. This assignment takes a  simple  form
in terms of the signature,
\bea
{\rm NS \ state \ traversing \ cycle \ } \alpha & \qquad 
& \< \alpha |\delta \>= +1
\no \\
{\rm R \ state \ traversing \ cycle \ } \alpha & \qquad 
& \<  \alpha |\delta \>= -1
\eea
To check this, it suffices to work out a specific twist assignment, e.g.
the standard twist $\alpha =\ep$ of (\ref{standardep}), 
whose only non-zero  entry is $\ep_2''=\12$. If $\delta '_2 = 0$, 
the state traversing the  $A_2$ cycle is  indeed NS and the 
signature  assignment comes out to be $+1$, while if 
$\delta '_2 = \12$,  the state is R and we get $-1$.

\medskip

The net effect in the sector $(a_I,b_I)$, parametrized by
two twists $\ep$ and $\alpha$, is to produce a ${\bf Z}_2$ 
reflection of the fields $x$ and $\psi _+$ on the cycle $\ep$, 
and an insertion of the phase $\<\alpha|\delta\>$. 
As $(a_I,b_I)$ will run over all sectors, 
$\lambda \equiv (\ep,\alpha)$ will run over all pairs of 
half-characteristics.
It follows that the set of chiral blocks for the 
$\bZ_2\times\bZ_2'$ theory is given by 
\be
d\mu_L[\delta;\ep,\alpha](p_L)
=
\<\alpha|\delta\>\ d\mu_L [\delta; \ep](p_L)
\ee
Assembling all ingredients, the GSO summed chiral measure for given
twists $\ep $ and $\alpha$ is obtained as follows,
\bea
d \mu _L[\ep,\alpha] (p _L )
=
\sum_{\delta}d\mu_L[\delta;\ep,\alpha](p_L)
=
\sum _\delta \< \alpha  |\delta \>\ d \mu _L [\delta , \ep ] 
(p _L )
\eea
Under modular transformations that leave the twist $\ep$ invariant, 
the measure transforms covariantly,
\bea
d \mu _L[\ep, \tilde \alpha] (\tilde p _L, \tilde \Omega )
\equiv
\varphi (\ep,\alpha,M) (c \tau _\ep + d)^{-2} d \mu _L[\ep , \alpha] (p _L, \Omega )
\eea
Here, the phase factor $\varphi (\ep,\alpha,M)= \< \alpha | (M)_0\> $ 
depends only on $M$ and $\alpha$.\footnote{To be precise, $(M)_0$ is
the inhomogeneous contribution in the modular transformation
of spin structures given in (\ref{modspin}), which is common 
to all spin structures.} The presence of the phase factor $\<\alpha |\delta\>$
represents a generalization of  discrete torsion \cite{vafa86} to the chiral case.

\medskip

By analyzing the action of the modular subgroup that leaves
{\sl two twists invariant}, as done in \S 6.4, one may show that 
the above phase assignment is the unique non-trivial
choice consistent with modular invariance. The proof is similar 
to the one given to determine  the unique phase assignment 
for the $\mathbf{Z_2}$ case in \S 8.5.
This will be proven  here for the case $\alpha \in \O_+[\ep]$ only; 
the other case is analogous.  Upon making the choice $\alpha =\ep_3$,
the stabilizer of both twists, $H_{\ep, \alpha}$,
is generated by $M_1,M_2$ and $M_3$. Acting on the spin structures
$\{\delta _1,\delta _2,\delta _3, \delta _4, \delta _7, \delta _8\}$,
$M_1$ and $M_3$ have fixed points. Therefore, $\varphi(\ep,\alpha;M_1)
=\varphi(\ep,\alpha;M_3)=1$ and thus 
$\eta _L[\delta _1;\ep,\alpha] = \eta _L[\delta _3;\ep,\alpha]$,
$\eta _L[\delta _2;\ep,\alpha] = \eta _L[\delta _4;\ep,\alpha]$, 
and $\eta _L[\delta _7;\ep,\alpha] = \eta _L[\delta _8;\ep,\alpha]$.
In view of the last equality, we must have $\varphi (\ep, \alpha;M_2)=1$,
and we therefore conclude that 
\bea
\eta _L[\delta _1;\ep,\alpha] = \eta _L[\delta _2;\ep,\alpha]= 
\eta _L[\delta _3;\ep,\alpha] = \eta _L[\delta _4;\ep,\alpha]
\qquad \quad
\eta _L[\delta _7;\ep,\alpha]= \eta _L[\delta _8;\ep,\alpha]
\eea
If all $\eta$ are equal, we recover the case of twisting by a single $\bZ_2$.
The only other linearly independent case is when 
\bea
\eta _L[\delta _i;\ep,\alpha] = +1,~ i=1,2,3,4
\qquad \quad
\eta _L[\delta _7;\ep,\alpha] = \eta _L[\delta _8;\ep,\alpha]=-1
\eea
but this assignment is readily seen to coincide with 
$\eta _L[\delta ;\ep,\alpha] = \<\alpha|\delta\>$, which
is what we set out to prove.

\medskip

\subsection{Analysis of the GSO resummed measure for KKS models}

For each sector  $\lambda =(\ep,\alpha)$, the chiral block 
$d\mu_C[\ep,\alpha](p_L,\Omega)$ may be analyzed by 
exploiting its modular properties. Using a first modular transformation, 
$\ep$ may be mapped into the standard form, familiar from
(\ref{standardep}),
\bea
\ep= \left (\matrix{0 \cr 0\cr} \bigg | \matrix{0 \cr \12 \cr} \right )
\eea
Under the modular subgroup $H_{\ep }$ (which leaves $\ep$ invariant), 
twists transform in 4  orbits, as was shown in \S 6.3. The contribution to the
GSO resummed measure must be analyzed orbit by orbit.

\bigskip

\noindent
$\mathbf{(1) \ \alpha \in \O _0 [\ep]}$. \\
This orbit corresponds to the untwisted sector and 
has only one element, $\alpha =0$.
Using the results of section 7.2, one establishes
$d \mu _L[\ep, 0] (p _L, \Omega ) =0$.

\bigskip

\noindent
$\mathbf{(2) \ \alpha \in \O_{\ep}[\ep]}$.\\
This orbit also has only one element, $\alpha = \ep$.
Using the results  (\ref{deltaepsig}), and (7.2),
one establishes $d \mu _L[\ep, \ep] (p _L, \Omega ) =0$.

\bigskip

\noindent
$\mathbf{(3) \ \alpha \in \O _-[\ep] }$. \\
The twists in this orbit are such that $\< \alpha| \ep \>=-1$.
One may choose the  representative
$$
\alpha = \left (\matrix{0 \cr \12\cr} \bigg | \matrix{0 \cr 0 \cr} \right )
$$
The signature with the spin structures becomes explicit :
$\< \delta ^\pm _i |\alpha \> = \pm 1$.
Therefore, the summation over $i=2,3,4$ vanishes
and we have  $d\mu _L ^{(1)}[\ep] (p_L , \Omega)=0$.
The measure $d\mu_L ^{(2)}[\ep](p_L,\Omega)$ involves the sum
$$
\sum _{i=2,3,4} \left ( \Xi _6 [\delta _i ^+ ] - \Xi _6 [\delta _i ^-] \right )
\tet [\delta _i ^+]^2 \tet [\delta _i ^- ]^2
$$
In the separating degeneration, this sum vanishes to order
$\tau ^2$ included. This may be seen from the following limit,
derived from (\ref{seplimits}) for each $j=3,4$,
\bea
\sum _{i=2,3,4}
\Xi _6  \left [\matrix{\mu_i \cr \mu_j} \right ] 
\tet \left [\matrix{\mu_i \cr 00} \right ]^2
\tet \left [\matrix{\mu_i \cr 0\12} \right ]^2
= \O (\tau^4)
\eea
using the Riemann identities of (\ref{Riemannid}) for the 
summation over $i=2,3,4$. It may be shown that the order
$\tau^4$ term, on the other hand is non-vanishing.

\bigskip

\noindent
$\mathbf{(4) \ \alpha \in \O _+[\ep] }$. \\
The twists in this orbit are such that $\< \alpha | \ep\>=+1$.
One may choose the  representative
$$
\alpha = \left (\matrix{0 \cr 0 \cr} \bigg | \matrix{\12 \cr 0 \cr} \right )
$$
so that $\< \delta | \alpha \> = (-)^{2 \delta '_1}$. The measure
$d\mu _L^{(1)}[\ep]  (p_L , \Omega )  $ manifestly vanishes, as the
contributions from $\alpha =+$ and $\alpha =-$ cancel one another out.
The measure $d\mu _L^{(2)}[\ep]  (p_L , \Omega )  $ on the
other hand vanishes only to leading order as $\tau \to 0$, but 
is non-vanishing to order $\tau^2$.
To see this, the following limit is
taken
\bea
&&
\sum _{\delta }
\< \alpha  | \delta \>\,  \Xi _6 [\delta ]\,  \tet [\delta ]^2
\, \tet [\delta + \ep ]^2
\\
&&
\quad
=
- 256 \pi ^2 \tau ^2 \eta (\tau_1)^{12} \eta (\tau_2 )^{12}
\left ( \tet ^8 _3 - \tet ^8 _4 \right )(\tau_1) \
\tet ^4 _2 \tet ^2 _3 \tet ^2 _4(\tau_2)
+ \O (\tau^4)
\no
\eea
Thus, the chiral measure $d\mu _L[\ep,\alpha](p_\ep)$ is 
non-vanishing on both orbits $\alpha \in \O_\pm [\ep]$.

\subsection{Conclusions on the KKS model}

The non-vanishing of the contribution from the orbit $\O_- [\ep]$
would appear to contradict the conclusions of \cite{KKS},
where, instead, it is claimed to be vanishing.
The arguments made in \cite{KKS} are based on the fact 
that, when viewed as elements of the full orbifold group $G$,
the generators $f$ and $g$  do not commute with one another.
Since their commutator is a pure translation, the 
elements $f$ and $g$ do commute, however, 
when viewed as elements of the point group $\bar P_G$. 

\medskip

Standard arguments for symmetric orbifolds,  
based on functional integral methods,
guarantee the equivalence between coseting $\bR^n$ by G
and coseting $T_G$ by $\bar P_G$. They also 
guarantee that no contributions will arise from 
any sector in which the homotopy relation 
$\prod _{I=1,2} a_I b_I a^{-1} _I b^{-1}_I =1$
fails to be satisfied, whether $a_I, b_I$ take values 
in $G$ or in $\bar P_G$.

\medskip

For asymmetric orbifolds, however, no direct functional 
integral formulation is available. Therefore, the 
arguments made in the case of symmetric orbifolds
may or may not hold for asymmetric orbifolds. 
In particular, it is unclear how the coset theory
$\bR^n /G$ should be constructed, since it is unclear
how the zero mode of the field $x$ should be handled.
Only the construction of the coset $T_G/\bar P_G$ is
well-defined. Its construction  guarantees that no 
contributions will arise from 
a sector in which the homotopy relation 
$\prod _{I=1,2} a_I b_I a^{-1} _I b^{-1}_I =1$
fails to be satisfied, only when $a_I, b_I$ take values 
in the space group $P_G$ (but not necessarily when
$f$ and $g$ take values  in the full group $G$). 

\medskip

To summarize the situation for orbit $\O_-[\ep]$ :
the GSO resummed chiral blocks in orbit
$\O_- [\ep]$ are non-vanishing. Yet, it is conceivable that 
assembling left and right blocks of the asymmetric 
orbifold will  effectively  enforce the vanishing of contributions 
arising from  sectors in which the homotopy relation 
$\prod _{I=1,2} a_I b_I a^{-1} _I b^{-1}_I =1$
fails to be satisfied, when $a_I, b_I$ take values 
in the  full group $G$. No evidence in favor
of this possibility is available, however, at this time.

\medskip

In the orbit $\O_+[\ep]$,  the chiral GSO resummed  
blocks are also  non-vanishing. 
As their contribution is $\O (\tau^2)$
in the separating degeneration limit, they dominate
in this limit (over the orbit $\O_-[\ep]$, which is
$\O(\tau ^4)$). This result contradicts the claim
of \cite{KKS}, where the corresponding chiral
contribution was found to vanish. Noticing that
the sectors for this block all satisfy the homotopy
relation, no cancellation should be expected
when assembling left and right blocks.

\medskip

To summarize, the non-vanishing of the contribution 
of orbit $\O_+[\ep]$ shows that the cosmological constant
{\sl cannot vanish pointwise on moduli space} and  
gives strong evidence that the full
cosmological constant  in the KKS models is non-vanishing.
The net effect of the correct treatment of the superstring
measure given here is that the cosmological constant
in KKS models involves the sum over the new modular
object $\Xi_6[\delta]$, discovered in \cite{IV},
which, in contrast to the measure derived from 
the BRST picture changing operator formalism
does not lead to a cancellation in the chiral blocks.

\bigskip
\noindent
{\bf Acknowledgments}

\medskip

We are happy to acknowledge conversations with Michael Gutperle, Per Kraus, 
and Edward Witten. We are especially grateful to Shamit Kachru and Eva 
Silverstein for several helpful discussions of various fine points in the 
theory of asymmetric orbifold compactifications and for detailed explanations 
of their specific models. 

\appendix

\section{Appendix : Genus 1 $\tet$-function identities}
\setcounter{equation}{0}

Let $\omega$ and $\omega '$ be the two half period of the torus and $\tau
= \omega ' / \omega$ its modulus. The Weierstrass function obeys
\bea
\wp ' (z) ^2 = 4 (\wp (z) - e_1)(\wp (z) - e_2)(\wp (z) - e_3)
\eea
Recall the prime form $E$ and the Szego kernel $S_\mu$
for even spin structure $\mu$ at genus 1,
\bea
E(z,w) = {\tet _1 (z-w|\tau) \over \tet _1 ' (0|\tau)}
\hskip 1in
S_\mu (z,w) =
{\tet [\mu] (z-w|\tau ) \over \tet [\mu ](0|\tau) E(z,w)}
\eea
The correspondence between the classical Jacobi
$\tet_i$-functions and the spin structures is given by
$\tet _i (z|\tau) \equiv \tet [\mu_i](z|\tau)$ with
\bea
\mu_1 = (\12 |\12) \qquad
\mu_2 = (\12 | 0) \qquad
 \mu_3 = (0|0) \qquad \mu_4 = [0\12 ]
\eea
The Dedekind $\eta$-function is related to the product 
of all $\tet$-constants,
\bea
\tet ' _1 (0| \tau) = - 2\pi \eta (\tau )^3 = 
-  \pi \tet _2 (0|\tau) \tet _3 (0|\tau) \tet _4 (0|\tau)
\eea
We have the following doubling formulas,
\bea
\tet _1 (2z| 2 \tau) & = &
\tet _1 (z| \tau) \tet _2 (z|\tau ) / \tet _4 (0|2 \tau)
\nonumber \\
\tet _2 (2z | 2 \tau) & = &
 (\tet _3 (z| \tau )^2 - \tet _4 (z|\tau )^2) / \tet _2 (0|2 \tau)
\nonumber \\
\tet _3 (2z | 2 \tau) & = &
(\tet _3 (z|\tau )^2 + \tet _4 ( z|\tau )^2) / \tet _3 (0|2 \tau)
\nonumber \\
\tet _4 (2z | 2 \tau) & = &
\tet _3 (z|\tau ) \tet _4 (z|\tau ) / \tet _4 (0|2 \tau)
\eea
The doubling formulas for the $\tet$-constants are as follows,
\bea
2 \tet _2 (0|2 \tau)^2 & = & \tet _3 (0|\tau)^2 - \tet _4 (0|\tau)^2
\nonumber \\
2 \tet _3 (0|2 \tau)^2 & = & \tet _3 (0|\tau)^2 + \tet _4 (0|\tau)^2
\nonumber \\
 \tet _4 (0|2 \tau)^2 & = & \tet _3 (0|\tau) \tet _4 (0|\tau)
\eea
We have the following derivation formula,
\bea
{\p \over \p z} \ln {\tet _1 (z|\tau)  \over \tet  [\mu](z|\tau )}
= - \pi  \tet  [\mu] ^2 (0|\tau )
{\tet  [\mu ' ] (z|\tau ) \tet  [\mu ''] (z|\tau ) \over
\tet _1 (z|\tau ) \tet  [\mu](z|\tau )}
\eea
This formula is valid for any even spin structure $\mu$; the spin structures
$\mu '$ and $\mu ''$ above are such that the triple $(\mu, \mu' , \mu'')$
is a permutation of $(\mu_2,\mu_3,\mu_4)$.

\section{Appendix : Genus 2 Riemann Surfaces}
\setcounter{equation}{0}

In this appendix, we review from \cite{IV} some fundamental facts about
genus 2 Riemann surfaces, their spin structures, $\tet$-functions
modular properties, and relations between the
$\tet$-function and the  hyperelliptic formulations.

\subsection{Spin Structures}

On a Riemann surface $\Sigma$ of genus 2, there are 16 spin structures, of
which 6 are odd (usually denoted by $\nu$) and 10 are even
(usually denoted by $\delta$). Each spin structure $\kappa$ can be
identified with a $\tet$-characteristic $\kappa = (\kappa '|\kappa'')$,
where $\kappa ', \kappa '' \in \{0,\half\}^2$, represented here by {\sl
column matrices}. The parity of the spin structure $\kappa$ is that of the
integer $4\kappa' \cdot \kappa''$. The signature assignment between 
(even or odd)
spin structures $\kappa$ and $\lambda$ is defined by
\be
\label{signature}
\< \kappa | \lambda \> \equiv \exp \{4 \pi i (\kappa ' \lambda '' -
\kappa '' \lambda ') \}
\ee
\begin{itemize}
\item If $\kappa _1$ and $\kappa _2$ have the {\sl same parity} (resp.
{\sl opposite parity}), we have
\bea
\<\kappa _1|\kappa _2\> = +1 &\Leftrightarrow & \kappa _1 - \kappa _2
\quad {\rm even \quad (resp. \ odd)}
\nonumber \\
\<\kappa _1|\kappa _2\> = -1 &\Leftrightarrow & \kappa _1 - \kappa _2
\quad {\rm odd \quad (resp. \ even)}
\eea

\item If $\nu_1$, $\nu_2$ and $\nu_3$ are odd and all distinct, then
\be
\<\nu _1|\nu_2\> \<\nu _2|\nu_3\> \<\nu _3|\nu_1\> = -1
\ee
\end{itemize}

It will be convenient to use a definite basis for the spin structures,
in a given choice of homology basis, as in \cite{IV}.
The odd spin structures may be labeled as follows,
\bea
\label{listodd}
2\nu_1 =\left (\matrix{0 \cr  1\cr} \bigg | \matrix{0 \cr 1\cr} \right )
\qquad
2\nu_3 =\left (\matrix{0 \cr  1\cr} \bigg | \matrix{1 \cr 1\cr} \right )
\qquad
2\nu_5 =\left (\matrix{1 \cr  1\cr} \bigg | \matrix{0 \cr 1\cr} \right )
\nonumber \\
2\nu_2 =\left (\matrix{1 \cr  0\cr} \bigg | \matrix{1 \cr 0\cr} \right )
\qquad
2\nu_4 =\left (\matrix{1 \cr  0\cr} \bigg | \matrix{1 \cr 1\cr} \right )
\qquad
2\nu_6 =\left (\matrix{1 \cr  1\cr} \bigg | \matrix{1 \cr 0\cr} \right )
\eea
The pairs for which $\<\nu_i|\nu_j\>=-1$ are 14, 16, 23, 25, 35, 46; all
others give $\<\nu_i|\nu_j\>=+1$.
The even spin structures may be labeled by,
\bea
\label{listeven}
2\delta_1 =\left (\matrix{0 \cr 0\cr} \bigg | \matrix{0 \cr 0\cr} \right )
\qquad
2\delta_2 =\left (\matrix{0 \cr 0\cr} \bigg | \matrix{0 \cr 1\cr} \right )
\qquad
2\delta_3 =\left (\matrix{0 \cr 0\cr} \bigg | \matrix{1 \cr 0\cr} \right )
\qquad
2\delta_4 =\left (\matrix{0 \cr 0\cr} \bigg | \matrix{1 \cr 1\cr} \right )
\nonumber \\
2\delta_5 =\left (\matrix{0 \cr 1\cr} \bigg | \matrix{0 \cr 0\cr} \right )
\qquad
2\delta_6 =\left (\matrix{0 \cr 1\cr} \bigg | \matrix{1 \cr 0\cr} \right )
\qquad
2\delta_7 =\left (\matrix{1 \cr 0\cr} \bigg | \matrix{0 \cr 0\cr} \right )
\qquad
2\delta_8 =\left (\matrix{1 \cr 0\cr} \bigg | \matrix{0 \cr 1\cr} \right )
\nonumber \\
2\delta_9 =\left (\matrix{1 \cr 1\cr} \bigg | \matrix{0 \cr 0\cr} \right )
\qquad
2\delta_0 =\left (\matrix{1\cr 1\cr} \bigg | \matrix{1\cr 1\cr} \right)
\eea
The pairs for which $\<\delta _i|\delta _j\>=-1$ are 25, 26, 29, 20, 37,
38, 39, 30, 45, 46, 47, 48, 58, 50, 67, 69, 70, 89; all others give $+1$.

\medskip

For genus 2 there exist many relations between even and odd spin
structures, some of which will be needed here.
First, the sum of all odd spin structures is a double period,
\be
\label{doubleperiod}
\nu _1 + \nu _2 + \nu _3  + \nu _4 + \nu _5 + \nu _6  = 4 \delta _0\, .
\ee
Second, each even spin structure $\delta$ can be written as $\delta =
\nu_{i_1}+\nu_{i_2}+\nu_{i_3}$ (modulo integral periods),
where the $\nu_{i_a}$, $a=1,2,3$ are odd
and pairwise distinct,
\bea
\label{evenodd}
\nu _1 + \nu _2 + \nu _3  & = & \delta _7  + 2 \nu _3
\qquad \qquad
\nu _1 + \nu _2 + \nu _4   =  \delta _5  + 2 \nu _4
\nonumber \\
\nu _1 + \nu _2 + \nu _5  & = & \delta _3  + 2 \nu _5
\qquad \qquad
\nu _1 + \nu _2 + \nu _6   =  \delta _2  + 2 \nu _6
\nonumber \\
\nu _1 + \nu _3 + \nu _4  & = & \delta _8  + 2 \nu _3
\qquad \qquad
\nu _1 + \nu _3 + \nu _5   =  \delta _0  + 2 \nu _1
\nonumber \\
\nu _1 + \nu _3 + \nu _6  & = & \delta _9  + 2 \nu _3
\qquad \qquad
\nu _1 + \nu _4 + \nu _5   =  \delta _4  + 2 \nu _5
\nonumber \\
\nu _1 + \nu _4 + \nu _6  & = & \delta _1  + 2 \delta _0
\qquad \qquad
\nu _1 + \nu _5 + \nu _6   =  \delta _6  + 2 \nu _5
\eea
Clearly, the mapping $\{\nu_{i_1},\nu_{i_2},\nu_{i_3}\} \to
\delta$ is 2 to 1, with $\nu_{i_1}+\nu_{i_2}+\nu_{i_3}$ and its complement
$4 \delta _0  - ( \nu_{i_1}+\nu_{i_2}+\nu_{i_3})$
corresponding to the same even spin structure, in view of
(\ref{doubleperiod}).

\subsection{$\tet$-functions}

The $\tet$-function is an entire function in the period matrix $\Omega$
and $\zeta \in {\bf C}^2$, defined by
\be
\tet [\kappa ] (\zeta, \Omega)
\equiv
\sum _{n \in {\bf Z}^2} \exp \{\pi i (n+\kappa ')\Omega (n+\kappa')
+ 2\pi i (n+\kappa ') (\zeta + \kappa '') \}
\ee
Here, $\tet$ is even or odd in $\zeta$ depending on the parity of the spin
structure. The following useful periodicity relations hold, in which
$m',m'' \in {\bf Z}^2$ and  $\lambda ', \lambda '' \in {\bf C}^2$,
\bea
\label{tetrelations}
\tet [\kappa ] (\zeta + m'' + \Omega m', \Omega )
&=&
\tet [\kappa ](\zeta , \Omega) \ \exp \{ -i \pi m' \Omega m'
- 2 \pi i m' (\zeta +\kappa '') + 2 \pi i \kappa ' m'' \}
\nonumber \\
\tet [\kappa ' +m', \kappa '' +m'' ] (\zeta , \Omega)
&=&
\tet [\kappa ',\kappa ''] (\zeta, \Omega) \ \exp \{ 2\pi i \kappa ' m'' \}
 \\
\tet [\kappa + \lambda ] (\zeta , \Omega )
&=&
\tet [\kappa ](\zeta + \lambda '' + \Omega \lambda' , \Omega) \ \exp \{ i
\pi \lambda ' \Omega \lambda ' + 2 \pi i \lambda ' (\zeta +\lambda '' +
\kappa '')  \}
\no
\eea
The $\tet$-function with vanishing characteristic is often denoted by
$\tet (\zeta, \Omega)= \tet[0] (\zeta, \Omega)$.
For each odd spin structure $\nu$ we have $\tet [\nu ](0,\Omega)=0$. For
each even spin structure $\delta$ one defines the particularly important
$\tet$-constants, $\tet[\delta]  \equiv  \tet [\delta] (0 , \Omega )$.
For every odd spin structure, there exists a Riemann relation for
$\tet$-constants,
\be
\sum _\delta \< \nu |\delta \> \tet [\delta ]^4=0
\ee

\subsection{The Action of Modular Transformations}

Modular transformations $M$ form the infinite discrete group $Sp(4,{\bf
Z})$, defined by
\be
M=\left ( \matrix{A & B \cr C & D \cr} \right )
\qquad \qquad
\left ( \matrix{A & B \cr C & D \cr} \right )
\left ( \matrix{0 & I \cr -I & 0 \cr} \right )
\left ( \matrix{A & B \cr C & D \cr} \right ) ^T
=
\left ( \matrix{0 & I \cr -I & 0 \cr} \right )
\ee
where $A,B,C,D$ are integer valued $2 \times 2$ matrices and the
superscript ${}^t$ denotes transposition.
To exhibit the action of the modular group on 1/2 characteristics,
it is convenient to assemble the 1/2
characteristics into a single column of 4 entries.
In this notation, the action of the
modular group on spin structures $\kappa$ and twists $\ep$
is then given by \cite{igusa}
\bea
\label{modspin}
\left (\matrix{ \tilde \kappa' \cr \tilde \kappa ''\cr}  \right )
& = &
\left ( \matrix{D & -C \cr -B & A \cr} \right )
\left ( \matrix{ \kappa ' \cr \kappa '' \cr} \right )
+ \half \ {\rm diag}
\left ( \matrix{CD^T  \cr AB^T \cr} \right )
\no \\
\left (\matrix{ \tilde \ep'  \cr \tilde \ep ''\cr}  \right )
& = &
\left ( \matrix{D & -C \cr -B & A \cr} \right )
\left ( \matrix{ \ep ' \cr \ep '' \cr} \right )
\eea
Here and below, ${\rm diag} (M)$ of a $n \times n$ matrix $M$ is an
$1\times n$ column vector whose entries are the diagonal entries on $M$.
On the period matrix, modular transformations act by
$\tilde \Omega = (A\Omega + B ) (C\Omega + D)^{-1}$, while on the
Jacobi $\tet$-functions,  we have
\bea
\tet [\tilde \kappa ] \biggl ( \{(C\Omega +D)^{-1} \}^T  \zeta , \tilde
\Omega \biggr ) =
\epsilon (\kappa, M) \det (C\Omega + D) ^\half
e^{ i \pi \zeta  (C\Omega +D)^{-1} C \zeta }
\tet [ \kappa ] (\zeta, \Omega)
\quad
\eea
where $\kappa = (\kappa ' |\kappa '')$ and $\tilde \kappa = (\tilde
\kappa ' | \tilde \kappa '')$. The phase factor $\epsilon (\kappa, M)$
depends upon both $\kappa $ and the modular transformation $M$ and obeys
$\epsilon (\kappa , M )^8=1$. Its expression was calculated in \cite{igusa},
and is given by 
$\epsilon (\delta, M) = \epsilon_0 (M) \exp \{ 2 \pi i \phi _\delta (M)\}$,
where
\bea
\epsilon _0 (M)^2 & = & \exp \{ 2 \pi i {1 \over 8} \tr (M-I) \}
\\
\phi _\delta (M) & = & - \half \delta ' D^T B \delta '
+ \delta ' B^T C \delta '' - \half \delta '' C^T A \delta ''
+ \half (\delta ' D^T - \delta '' C^T) {\rm diag} (AB^T)
\no
\eea
The modular group is generated by the
following elements
\bea
M_i &=& \left ( \matrix{I & B_i \cr 0 & I \cr} \right )
\qquad \ \ \
B_1 = \left ( \matrix{1 & 0 \cr 0 & 0 \cr} \right )
\quad
B_2 = \left ( \matrix{0 & 0 \cr 0 & 1 \cr} \right )
\quad
B_3 = \left ( \matrix{0 & 1 \cr 1 & 0 \cr} \right )
\nonumber \\
S &=& \left ( \matrix{0 & I \cr -I & 0 \cr} \right )
\\
\Sigma &=& \left ( \matrix{\sigma & 0 \cr 0 & -\sigma \cr} \right )
\qquad \ \ \
\sigma = \left ( \matrix{0 & 1 \cr -1 & 0 \cr} \right )
\nonumber \\
T &=& \left ( \matrix{\tau _+ & 0 \cr 0 & \tau _- \cr} \right )
\qquad \quad
\tau _+ = \left ( \matrix{1 & 1 \cr 0 & 1 \cr} \right )
\quad
\tau _- = \left ( \matrix{1 & 0 \cr -1 & 1 \cr} \right )
\nonumber
\eea
The transformation laws for even spin structures under these generators
are given in Table \ref{table:7}; those for odd spin structures will not be
needed here and may be found in \cite{IV}.

\medskip

We shall be most interested in the modular transformations of
$\tet$-constants $\tet ^2 [\delta]$ and thus in even spin structures
$\delta$ and the squares of $\epsilon$, which are given by
\bea
\label{eps}
\left \{ \matrix{
\epsilon (\delta , M_1) ^2 =
\exp \{ 2 \pi i \delta _1' (1 - \delta _1') \}
\cr
\epsilon (\delta , M_2) ^2 =
\exp \{ 2 \pi i \delta _2' (1 - \delta _2') \}
\cr
\epsilon (\delta , M_3) ^2 =
\exp \{ -4 \pi i \delta _1 ' \delta _2' \}
\cr} \right .
\hskip 1in
\left \{ \matrix{
\epsilon (\delta , S) ^2 = - 1
\cr
\epsilon (\delta , \Sigma) ^2 = -1
\cr
\epsilon (\delta , T) ^2 = + 1\cr} \right .
\eea
A convenient way of establishing these values is by first analyzing the
case of the shifts $M_i$, whose action may be read off from the
definition of the  $\tet$-function.
The values for the transformations $S$, $\Sigma$ and $T$ may be obtained
 by letting the
surface undergo a separating degeneration $\Omega _{12} \to 0$, and
using the sign assignments of genus 1 $\tet$-functions.
 The non-trivial entries for $\epsilon ^2$ are listed in Table \ref{table:7}.

\begin{table}[htb]
\begin{center}
\begin{tabular}{|c||c|c|c|c|c|c||c|c|c|} \hline
 $\delta$ & $M_1$  & $M_2$ & $M_3$ & $S$ & $\Sigma$ &
$T$ & $\epsilon ^2 (\delta, M_1)$ & $\epsilon ^2(\delta, M_2)$ & 
$\epsilon ^2(\delta, M_3)$
                \\ \hline \hline
             $\delta _1$
            & $\delta _3$
            & $\delta _2$
            & $\delta _1$
            & $\delta _1$
            & $\delta _1$
            & $\delta _1$ & $1$ & $1$ & $ 1$
 \\ \hline
             $\delta _2$
            & $\delta _4$
            & $\delta _1$
            & $\delta _2$
            & $\delta _5$
            & $\delta _3$
            & $\delta _4$ & $1$ & $1$& $ 1$
 \\ \hline
             $\delta _3$
            & $\delta _1$
            & $\delta _4$
            & $\delta _3$
            & $\delta _7$
            & $\delta _2$
            & $\delta _3$ & $1$ & $1$ & $ 1$
 \\ \hline
             $\delta _4$
            & $\delta _2$
            & $\delta _3$
            & $\delta _4$
            & $\delta _9$
            & $\delta _4$
            & $\delta _2$ & $1$ & $1$ & $ 1$
 \\ \hline
             $\delta _5$
            & $\delta _6$
            & $\delta _6$
            & $\delta _6$
            & $\delta _2$
            & $\delta _7$
            & $\delta _5$ & $1$ & $i $ & $ 1$
 \\ \hline
              $\delta _6$
            & $\delta _5$
            & $\delta _6$
            & $\delta _5$
            & $\delta _8$
            & $\delta _8$
            & $\delta _6$ & $1$ & $i$ & $ 1$
 \\ \hline
                $\delta _7$
            & $\delta _7$
            & $\delta _8$
            & $\delta _8$
            & $\delta _3$
            & $\delta _5$
            & $\delta _9$ & $i$ & $1$ & $ 1$
 \\ \hline
               $\delta _8$
            & $\delta _8$
            & $\delta _7$
            & $\delta _7$
            & $\delta _6$
            & $\delta _6$
            & $\delta _0$ & $i$ & $1$ & $ 1$
 \\ \hline
                $\delta _9$
            & $\delta _9$
            & $\delta _9$
            & $\delta _0$
            & $\delta _4$
            & $\delta _9$
            & $\delta _7$ & $i$ & $i$ & $ -1$
 \\ \hline
              $\delta _0$
            & $\delta _9$
            & $\delta _0$
            & $\delta _9$
            & $\delta _0$
            & $\delta _0$
            & $\delta _8$ & $i $ & $i$  & $ -1$
 \\ \hline
\end{tabular}
\end{center}
\caption{Modular transformations of even spin structures }
\label{table:7}
\end{table}

\subsection{The hyperelliptic representation}

Every genus 2 surface admits a hyperelliptic representation, given by a
double cover of the complex plane with three quadratic branch cuts
supported by 6 branch points, which we shall denote $u_i$, $i=1,\cdots,6$.
The full surface $\Sigma$ is obtained by gluing together two copies of
${\bf C}$ along, for example, the cuts from $u_{2j-1}$ to $u_{2j}$,
$1\leq j\leq 3$.   The surface is then parametrized by\footnote{It is
customary to introduce a local coordinate system $z(x)=(x,s(x))$, which is
well-defined also at the branch points. Throughout, the formulas in the
hyperelliptic representation will be understood in this way. However, to
simplify notation, the local coordinate $z(x)$ will not be exhibited
explicitly.}
\be
\label{hyperelliptic}
s^2 = \prod _{a=1} ^6 (x-p_a)
\ee
In the hyperelliptic representation, there is another
convenient way of identifying spin structures. Each spin structure can be
viewed then as a partition of the set of branch points $p_a$, $a=1,\cdots
,6$ into two disjoint subsets, in the following way.
\bea
\label{partition}
\nu \ {\rm odd } & \Leftrightarrow & {\rm branch \
point } \ p_a
\\
\delta \ {\rm even } & \Leftrightarrow & {\rm
partition} \ A \cup B, \qquad
A = \left \{ p_a , p_b, p_c \right \}, \
B= \left \{ p_d , p_e, p_e \right \}
\nonumber
\eea
where $(a, b,c,d,e,f )$ is a permutation of
$(1,2,3,4,5,6)$.

\subsection{Thomae-type formulas via $\M_{ab}$}

As shown in \cite{IV}, Thomae relations may be derived most easily in
terms of the {\sl bilinear $\tet$-constants} $\M_{ab} \equiv \M _{\nu _a
\nu _b} $ which were defined in \cite{IV},
\be
\M _{ab}
\equiv
\p _1 \tet [\nu _a] (0, \Omega) \p _2 \tet [\nu _b](0, \Omega ) -
\p _2 \tet [\nu _a] (0, \Omega) \p _1 \tet [\nu _b](0, \Omega)
\ee
The abbreviation $\p_I \tet [\nu]\equiv \p_I \tet [\nu]
(0,\Omega)$ will be convenient. The holomorphic 1-form
$\omega _{\nu _a}(z)$, defined for any odd spin structure $\nu_a$ by
$\omega _{\nu _a} (z) \equiv \omega _I (z) \p _I \tet [\nu_a ] $,
has a unique double zero at $p_a$, which is the branch point that may be
canonically associated with $\nu_a$. The following fundamental results
may be derived from a comparison of holomorphic forms in the
$\tet$-functions and hyperelliptic formulations,
\be
\label{Mratio}
{\N _{\nu _b} (p_a - p_b) \over \N _{\nu _c}
(p_a - p_c)}
={\omega _{\nu _b} (p_a) \over \omega _{\nu_c} (p_a)}
=
{\M _{\nu_a \nu _b} \over \M _{\nu _a \nu _c}}
\ee
where $\N_{\nu_a}$ depends only on $a$ and not on $b$.
Taking the cross ratio of four branch points (with $a,d \not=b,c$), the
normalization factors $\N$ cancel out and we get the desired
identity
\be
\label{thomaeM}
{p_a - p_b \over p_a - p_c}
\cdot
{p_c - p_d \over p_b - p_d}
=
{\M _{\nu _a \nu _b} \M _{\nu _c \nu _d} \over
 \M _{\nu _a \nu _c} \M _{\nu _b \nu _d}}
\ee
This is a Thomae-type formula, relating $\tet$-constants
to rational expressions of branch points.

\subsection{$\M_{ab}$ in terms of $\tet$-constants}

In \cite{IV}, it was shown that the following relation holds between
$\M_{ab}$ and the $\tet$-constants for even spin structure,
\bea
\M _{ab} =  m_{ab} \ \pi^2 \prod _{c\not= a,b} \tet [\nu_a + \nu_b +
\nu_c]
\eea
where $m_{ab}^2=1$. The sign factor $m_{ab}$ is not intrinsic, for two
reasons. First, $\M_{ab}$ is odd under $a\leftrightarrow b$, while the
product of $\tet$-constants on the rhs is even. Second, the
$\tet$-constants on the lhs and on the rhs arise to the first power, so
that $\M_{ab}$ is actually sensitive to shifts in the $\nu$'s by single
periods (although they are invariant under shifts by any double period).
Here, the convention will be taken that the odd spin structures $\nu$ are
normalized as given in (\ref{listodd}), while the even spin structures
$\nu _a + \nu_b + \nu_c$ are {\sl truncated to the standard normalization}
given in (\ref{listeven}),
\bea
\nu _a + \nu_b + \nu_c \equiv \delta _i, \qquad i=0,1,\cdots 9
\eea
In the calculation of $\Gamma [\delta; \ep]$ above, the signs are in fact
needed. The sign factors are then given as follows,
\bea
\label{littlem}
m_{ab} & = & - m_{ba}
\\
m_{ab} & = & +1 \qquad ab = 15, ~ 23, ~25, ~26, ~35, ~45, ~ 46
\no \\
m_{ab} & = & -1 \qquad ab = 12, ~ 13, ~14, ~16, ~24, ~34, ~ 36, ~56
\no
\eea
The derivation was given in \cite{IV}. To obtain the results in the above
form, one proceeds as follows. Let $\nu_0$ be the unique genus 1 odd spin
structure and let $\mu_1,\mu_3$ and $\mu_5$ (and denote an independent
set by $\mu_2,\mu_4$ and $\mu_6$) denote the three distinct genus 1 even
spin structures, obeying $\mu_1+\mu_3+\mu_5=\nu_0$, then any pair of
genus 2 odd spin structures may be expressed in the following basis,
\bea
\nu _1 = \left [ \matrix{\mu_1 \cr \nu _0 \cr } \right ]
\qquad
\nu _2 = \left [ \matrix{\nu_0 \cr \mu _2 \cr } \right ]
\qquad
\nu _3 = \left [ \matrix{\mu_3 \cr \nu_0 \cr } \right ]
\eea
The result of \cite{IV} is then,
\bea
\M_{\nu_1 \nu_2} & = & - \pi ^2 ~
\tet  \left [ \matrix{\mu_3 \cr \mu _2 \cr } \right ]
\tet  \left [ \matrix{\mu_5 \cr \mu _2 \cr } \right ]
\tet  \left [ \matrix{\mu_1 \cr \mu _4 \cr } \right ]
\tet  \left [ \matrix{\mu_1 \cr \mu _6 \cr } \right ]
 \\
\quad
\M_{\nu_1 \nu_3} & = & - \pi ^2 ~ \sigma (\mu_1,\mu_3) ~
\tet  \left [ \matrix{\nu_0 \cr \nu_0 \cr } \right ]
\tet  \left [ \matrix{\mu_5 \cr \mu _2 \cr } \right ]
\tet  \left [ \matrix{\mu_5 \cr \mu _4 \cr } \right ]
\tet  \left [ \matrix{\mu_5 \cr \mu _6 \cr } \right ]
\no
\eea
with the sign factor $\sigma(\mu_1,\mu_3) = - \sigma (\mu_3,\mu_1)$ given as
follows,
\bea
\sigma ((0|0),(0|\12)) = \sigma ((\12| 0),(0|0)) = \sigma ((\12| 0),(0|\12))=+1
\eea
Notice that the ordering of $\mu_1$ and $\mu_2$ determines the sign on the
right hand side. Going through all cases, (\ref{littlem}) is readily
derived.


\end{document}